\documentclass[useAMS,usenatbib]{mn2e}
\usepackage[english]{babel}

\usepackage{fancyhdr}
\usepackage{amsfonts}
\usepackage{amsmath}
\usepackage{amssymb}
\usepackage[colorlinks,linkcolor=blue,citecolor=blue,urlcolor=blue,pdftitle={Means of confusion: how pixel noise affects shear estimates for weak gravitational lensing},pdfauthor={Viola}]{hyperref}
\usepackage{multicol}
\usepackage{layout}
\usepackage{graphicx}
\usepackage{subfigure}

\usepackage{times}
\usepackage{natbib}

\newif\ifAMStwofonts
\AMStwofontstrue

\renewcommand{\vec}[1]{\bmath{#1}}
\newcommand{\be}{\begin{equation}}
\newcommand{\ee}{\end{equation}}
\newcommand{\ba}{\begin{eqnarray}}
\newcommand{\ea}{\end{eqnarray}}
\newcommand{\brr}{\begin{array}}
\newcommand{\err}{\end{array}}
\newcommand{\bc}{\begin{center}}
\newcommand{\ec}{\end{center}}

\newcommand{\mincir}{\raise
  -2.truept\hbox{\rlap{\hbox{$\sim$}}\raise5.truept \hbox{$<$}\ }}
\newcommand{\magcir}{\raise
  -2.truept\hbox{\rlap{\hbox{$\sim$}}\raise5.truept \hbox{$>$}\ }}
\newcommand{\siml}{\raise
  -2.truept\hbox{\rlap{\hbox{$\sim$}}\raise5.truept \hbox{$<$}\ }}
\newcommand{\simg}{\raise
  -2.truept\hbox{\rlap{\hbox{$\sim$}}\raise5.truept \hbox{$>$}\ }}

\newcommand{\eq}[1]{\begin{equation}  #1 \end{equation}}

\newcommand{\eqa}[1]{\begin{eqnarray}   #1 \end{eqnarray}}

\newcommand{\br}[1]{\left( #1 \right)}
\newcommand{\bbc}[1]{\left\{ #1 \right\}}
\newcommand{\bba}[1]{\left\langle #1 \right\rangle}
\newcommand{\bb}[1]{\left[ #1 \right]}

\newcommand{\nn}{\nonumber\\}

\newcommand{\dd}{{\rm d}}
\newcommand{\expo}[1]{~{\rm e}^{ #1 }}
\newcommand{\vek}[1]{\mbox{\boldmath $#1$}}
\newcommand{\ic}{{\rm i}}

\newcommand{\cov}[2]{{\rm Cov}\left[ #1; #2 \right]}

\newcommand {\apgt} {\ {\raise-.5ex\hbox{$\buildrel>\over\sim$}}\ }
\newcommand {\aplt} {\ {\raise-.5ex\hbox{$\buildrel<\over\sim$}}\ }

\title[Probability Distributions of Ellipticity]{On the Probability Distributions of Ellipticity}
\author[Viola, Kitching, Joachimi] 
{M. Viola$^{1}$ \thanks{viola@strw.leidenuniv.nl}, T. D. Kitching$^{2}$ \thanks{t.kitching@ucl.ac.uk}, B. Joachimi$^{3,4}$\thanks{b.joachimi@ucl.ac.uk}\\
  $^1$Leiden Observatory, Leiden University, Niels Bohrweg 2, 2333 CA Leiden, The Netherlands \\
  $^2$University College London, Mullard Space Science Laboratory, Holmbury St. Mary, Dorking, Surrey, RH5 6NT, UK\\
  $^3$Scottish Universities Physics Alliance, Institute for Astronomy, University of Edinburgh, Royal Observatory, Blackford Hill, Edinburgh, EH9 3HJ, UK \\
  $^4$Department of Physics \& Astronomy, University College London, Gower Place, London WC1E 6BT, UK.
}

\pubyear{2013}

\begin{document}
\label{firstpage}
\maketitle

\begin{abstract}
In this paper we derive an exact full expression for the 2D probability distribution of the ellipticity of an object measured from data, only assuming Gaussian noise in pixel values. This is a generalisation of the probability distribution for the ratio of single random variables, that is well-known, to the multivariate case. This expression is derived within the context of the measurement of weak gravitational lensing from noisy galaxy images. 
We find that the third flattening, or $\epsilon$-ellipticity, has a biased maximum likelihood but an unbiased mean; and that the third eccentricity, or normalised polarisation  $\chi$, has both a biased maximum likelihood and a biased mean. 
The very fact that the bias in the ellipticity is itself a function of the ellipticity requires an accurate knowledge of the intrinsic ellipticity distribution of the galaxies in order to properly calibrate shear measurements.
We use this expression to explore strategies for calibration of biases caused by measurement processes in weak gravitational lensing. We find that upcoming weak lensing surveys like KiDS or DES require calibration fields of order of several square degrees and 1.2 magnitude deeper than the wide survey in order to correct for the noise bias. Future surveys like Euclid will require calibration fields of order 40 square degree and several magnitude deeper than the wide survey.
We also investigate the use of the Stokes parameters to estimate the shear as an alternative to the ellipticity. We find that they can provide unbiased shear estimates at the cost of a very large variance in the measurement. 
The python code used to compute the distributions presented in the paper and to perform the numerical calculations are available on request.
\end{abstract}

\begin{keywords}
Cosmology: theory; Cosmology: dark matter; Physical data and processes: gravitational lensing; Methods: analytical
\end{keywords}

\section{Introduction}
Weak gravitational lensing has become a  powerful and standard tool to probe the formation of structures in the 
Universe \citep[e.g][]{Schrabback10, Kilbinger13} and properties of the gravitational field of massive structures 
such as galaxies or galaxy clusters \citep[e.g.][]{Hoekstra13}. Moreover, because the correlation of the weak lensing 
signal depends on both the geometry of the Universe and the growth of structure, it has the power to test 
the acceleration of the expansion history of the Universe, thereby shedding light on the nature of dark energy \citep[e.g.][]{Bartelmann01}. 

The weak lensing effect generates spin-2 distortions (that have a 180 degree symmetry), referred to as `shear', on the observable shape of distant 
galaxies induced by some intervening gravitational tidal field. Because galaxies are not intrinsically spherical in nature, but in general 
are elliptical, the gravitationally induced spin-2 distortion cannot be disentangled from the intrinsic ellipticity of a galaxy on an 
object-by-object basis. Hence statistics over spin-2 distortion measurements from a large number of galaxies have to be taken in order 
to isolate the averaged distortion induced by the gravitational tidal field, or the higher-order moments of the gravitational tidal field. 
Weak lensing therefore requires a large number of galaxies to constrain properties of dark matter and dark energy, in order to overcome the 
shape noise introduced in any statistic used, caused by the intrinsically elliptical nature of galaxies. 

Traditionally the spin-2 distortion in the light distribution of distant galaxies is measured in terms of a 
galaxy `ellipticity'. This is a very good estimator of the shear field, and in the limit that a galaxy is measured at 
infinite signal-to-noise it can be proved that the average ellipticity of many galaxies in a given area of the sky is an 
unbiased estimator for the shear in that particular region -- assuming that the shear is coherent across such a region. 
There are many advantages in using the ellipticity as a proxy for the shear. Firstly it is an `intuitive' quantity, that has 
a simple geometric interpretation on an image, and it is relatively straightforward to measure from the data. 
Secondly it is fully specified by only two numbers, an amplitude and a phase, which are also bounded between 0 and 1, and 0 and $2\pi$, respectively. Both of these properties are desirable mathematically 
and from a perspective of efficient computation.

The ellipticity is always defined as a ratio of two quantities (the polarisation and a measurements of the galaxy size, 
or the semi-major and semi-minor axis of the galaxy to mention just two possibilities) and therefore requires 
some non-linear combination of the image pixels. This leads to come subtleties with the use of ellipticity in any realistic case, 
where noise is present in the image.  
In particular it has been shown by \cite{Melchior12} that the very fact that the ellipticity is defined as a ratio of two correlated noisy quantities 
undermines the possibility of measuring the ellipticity of a galaxy in an unbiased way.
Similar results were derived independently by \cite{Refregier12,Miller13}. 
However `noise bias' has been known for more than a decade in weak-lensing literature. Early works were done by \cite{Kaiser00} 
and \cite{Bernstein02} who briefly discussed the problem and proposed some approximate corrections, 
and by \cite{Hirata04} who obtained an analytic description of the dependence of this bias on the size of the galaxy. 

We remind the reader in this paper that the noise bias is not a novel problem, or peculiar to weak lensing but is  
in fact a well known effect, and has been discussed for over $50$ years in the literature. 
Such an effect on stellar polarisation measurements was 
first discussed and measured in \cite{Serkowski58} in a study of the polarisation and reddening of the double cluster Perseus. Similar effects are well known in the radio astronomy community where a correction scheme proposed by \cite{Wardle74} is commonly used. The distribution function for normalised Stokes parameters was first derived in \cite{Clarke83}, and the probability distribution for a ratio of two random Gaussian distributed variables derived in \cite{Marsaglia65} and \cite{Tin65}. In this paper we generalise the Marsaglia-Tin distribution to the case of two ratios constructed from three, correlated, random variables; the case that is relevant for ellipticity measurements. 

In this paper we explore the probability distribution for ellipticity, and examine the magnitude of any bias in shear measurement that this causes. 
We will explore possible ways to calibrate bias for current and future generations of survey. Moreover we will explore different possible ways 
to measure the spin-2 distortions induced by the gravitational tidal field which do not require a definition of an ellipticity, namely the use 
of the Stokes parameters directly, and we will compare them to the standard approach. 

In Section 2 we outline the problem and explain why, in general, ellipticity measurements are biased in presence of noise in the pixels, in Section 3 we provide an analytical expression for the 2D noisy ellipticity distribution and we investigate its properties. In Section 4 we present a way to propagate requirements on the knowledge of the shear bias into requirements on the knowledge of the intrinsic ellipticity distribution and in Section 5 we discuss possible calibration strategies for current and future surveys. 
In Section 6 we show how the noise bias could be avoided by using the Stokes parameters rather than ellipticity to measure the shear at the price of at higher variance in the final measurement. We conclude in Section 7.

\section{The Problem}

Measurements of the shear field have always relied on measurements of the ellipticity of the galaxies because 
the galaxy ellipticity is a direct tracer of the gravitational tidal field along the line of sight \citep{Bartelmann01}. 
In fact there is a simple relation between the third flattering ($\epsilon$-ellipticity hereafter) defined as $\epsilon=[(a-b)/(a+b)] \exp (2i\phi)$ in terms of the semi-major and semi-major axis, and the shear $g$:

\begin{equation}
\label{eq:shear2el}
\epsilon^{s}=\frac{\epsilon-g}{1-g^{\star}\epsilon}.
\end{equation}
where $\epsilon^s$ is the intrinsic ellipticity of the object and $\epsilon$ the observed one \citep{Seitz97}. This relation is valid for $|g| <1$, but a similar one can be written for the case $|g| > 1$ 
All quantities in the above equation are complex numbers and $g^{\star}$ is the conjugate of $g$. 

A similar relation existst between the third eccentricity (herafter normalised polarisation $\chi$) and the shear \citep{Schneider95}.
Under the assumption that galaxies in the Universe do not have any preferred orientation, 
the shear can then be estimated by averaging over many galaxies in a region where the gravitational tidal field can be considered constant:
\begin{equation}
\label{eq:unbiaseddis}
g=\int \epsilon p(\epsilon)\mathrm{d^2}\epsilon=\int \epsilon p(\epsilon^{s})\mathrm{d^2}\epsilon^{s}
\end{equation}
where $p(\epsilon)$ is the observed $\epsilon$-ellipticity distribution, and we write this in the continuous case. 

At this point a semantic clarification is required. In the rest of the paper we will use the term \textit{ellipticity} and the letter $e$ when generically referring to some spin-2 dimensionless property of the object, otherwise we will make the distiction between $\epsilon$-ellipticity and normalised polarisation $\chi$. A summary of the notation and terminologu used in the paper is provided in Table 1.

\begin{table*}
\caption{Summary of the different ellipticity definitions used throught the paper (and in the lensing literature). In the first column we report the symbol used in the paper, in the second column the name, in the third column the relation with the semi-major axis $a$, semi-minor axis $b$ and orientation $\phi$ of the ellipse and in the fourth column the geometrical name.}
\label{tab:summary}
\begin{tabular}{llll}
\hline
Symbol & Name & Geometrical definition & Geometrical name\\
\hline
$\epsilon$ & $\epsilon$-ellipticity  & $\frac{a-b}{a+b}\exp(2i\phi)$ & third flattering \\[1ex]
$\chi$ & Normalised polarisation & $\frac{a^2-b^2}{a^2+b^2}\exp(2i\phi)$  & third eccentricity \\
\hline
\end{tabular}
\end{table*}
 
If the $\epsilon$-ellipticity of an object is \emph{perfectly} measured (i.e. with zero error) 
then the shear can be perfectly recovered \textit{independently} of the intrinsic ellipticity distribution $p(\epsilon^{s})$ as shown in \citet{Seitz97}. 
In Appendix A, for completeness, we also include a derivation of the full posterior for shear $p(g)$ that equation (\ref{eq:shear2el}) implies. 

However in practice one can never perform a perfect measurement, and many things undermine the ability of perfectly measuring the object's ellipticity such as poor knowledge of the point-spread-function (PSF hereafter) with which the galaxy profile is convolved when observed through a telescope, a wrong determination of the galaxy model or more generally, and inescapably, the presence of noise in the data. Such effects in general cause biases in the measured ellipticity, and hence the shear. The bias in the estimation of the ellipticity of an object is usually parametrised \citep{Heymans06, Massey07b, Bridle09, Kitching12} in terms of a \textit{multiplicative} ($m$) and an \textit{additive} ($c$) bias\footnote{In fact \citep{Heymans06, Massey07b, Bridle09} only used such a relation to parameterise bias in the shear of an object $g$, but more generally it is applied to ellipticity measurements.}:
\begin{equation}
\label{eq:mandc}
e^{obs}\simeq (1+m)e^{s} + c.
\end{equation}
These biases can be propagated into the power spectrum of shear estimates 
\citep{Kitching12} and relate to properties of instrument PSF and detector effects \citep{Massey13}.  
Any bias in the ellipticity translates immediately into a bias in the shear, as is clear from equation (\ref{eq:unbiaseddis}), and 
eventually in a bias in the value of cosmological parameters \citep{Bartelmann12}.

\subsection{The Measurement}

Methods to measure weak lensing are generically and colloquially referred to as `shape measurement' methods or techniques. To date there are two general classes of method for extracting ellipticity information from images, that we refer to as `model-based' and `moment-based'. In general we refer to `moment-based' as those methods that directly measure moments of a galaxy pixel distribution i.e. there is an explicit single-valued function that maps pixels to an estimate of a quadrupole moment. Note that in a statistical sense they are not free (independent) of any model assumption since at a minimum a weight function must be used to regularise the measurement process. The most general definition we define here is that such methods are `many-to-one' in that from many pixel values only the moments are extracted in a direct way. Such algorithms are a mapping from pixel-space to moments-space. Approaches such as KSB \citep{KSB95} and DEIMOS \citep{Melchior10} are examples of such algorithms. 

The alternatives are `model-based' approaches that we refer to as any method that finds the extremum of a loss-function $L(f|\{p\})$ over the pixel values, given a space of alternative functions $\{f\}$ (or models) that may represent the data. For each candidate function there is a value representation in the loss-function i.e. $L(f|\{p\})$ is defined for all $f$. A maximum likelihood fitting routine would be an example of such an approach. 
The most general definition here is that such methods are `many-to-many' in that from the set of pixel values $\{p\}$ 
a set of probabilistic values are extracted via the loss-function, where the number of probability values can be much larger than the number of generating pixels. These algorithms can then be written as a mapping from pixel-space to probability space (given a set of functions $\{f\}$ or models): $\{p\}\rightarrow L$. Lensfit \citep{Miller07, Kitching08} is an example of such an algorithm. 

\subsection{Sources of Bias}

There are in general three sources of bias, with completely different origins. These are model bias, method bias and noise bias:
\begin{itemize}
\item
{\bf Model bias} arises when a wrong galaxy model is used to describe the data \citep[e.g.][]{Voigt10}.  
A trivial example might be using a Gaussian model to describe a model with an exponential profile. 
This bias may be common to fitting techniques, in which the model is explicitly fit to the data, and to moment-based techniques, in which the ellipticity is computed as a suitable combination of the second-order moments of the object surface brightness. In the latter case the model bias is caused by measuring the moments employing a weighting function which is different from the actual object's surface brightness. 
\item
{\bf Method bias} is specific to each shape measurement technique and it comes from particular algorithmic or computational choices and approximations 
made in the implementation. 
A classical example might be the PSF correction performed by the KSB algorithm \citep{KSB95,Hoekstra98,Viola10b}: the correction holds exactly only 
in the case of a circularly symmetric PSF with a small anisotropy; if the real PSF does not fulfil those requirements a bias is expected. 
\item 
{\bf Noise bias} is the the bias that is introduced due to the presence of noise in the data. This is caused by the fact that the parameters 
describing galaxy morphology (such as size and ellipticity) are non-linear quantities in the image pixels, and is present even if the galaxy 
profile is perfectly known \citep{Melchior12, Refregier12, Miller13}. It is important to note that this is present even at high signal-to-noise 
(the noise is never zero) to some degree, hence it is simply a feature of ellipticity measurement \emph{not} a separable bias term \emph{per se}. 
In this paper we show that it is a fundamental feature of the probability distribution expected for any ellipticity measurement.  
\end{itemize}
In practice these three effects act simultaneously and the resulting bias a is a combination of the three. We refer to \cite{Kacprzak13} for a recent investigation of the interaction between model bias and noise bias.

All these three effects can cause both an additive and a multiplicative bias. However in the case the elliptcity of the PSF is perfectly known at the position of galaxies and the detector effects (such as charge transfer inefficency) are perfectly corrected the bias in shear measurements is purely multiplicative \citep{Massey13}. In this work we are not interested in investigating the impact of a poor PSF model on shape measurements nor of detector effects. Hence in the following we will investigate only multiplicative biases.

Different shape measurement techniques might be more or less prone to the model and the noise bias depending on their complexity and 
specific implementation. We refer to \cite{Kitching12} for a recent analysis of systematics errors 
associated with different shape measurement algorithms.

\subsection{Notation}
We start by defining the $i+j$ order moments of the object surface brightness $I(x,y)$:
\begin{equation}
\{Q\}_{i,j}=\int I(x,y)x^{i}y^{j}\mathrm{d}x\mathrm{d}y
\end{equation} 
The zero-th moment $\{Q\}_{00}$ is the object's flux, the first order moments $\{Q\}_{01}$ and $\{Q\}_{10}$ correspond to the object's centroid, while the second-order moments can be used to characterise the object's  
\textit{polarisation}. In particular it is convenient to map the three second order moments into the so-called \textit{Stokes parameters:}
\be 
\left( \begin{array}{c}
u  \\
v  \\
s \end{array} \right)= \underbrace{
\left( \begin{array}{ccc}
1 & -1 & 0 \\
0 & 0 & 2 \\
1 & 1 & 0 \end{array} \right)}_M
\left( \begin{array}{c}
\{Q\}_{20}  \\
\{Q\}_{02}  \\
\{Q\}_{11} \end{array} \right),
\ee
They characterise the polarisation of the light along the x-axis ($u$), along an axis rotated by $\pi/4$ with respect to the x-axis ($v$) and the total intensity of the polarisation ($s$). This later quantity is tightly related to the area of the 2D-surface brightness of the object.

The second-order moments of the light distribution (or analogously the Stokes parameters) 
can also be used in order to define the normalised polarisation $\chi$ and the $\epsilon$-ellipticity of the object:
\begin{subequations}
\label{eq:ellipticities}
\begin{align}
\label{eq:chi}
\bchi &:= \frac{\{Q\}_{20} - \{Q\}_{02} + 2\mathrm{i}\{Q\}_{11}}{\{Q\}_{20} + \{Q\}_{02}} \equiv \frac{u+iv}{s} \ \ \ \text{or}\\
 \label{eq:ellipticity}
\bepsilon &:= \frac{\{Q\}_{20}-\{Q\}_{02}+2\mathrm{i}\{Q\}_{11}}
   {\{Q\}_{20}+\{Q\}_{02}+2\sqrt{\{Q\}_{20}\{Q\}_{02}-\{Q\}_{11}^2}}.
\end{align}
\end{subequations}
The two definitions are related through:
\begin{equation}
\chi=\frac{2\epsilon}{1+|\epsilon |^2}.
\end{equation}
The $\chi$-normalised polarisation has a simple definition in terms of $Q_{ij}$: it is in fact the only dimensionless spin-2 combination of second-order moments which does not require products of moments. Furthermore normalised polarisation $\chi$ can be written as a simple ratio of the Stokes parameters. 
This definition of an object's ellipticity is commonly used in moment-based methods, in particular those that include variations of the KSB algorithm. However it is not an unbiased shear estimator in the sense of equation (\ref{eq:unbiaseddis}) \citep{Schneider95}. \footnote{In fact in order to turn $\chi$ into a shear estimate knowledge of the intrinsic ellipticity distribution is required.}
On the other hand the $\epsilon$-ellipticity requires products of quadrupole moments in its definition, but it is an unbiased shear estimator. This definition of the ellipticity is commonly used by model-based methods such as lensfit \citep{Miller07, Kitching08,Miller13} as well by DEIMOS \citep{Melchior10}. 

We assume in this work that the pixel noise is homoscedastic, that is uncorrelated and Gaussian (e.g. in the case of uniform sky noise):
\begin{equation}
\langle N(x_p,y_p)N(x_q,y_q)\rangle =\sigma^2_{n}\delta_{pq}.
\end{equation}
This means that the noise $N$ in the image's pixels is independently drawn from a Gaussian distribution with variance $\sigma^2_n$ for any two positions $(x_p,y_p)$ and $(x_q,y_q)$.
Hence we can write the observed moments as:
\begin{equation}
\{Q\}^{obs}_{i,j}=\int W(x,y)[I(x,y)+N(x,y)+N_p(x,y)]x^{i}y^{j}\mathrm{d}x\mathrm{d}y
\end{equation} 
where $W(x,y)$ is an arbitrary weighting function which is employed to suppress noise at large scales.
The first term in the square bracket is the true weighted $i+j$ order moment, the second term is contribution to the measurement from the background noise and the third term is the contribution from the Poisson noise (shot-noise) due to the image.
Using the above equation we can compute the moment error: 
\ba
\sigma^2_{ij}=\sigma_n^2 \int W^2(x,y)x^{2i}y^{2j}\mathrm{d}x\mathrm{d}y + \nn
\int I(x,y)W^2(x,y)x^{2i}y^{2j}\mathrm{d}x\mathrm{d}y.
\label{eq:errMom}
\ea
The first term dominates at low signal-to-noise, while the second term, being the expected variance assuming Poisson photon noise, becomes important when the signal-to-noise becomes higher.
In the following calculations we will need to make use of the moments and the Stokes parameter's covariance matrix. 
This has in general a quite complicated form. It is possible however to simplify the calculation by rotating 
first the object by an angle $\theta$ such that $v=0$ after rotation:
\be 
\left( \begin{array}{c}
u^{rot}  \\
v^{rot}  \\
s^{rot} \end{array} \right)=\underbrace{
\left( \begin{array}{ccc}
\cos(\theta) & \sin(\theta) & 0 \\
-\sin(\theta) & \cos(\theta) & 0 \\
0 & 0 & 1 \end{array} \right)}_{R}
\left( \begin{array}{c}
u  \\
v  \\
s \end{array} \right).
\ee
In the rotated frame the moment's covariance matrix can be written as:
\ba
\Sigma_{Q}\equiv \left( \begin{array}{ccc}
\sigma^2_{20} & \sigma_{11}^2 & 0 \\
\sigma_{11}^2 & \sigma^2_{02} &  0 \\
0  & 0 & \sigma^2_{11} \end{array} \right)=\nn
\left( \begin{array}{ccc}
\sigma^2_{20} & \rho^{(1)}\sigma_{20}\sigma_{02} & \rho^{(2)}\sigma_{20}\sigma_{11} \\
\rho^{(1)}\sigma_{02}\sigma_{20} & \sigma^2_{02} &  \rho^{(3)}\sigma_{02}\sigma_{11} \\
\rho^{(2)}\sigma_{11}\sigma_{20}  & \rho^{(3)}\sigma_{11}\sigma_{02}  & \sigma^2_{11} \end{array} \right),
\ea
where $\rho^{i}$ are the correlation coefficients between the second-order moments. From the equation above we can read $\rho^{(2)}=\rho^{(3)}=0$ and:
\be
\rho^{(1)}=\frac{\sigma^2_{11}}{\sigma_{20}\sigma_{02}}
\ee
in the case of Gaussian object and Gaussian weighting function $\rho^{(1)}=1/3$ independently of the size and the ellipticity of the object. Moreover this correlation is also largely unaffected by changes to the radial profile of the source \citep{Melchior12}.


Finally the covariance matrix of the Stokes parameters in the original reference frame can be easily computed as:
\be
\Sigma_{u,v,s}=R^{-1}M\Sigma_{Q}(R^{-1}M)^T.
\label{eq:CovStokes}
\ee
When investigating effects that depend on the noise properties of the data it is of paramount importance to define the signal-to-noise of 
an object clearly. The signal-to-noise of an objects is defined in this paper as the ratio between the observed flux and the error on the observed flux:
\begin{equation}
\label{eq:sn}
\nu=\frac{\int I(\bar{x})W(\bar{x})\mathrm{d}\bar{x}}{\sqrt{\sigma_n^2\int W(\bar{x})^2\mathrm{d}\bar{x}+\int I(\bar{x})W(\bar{x})^2\mathrm{d}\bar{x}}}
\end{equation}
where $\bar{x}=(x,y)$. Typically weak lensing objects are faint and small objects, hence the noise comes mostly from the background. However for the sake of completness we include in the following calculations also the shot-noise contribution.

\section{The probability distributions of ellipticity}
\label{The probability distribution of ellipticity}
We discuss in this section how it is possible to characterise mathematically the distribution of ellipticity measured from data that contain some level of noise. We assume here that the noise is homoscedastic, uncorrelated and Gaussian (those are the same assumptions entering in the signal-to-noise definition in equation \ref{eq:sn}).

\subsection{Probability distribution of the normalised Stokes parameter: The Marsaglia-Tin distribution}

The probability distribution function of the Stokes parameters in presence of noise can be described in terms of a 
trivariate Gaussian with correlation coefficients $\rho_{ij}$ between each of the variables such that the covariance matrix is 
\be 
\label{covsigma}
\Sigma_{u,v,s}\equiv
\left( \begin{array}{ccc}
\sigma_u^2 & \rho_{uv}\sigma_u\sigma_v & \rho_{us}\sigma_u\sigma_s \\
\rho_{uv}\sigma_u\sigma_v & \sigma_v^2 & \rho_{vs}\sigma_v\sigma_s \\
\rho_{us}\sigma_u\sigma_s & \rho_{vs}\sigma_v\sigma_s & \sigma_s^2 \end{array} \right)
\ee
and the mean in each dimension is given by $(\mu_u,\mu_v,\mu_s)$. 
The full expression for the correlated trivariate Gaussian is:
\ba 
\label{triv}
p_{u,v,s}(u,v,s)=\frac{1}{(2\pi)^{\frac{3}{2}}\sigma_u\sigma_v\sigma_s D^{\frac{1}{2}}}\exp \Bigg(-\frac{1}{2D}[a\Delta_u^2+d\Delta_v^2+ \nn
f\Delta_s^2+2b\Delta_u\Delta_s+2c\Delta_u\Delta_v+2e\Delta_v\Delta_s]\Bigg)
\ea
where we define the following variables 
\ba 
\label{vars}
\Delta_x&=&(x-\mu_x);\nn
D&=&1+2\rho_{uv}\rho_{us}\rho_{vs}-\rho_{us}^2-\rho_{vs}^2-\rho_{uv}^2; \nn
a&=&\frac{1-\rho_{vs}^2}{\sigma_u^2};\nn
d&=&\frac{1-\rho_{us}^2}{\sigma_v^2};\nn
f&=&\frac{1-\rho_{uv}^2}{\sigma_s^2};\nn 
b&=&\frac{\rho_{uv}\rho_{vs}-\rho_{us}}{\sigma_u\sigma_s};\nn
c&=&\frac{\rho_{us}\rho_{vs}-\rho_{uv}}{\sigma_s\sigma_v};\nn 
e&=&\frac{\rho_{us}\rho_{uv}-\rho_{vs}}{\sigma_v\sigma_s} 
\ea
the variables $a,\dots,f$ are the components of the inverse covariance, which is symmetric.

The 2-dimensional probability distribution for the normalised polarisation $\chi$ defined as $(u/s, v/s)$ can be 
derived starting from the three-dimensional probability distribution for the Stokes parameters.
First of all we transform the distribution $p_{u,v,s}(u,v,s)$ into $p(\chi_1,\chi_2,s)$ by a change of 
variable and then we marginalise over $s$
\be
\label{mq}
p_{\chi}(\chi_1,\chi_2)=\int_{-\infty}^{\infty} {\rm d}s s^2 p_{u,v,s}(\chi_1s,\chi_2s,s)\;;
\ee
this is the form of a two dimensional quotient distribution. 

The derivation is then a matter of substituting the Gaussian trivariate distribution in equation (\ref{triv}) 
into the multivariate quotient equation (\ref{mq}) and performing the integration over the range $(-\infty,\infty)$.\footnote{In fact $s$ can have  negative values in presence of noise. This can happen for example if the intensity of a pixel is negative.}
This results in: 
\ba
p_{\chi}(\chi_1,\chi_2)=\frac{1}{(2\pi)^{\frac{3}{2}}\sigma_u\sigma_v\sigma_s D^{\frac{1}{2}}}
\left(\frac{2\pi D}{\alpha^5(\chi_1,\chi_2)}\right)^{1/2} \times \nn
{\exp}{\left(\delta+\frac{\phi^2(\chi_1,\chi_2)}{2D\alpha(\chi_1,\chi_2)}\right)}\left[\phi(\chi_1,\chi_2)^2+D\alpha(\chi_1,\chi_2)\right]
\label{eq:MarsagliaTin}
\ea
where we simplify the expression by defining the following variables
\ba
\label{alpha}
\alpha(z,z')&=&az^2+dz'^2+2czz'+2bz+2ez'+f\\
\phi(z,z')&=&(a\mu_u+c\mu_v+b\mu_s)z \nn
&+&(c\mu_u+d\mu_v+e\mu_s)z'\nn
&+&(b\mu_u+e\mu_v+f\mu_s)\\
\delta&=&-\frac{1}{2D}\left(a\mu_u^2+d\mu_v^2+f\mu_s^2\right)\nn
&-&\frac{1}{2D}\left(2c\mu_u\mu_v+2b\mu_u\mu_s+2e\mu_v\mu_s\right)\\
\beta(z,z')&=&\frac{\phi^2}{\alpha D},
\ea
where the other variables are defined in equation (\ref{vars}). 
Note that the $\mu_{i}$ have to be considered as the \textit{weighted} `true' values of the Stokes parameters (those which would be measured if there was 
zero noise). We note that the 1D Marsaglia-Tin distribution was derived by \cite{Marsaglia65} and \cite{Tin65}, and that this 1D expression,   
used in \cite{Melchior12} in the weak lensing context, can be derived from equation (\ref{eq:MarsagliaTin}) 
by a further marginalisation over $\chi_2$.
We remind the reader that in order to derive this pdf we only assume that the pixel noise is Gaussian and that the centroid is known.
We further assume a Gaussian weighting function later in this paper in order to get an analytic expression for the $\sigma_i$ 
themselves. We refer to this expression as the {\bf (multivariate) Marsaglia-Tin distribution} or $\chi$ probability distribution. 
\begin{figure*}
\includegraphics[width=\columnwidth, angle=0]{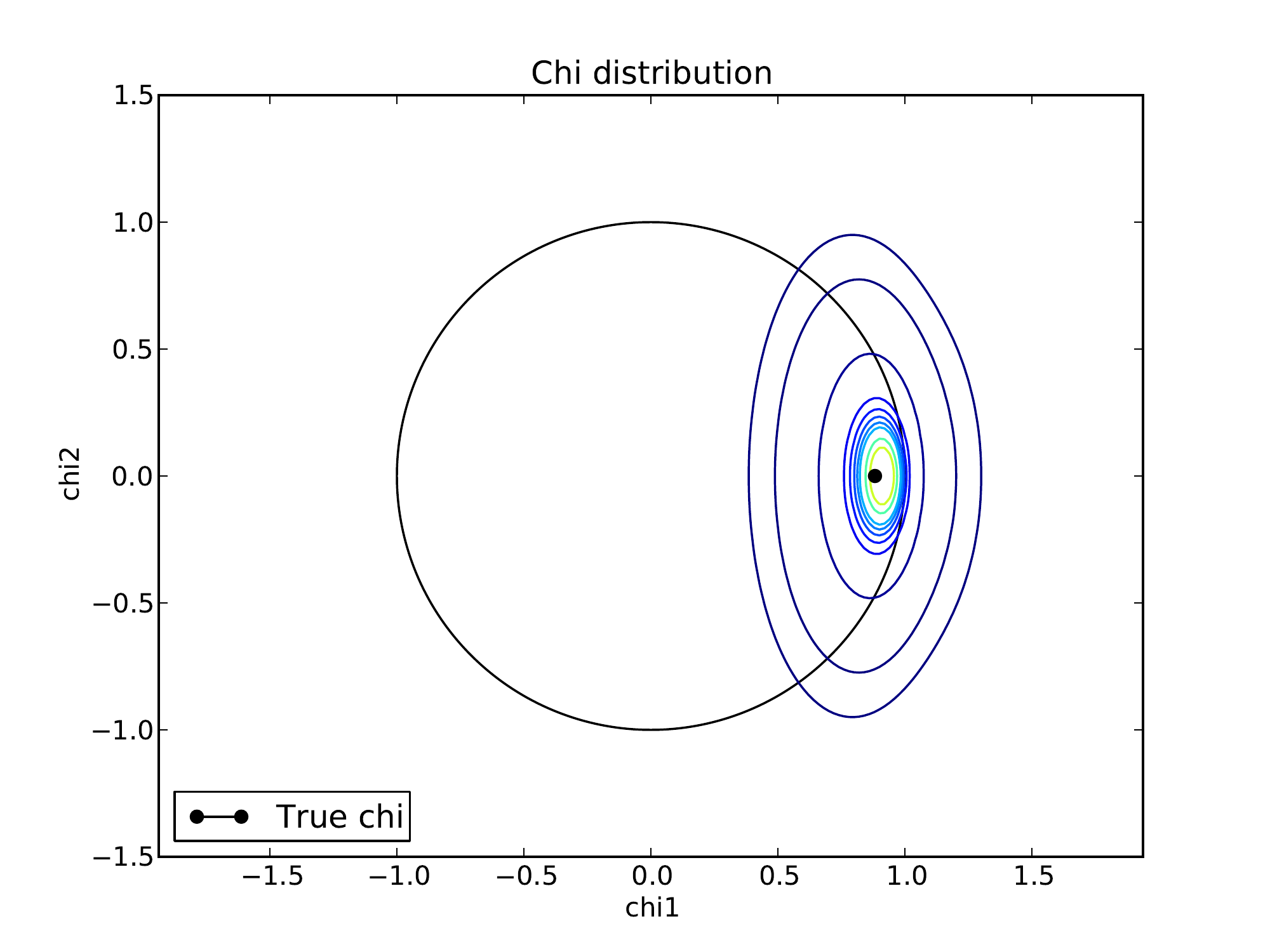}
\includegraphics[width=\columnwidth, angle=0]{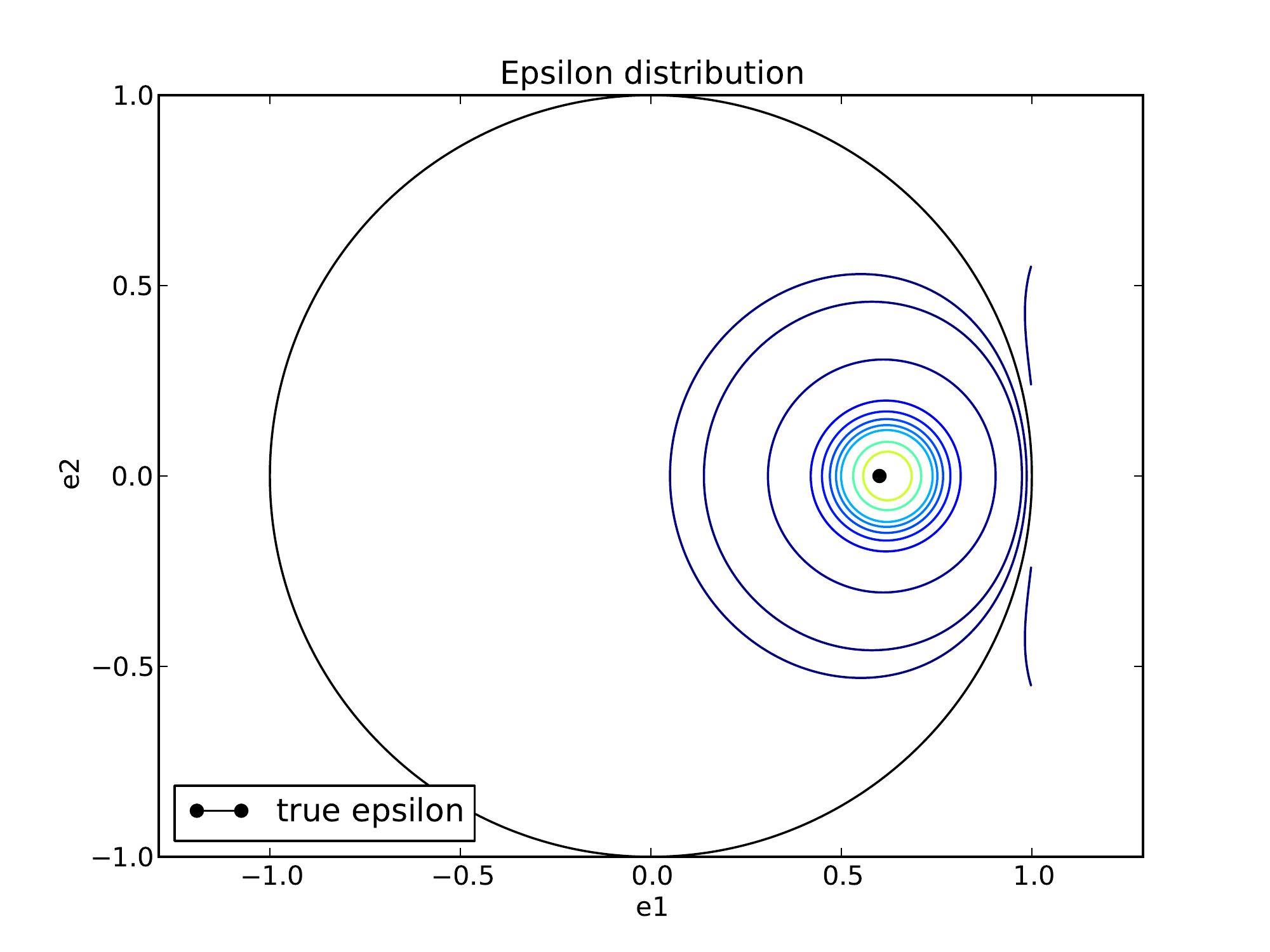}
\caption{Ellipticity distribution caused by noise ($\nu=5$) for an elliptical Gaussian object with $\epsilon=(0.6,0.0)$. In the left panel the distribution of the normalised polarisation is shown $\chi$ while in the right panel the distribution of the $\epsilon$-ellipticity. The $\chi$ probability distribution is not bounded by the unit circle (black solid line), while the $\epsilon$ probability distribution is. In both cases the true value does not correspond to the maximum of the distribution. }
\label{fig:Marsaglia9}
\end{figure*}
 
To transform $p_{\chi}(\chi_1,\chi_2)$ to other expressions for the ellipticity one simply applies a Jacobian 
to the distribution 
\be 
\label{eq:transfo}
p(n_1,n_2)=\left|\frac{\partial\chi_i}{\partial n_j}\right|_{i,j=\{1,2\}}p_{\chi}(\chi_1,\chi_2)
\ee
where $n_i$ are variables of the new distribution. 

The $\chi$ probability distribution is defined on ${\mathbf R}^2$ and not bounded to the unit circle despite $\chi$ being theoretically bounded, in the sense that no object can exist that has $\chi>1$. 
This is a consequence of equation (\ref{eq:chi}) and in particular it comes from the fact that the values that moments can assume in presence of noise, are unbounded. In particular a random negative fluctuation can cause values of $\chi>1$ to be measured. However those measurements are purely noise driven since no such object can exist in reality.\footnote{Mathematically it is possible to construct a surface brightness profile such that $\chi >1$. For example:
\begin{equation}
f(x,y)=(1 -2(1 - e)x^2 + (1 + e)y^2)\exp{(-(1 - e)x^2 - (1 + e)y^2)}
\end{equation}
where $e$ is an arbitrary number between $[-1..1]$.
However we do not expect the galaxy surface brightness to change sign and being non-monotonic in its domain, which are necessary conditions such that $\chi >1$. Those conditions might be fulfilled in real data in presence of noise (as exensively discussed in the paper) or if mistakes in the data reduction have been made (e.g. a wrong background subtraction or deblending).
}
Hence the part of the probability distribution outside the unit circle 
has to be regarded as unphysical. This fact becomes relevant when the pdf of $\epsilon$ is derived from the pdf of $\chi$. 


Computing the pdf of a variable $y$ which is a function of a variable $x$, for which the pdf is known, as shown in equation 25, implies that there is a bijective mapping between $x$ and $y$. In the case of $\epsilon$ and $\chi$ this is true only inside the unit circle. In fact a point on the unit circle in $\epsilon$-space can be mapped either outside or inside the unit circle in $\chi$ space.  A point outside the unit circle in $\chi$-space is mapped on the unit in $\epsilon$-space and hence the mapping is also not bijective. This is a consequence of the fact that $|\epsilon|$ cannot be larger than 1 even in the presence of noise: the reason is the second non-linear term in the denominator of equation (\ref{eq:ellipticity}) that restricts the measurable values.

Hence to compute the $\epsilon$ pdf we first truncate the $\chi$ pdf such that it vanishes outside the unit circle, we then apply equation (\ref{eq:transfo}), and finally we re-normalise the pdf (this last step is required due to the truncation of the $\chi$ pdf):
\be 
\label{eq:transfo2}
p(\epsilon_1,\epsilon_2)=\frac{4(1-|\epsilon|^2)}{(1+|\epsilon|^2)^3}p_{\chi}\left(\frac{2\epsilon_1}{1+|\epsilon|^2},\frac{2\epsilon_2}{1+|\epsilon|^2}\right).
\ee

The reason to truncate the part of the pdf of $\chi$ outside the unit circle is then a mathematical necessity rather than a practical issue (which would be eliminating part of the pdf inside the unit circle because for example very elliptical objects cannot be measured by a certain algorithm).

Similar transformations can be easily calculated for the amplitude and position angle parameterisation of ellipticity $(|\epsilon|,\theta)$:
\ba
\chi_1 &=&\frac{|\epsilon|}{\sqrt{1+\tan^2(2\theta)}} \nn
\chi_2 &=&\frac{|\epsilon|}{\sqrt{1+[\tan^2(2\theta)]^{-1}}} 
\ea
from which
\ba 
\label{eq:transfo3}
p(|\epsilon|,\theta)=\frac{|\epsilon|(2\cos(\theta)\sin(\theta)-\sin(6\theta))}{\cos(2\theta)}\times \nn
p_{\chi}\left(\frac{|\epsilon|}{\sqrt{1+\tan^2(2\theta)}},\frac{|\epsilon|}{\sqrt{1+[\tan^2(2\theta)]^{-1}}} \right).
\ea

In Figure \ref{fig:Marsaglia9} we show an example of the normalised polarisation $\chi$ and $\epsilon$-ellipticity distributions for an object with $\epsilon=(0.6,0.0)$ 
and a noise level $\nu=5$ (as defined in equation \ref{eq:sn}).

The Marsaglia-Tin distribution is applicable for moment-based methods, that are necessarily governed by it. However, it is also applicable for model-based methods, in which any model parameters can be projected into the space of the moments. If such a projection results in a trivariate Gaussian, then the Marsaglia-Tin is exactly applicable, but if the projection results in a non-Gaussian distribution the final probability distribution for ellipticity will be more involved. For example if the moments of the model are fitted to the data, and the centroid is known, the distribution of the moments will be a trivariate Gaussian and the ellipticity distribution will follow the Marsaglia-Tin distribution. However in the case a model is fitted to the data, it is easy to see that the moments are non-linear parameters in the model and hence their probability distribution may not be a trivariate Gaussian.

\subsection{Properties of the Marsaglia-Tin distribution} 
\begin{figure}
\includegraphics[width=9cm, angle=0]{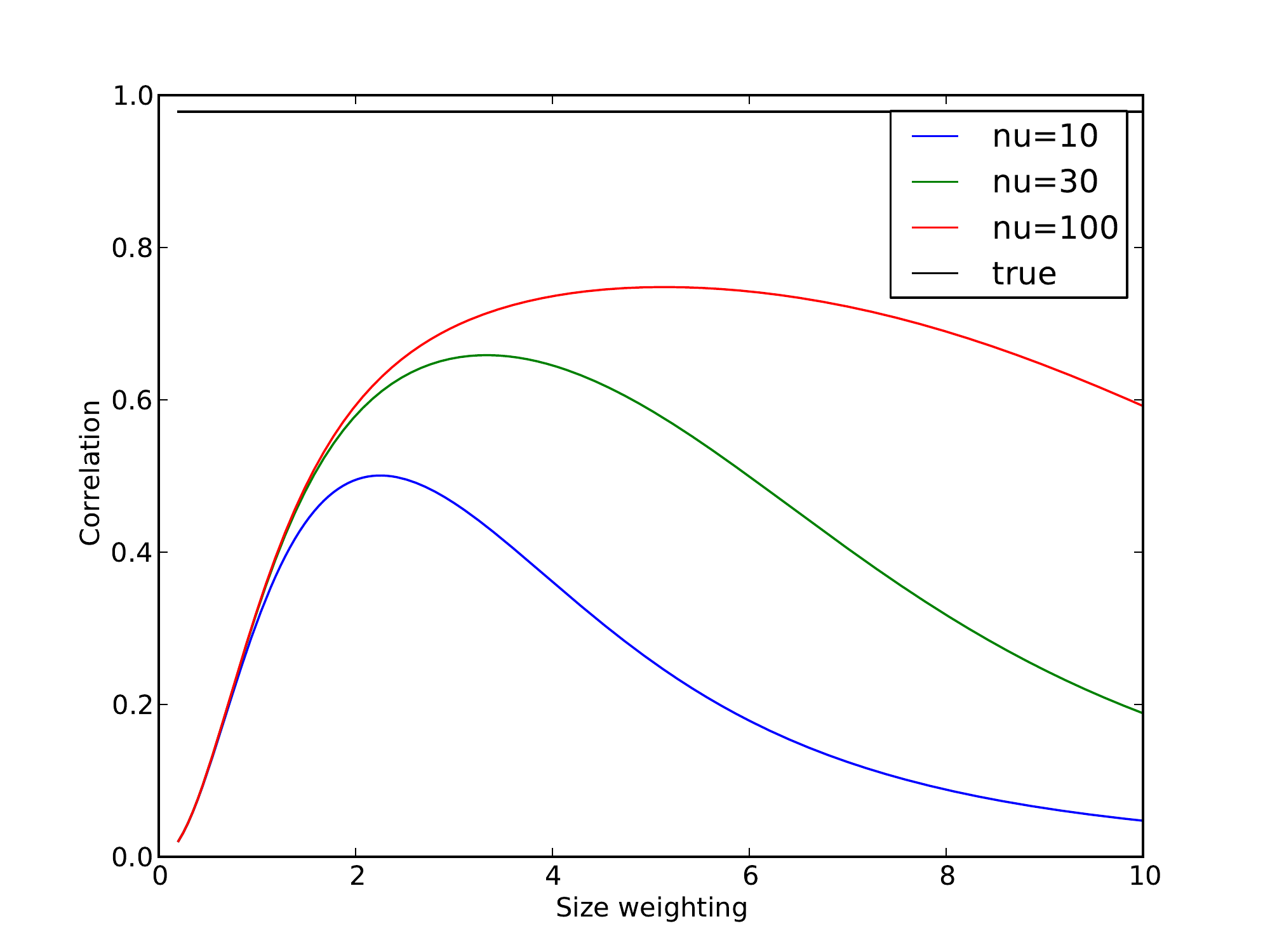}
\caption{Correlation between $u$ and $s$ for a Gaussian galaxy with ellipticity $\epsilon=(0.5,0.0)$ and size $sg=1$ as a function of the size of the weighting function for three values of signal-to-noise $\nu =10,30,100$. The solid black line corresponds to the true correlation between $u$ and $s$. The correlation between the Stokes parameters vanishes at low signal-to-noise and for large size of the weighting function.}
\label{fig:CorrSize}
\end{figure}

The Marsaglia-Tin distribution, derived in the previous section, is the probability of measuring an ellipticity given a 
noise level and a true ellipticity value. It depends on several quantities:
\begin{itemize}
\item 
The root mean square $\sigma_n$ of the image background. This number can be measured from the data.
\item 
The weighting function used to compute the moments. 
The shape and the size of the weighting function are arbitrary. The weighting function is used to compute the errors on the moments. The form of the weighting function determines the \textit{measured} correlation between the Stokes parameters (via the error on the moments, as in equation \ref{eq:errMom}), while the \textit{true} or \textit{intrinsic} correlation is determined by the object ellipticity. In the case that the noise in the image is dominated by the background (low signal-to-noise) the measured correlation between the Stokes parameters vanishes, as well as in the case that the size of the weighting function is very small compared to the size of the galaxy. In general the correlation depends on the signal-to-noise level and on the weighting function size as we show in Figure \ref{fig:CorrSize}.  
\item 
The values of the weighted Stokes parameters $\mu_i$. Those can be calculated once the weighting function, the galaxy profile and morphology are specified.
\end{itemize}
In its original derivation the Marsaglia-Tin distribution represents the probability of measuring a normalised polarisation $\chi$. 
However we showed in the previous section how it can be transformed into the probability of measuring an $\epsilon$-ellipticity. 
Depending on how the ellipticity is defined in terms of moments, or semi-major and semi-minor axis, the form and the properties of the Marsaglia-Tin are different as we will show.
We investigate in the following section the bias in the mean and maximum of the Marsaglia-Tin distribution as a function of the ellipticity. The bias is here defined as the difference between the mean (or the maximum) of the Marsaglia-Tin distribution and the true ellipticity. 

\begin{figure}
\includegraphics[width=9cm, angle=0]{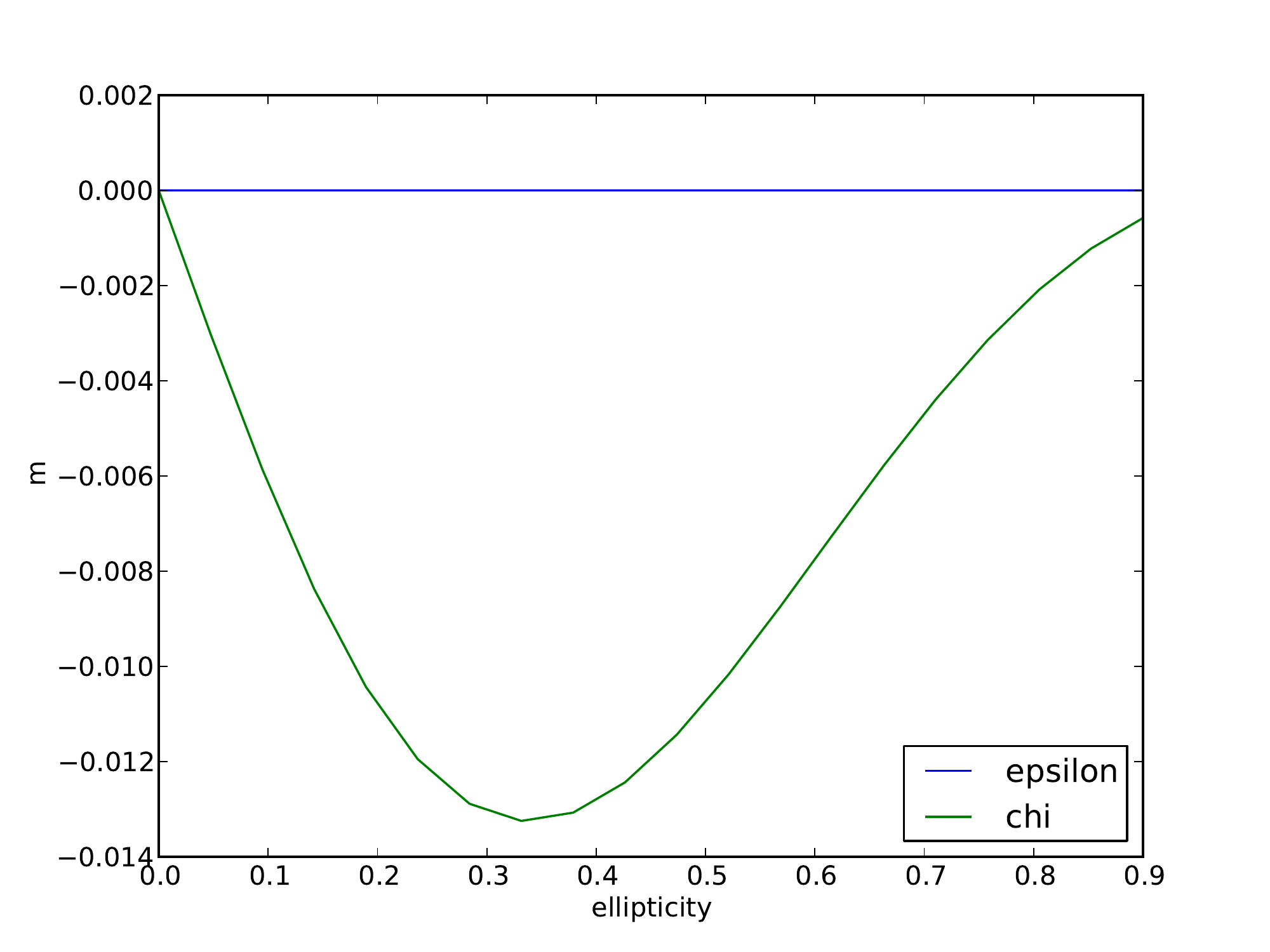}
\includegraphics[width=9cm, angle=0]{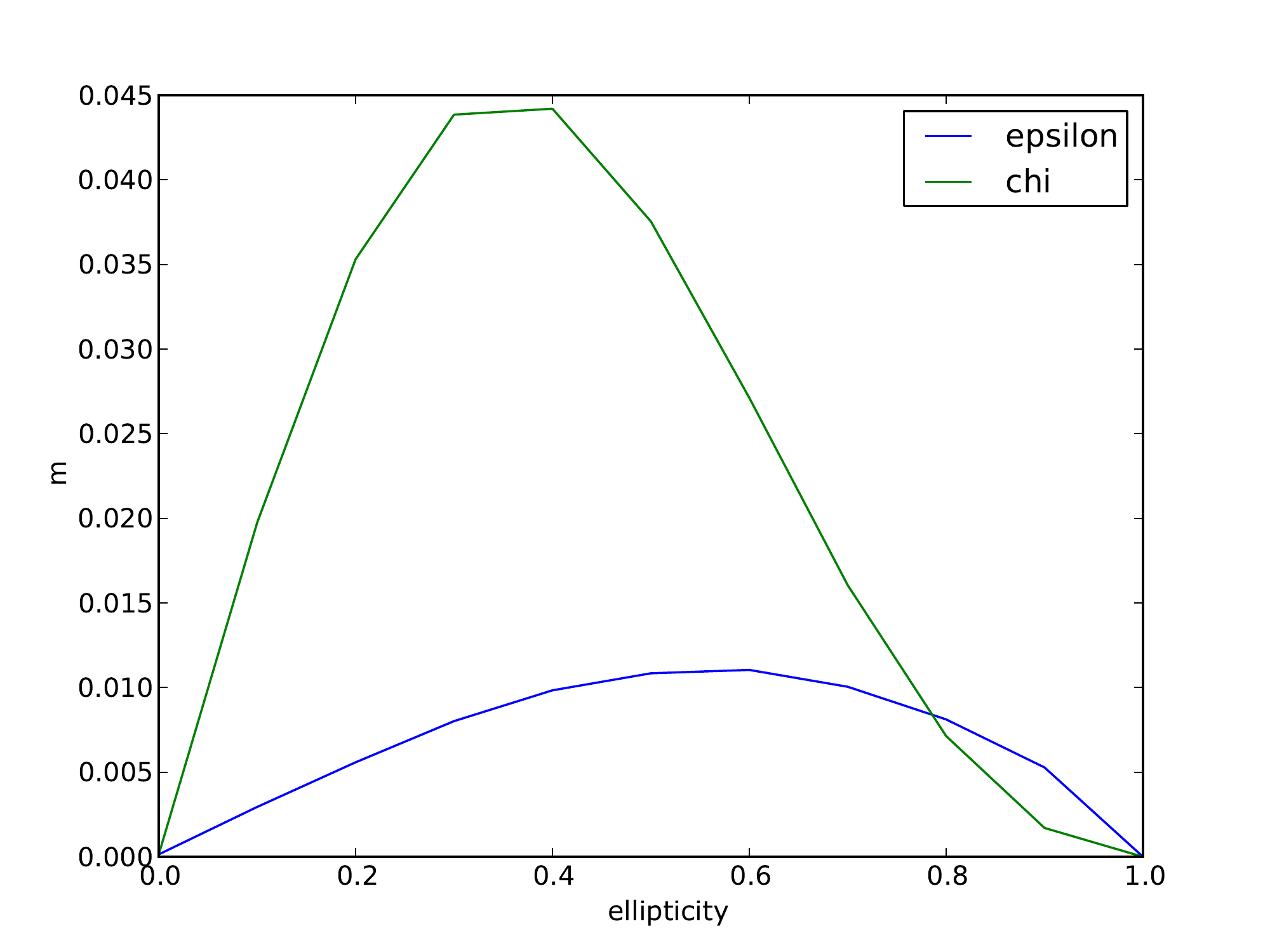}
\caption{Multiplicative bias $m$ as a function of the object's ellipticity absolute value. The object has a here a Gaussian profile. In the \textit{upper panel} the observed ellipticity is defined as the mean of the Marsaglia-Tin distribution, while in the \textit{lower panel} as the maximum of the distribution. The \textit{blue} curves refer to the case in which the $\epsilon$-ellipticity is used, while the \textit{green} curves refer to the case in which the normalised polarisation $\chi$ is used. The mean of the $\epsilon$-ellipticity is always unbiased if the profile, size and ellipticity of the weighting function are matched to the ones of the galaxy, while the maximum of the distribution is generally biased. On the contrary both the mean and the maximum of the $\chi$-distribution are always biased. The angle $(1/2)\tan^{-1}(\chi_2/\chi_2)$ is always unbiased. }
\label{fig:BiasEl_pm}
\end{figure}

We start by choosing an `optimal' weighting function that exactly matches the radial profile, ellipticity and size of the object. 
This is an ideal scenario that is in practice never possible to reach. In this case, even in presence of noise, the correlation between the Stokes parameters is the right one given the size and an ellipticity of the object. 
Furthermore \cite{Melchior11} noted that as long as the weighting function used to measure the moments has the the same radial profile, size and ellipticity as the galaxy, the correlations between the Stokes parameters (and hence the shape of the Marsaglia-Tin distribution) is virtually unchanged. For this reason and sake of simplicity we will use in the following a Gaussian profile for all the calculations. We do not investigate the cross-talk between model and noise bias here, which would correspond to the case of employing in the measurement of the moments a weighting function with a different profile than the galaxy.  We refer to \cite{Kacprzak13} for a recent investigation of this matter.
  In Figure \ref{fig:BiasEl_pm} we show the multiplicative bias on the maximum likelihood and mean of the ellipticity probability distributions as a function of the amplitude of the ellipticity for this optimal case. We will show 
biases only as a function of the amplitude of the ellipticity because the angle $(1/2)\tan^{-1}(\chi_2/\chi_2)$ is always unbiased \citep{Wardle74}.  
We note that when $\chi$ is used as a definition for the ellipticity both the mean and the maximum of the Marsaglia-Tin distribution are biased, while in the case that $\epsilon$ is used only the maximum is biased while the mean is unbiased \textit{independent} of the signal-to-noise level.

\subsection{Properties of $\epsilon$-ellipticity}

\subsubsection{Bias from truncation}

\begin{figure}
\includegraphics[width=9cm, angle=0]{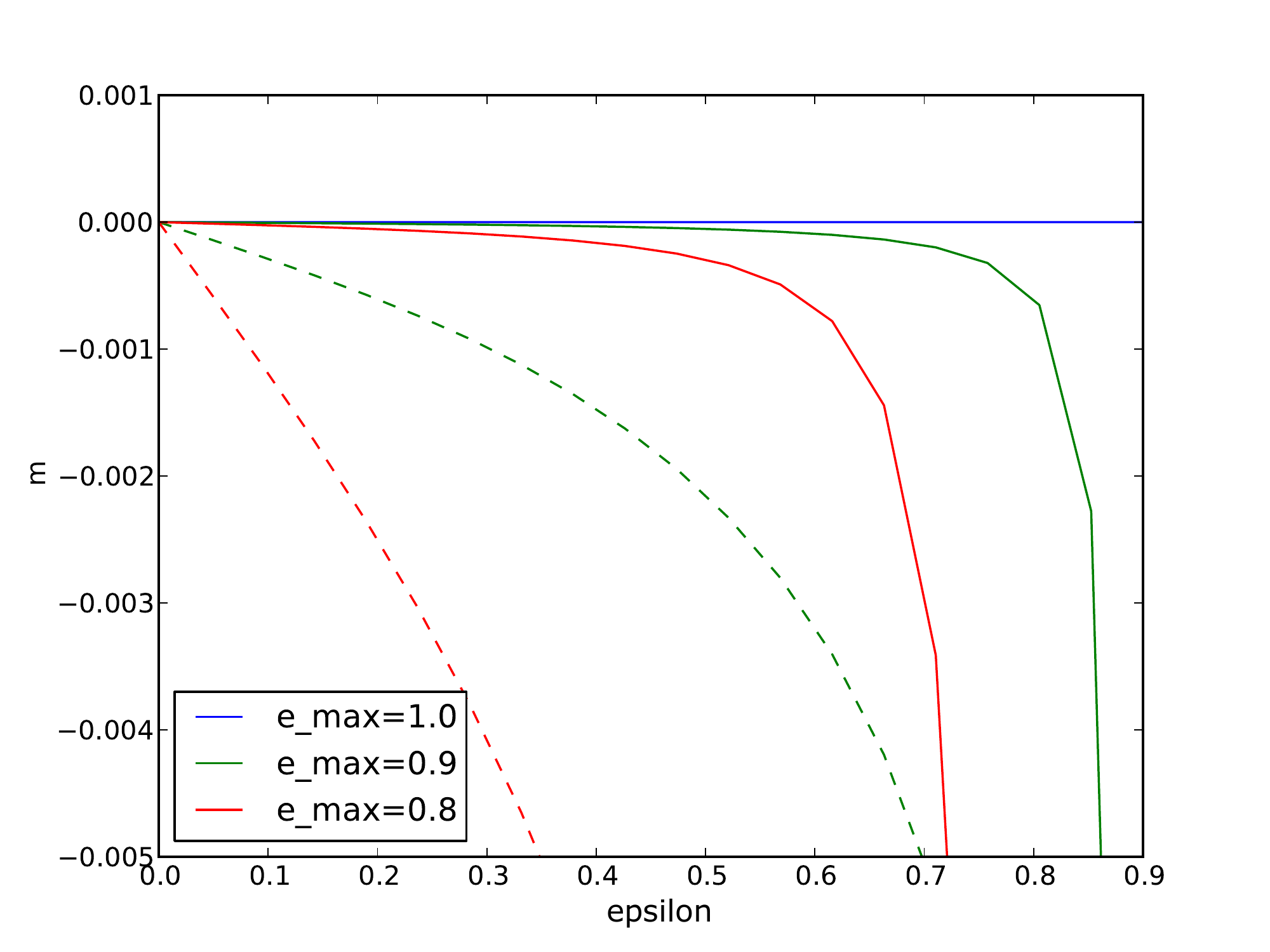}
\caption{Multiplicative bias $m$ as a function of the object $\epsilon$-ellipticity for two different cuts in ellipticity space, $\epsilon_{max}=0.9$ (green line) and $\epsilon_{max}=0.8$ (red line). The object has here a Gaussian profile. The blue line corresponds to the case of no cuts in $\epsilon$-ellipticity.The solid lines correspond to a signal-to-noise $\nu=10$, while the dashed line to a signal-to-noise $\nu=5$.}
\label{fig:BiasEl_cut}
\end{figure}

The mean of the $\epsilon$ distribution is unbiased \emph{only} if the average is computed as an integral 
over the full ellipticity range: from 0 to 1. However in practice highly elliptical objects could either be not detected or heavily affected by pixelisation. Hence the average is commonly computed as an integral between 0 and some 
$\epsilon_{max}$. This truncation introduces a bias in the measurements of the mean $\epsilon$ even in the ideal case 
of a weighting function that matches perfectly the galaxy profile. The amplitude of this truncation-bias is shown in Figure \ref{fig:BiasEl_cut} as a function of ellipticity. It is typically lower than  $10^{-3}$ for ellipticities smaller than $\epsilon \sim 0.5$ at signal-to-noise $\nu=10$. At lower signal-to-noise level the bias coming from truncation of the ellipticity space becomes larger due to the fact that the Marsaglia-Tin distribution becomes wider exceeding the unit-circle, and at $\nu=5$ it exceeds $10^{-3}$ even for small values of the ellipticity.

\subsubsection{Bias from using a weighting function}

In practical applications the weighting function used to measure the moments of the light distribution is either chosen to be a circular Gaussian with a size matching the size of the object (as is implemented in KSB for example), or it is matched iteratively to the galaxy profile (as is implemented in DEIMOS for example). We consider here the effect of using a circular weighting in a moment-based method, since the effect of an iterative matching would translate to a very complicated observed ellipticity distribution which is impossible to describe analytically.

Employing a weighting function which does not match the galaxy profile changes the measured ellipticity. This effect 
does not depend on the noise but it is simply a mathematical consequence of the multiplication of the galaxy profile with the weighting function. Using higher order moments of the object it is possible to recover the original un-weighted ellipticity. This is a common practice in moment-based method. However in the low-signal-to-noise regime those high order moments are typically very noisy and hence the correction for the application of the weighting function can be done only approximately. We are not interested here in investigating this particular problem (which in a moment-based method can be accounted for as model bias), therefore we always correct the measured ellipticity to account for the weighting function. 
The correction can be exactly computed if the galaxy profile is known. In fact the weighted quadrupole moments can be written as a sum of the unweighted quadrupole moments and weighted high-order moments. The latter can be precisely computed if the galaxy profile is known. 
By using the relation between weighted and unweighted quadrupole moments it is possible to derive a relation between the weighted and the unweighted ellipticity that can be inserted directly into the Marsaglia-Tin distribution. 
However it is much easier from a practical point of view to compute the unweighted Marsaglia-Tin distribution numerically. We summarise here the main steps:
 
\begin{itemize}
\item We assign an ellipticity and a size to the object (having an elliptical Gaussian profile in this work) and we specify the object's signal-to-noise and a weighting function;
\item We rotate the object in a reference frame such that $e_2=0$, and we keep the record of the original phase.
\item We compute the weighted moments and their errors using a circular weighting function.
\item We sample values of the weighted moments $Q_{20}$ and $Q_{02}$ from a correlated bi-variate Gaussian  having the correct variances and correlations as per equation (\ref{eq:CovStokes}), while $Q_{11}$ is sampled from a Gaussian distribution with zero mean (for a more technical discussion we refer the interested reader to Appendix B in \citealp{Melchior12}).
\item We de-weight the noisy moments. This is done numerically as described in Appendix B.
\item We define an ellipticity using the de-weighted moments and we finally rotate back the galaxy in its original reference frame.
\item The observed ellipticity is computed as an average over $10^8$ noise realisations.
\end{itemize}

\begin{figure*}
\centering
\begin{tabular}{cc}
\includegraphics[width=9cm, angle=0]{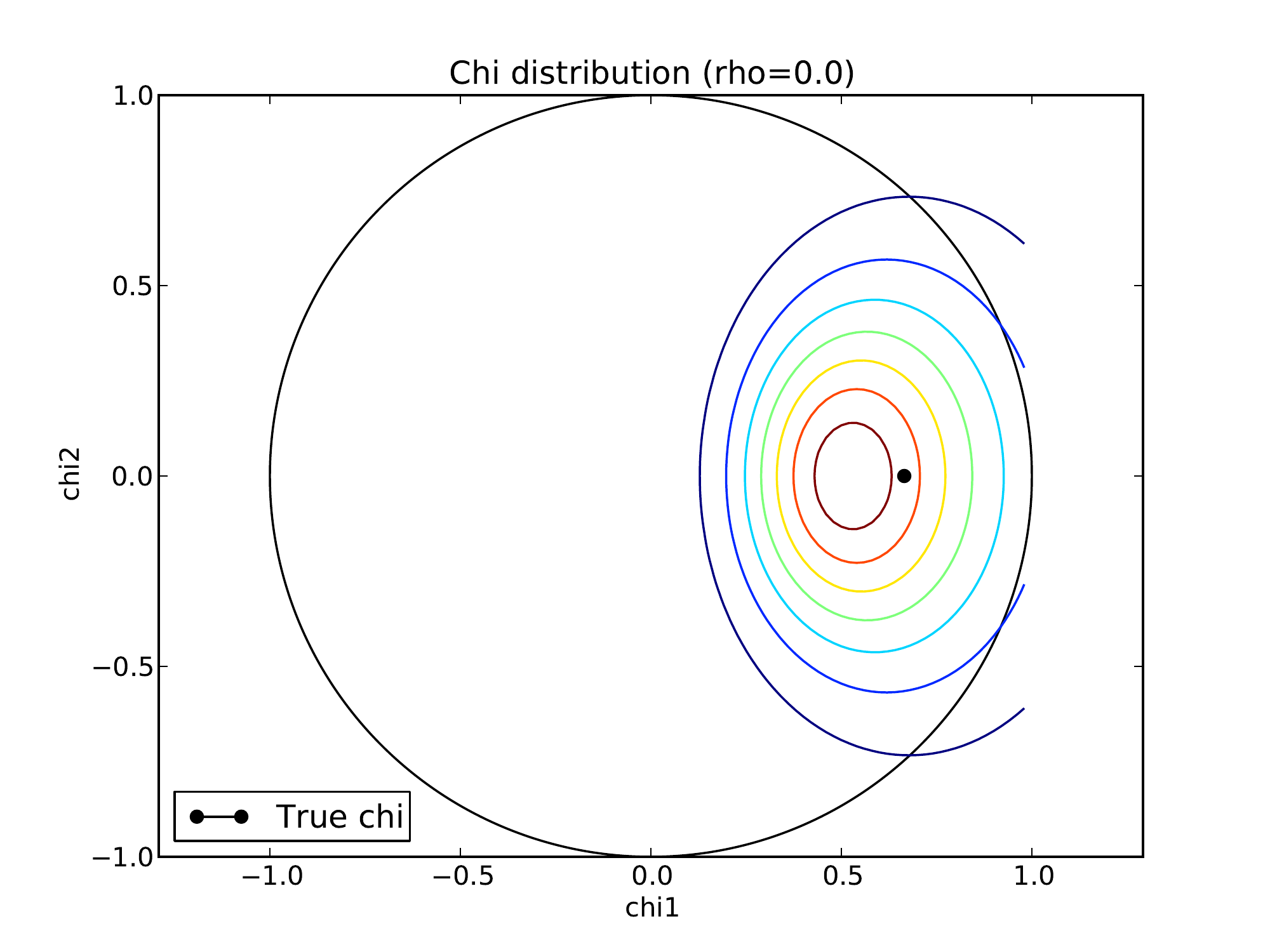}&
\includegraphics[width=9cm, angle=0]{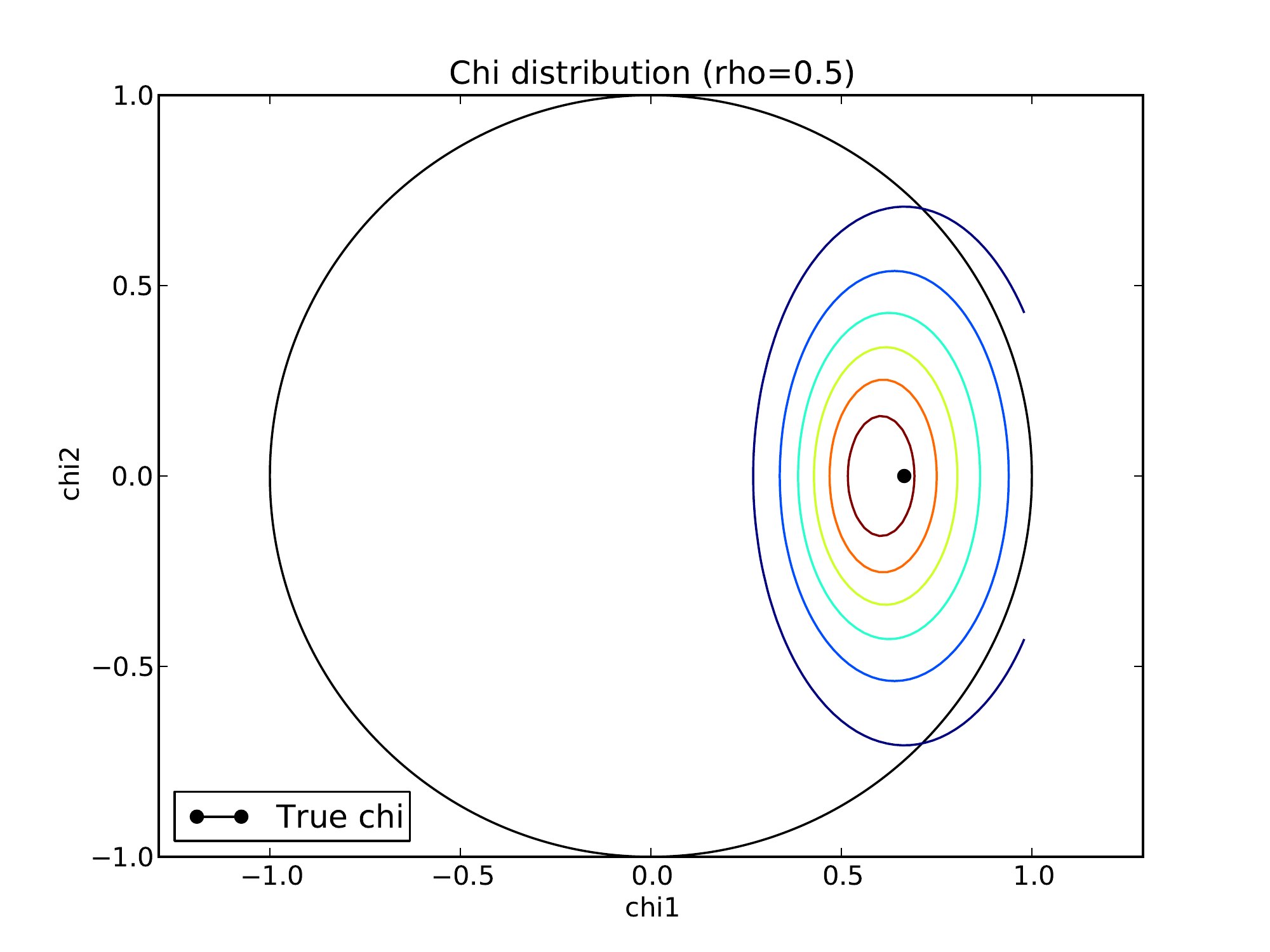}\\
\includegraphics[width=9cm, angle=0]{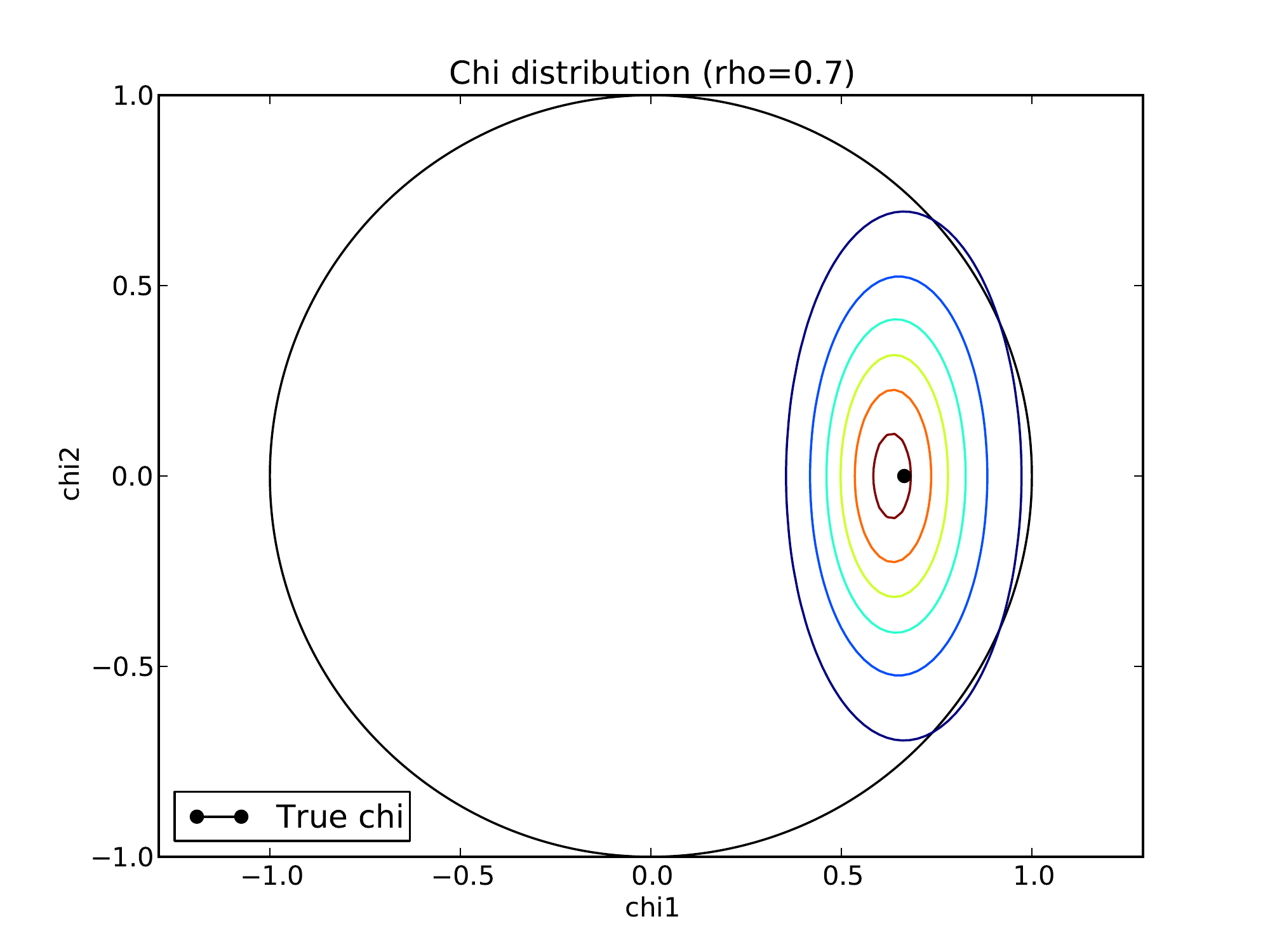}&
\includegraphics[width=9cm, angle=0]{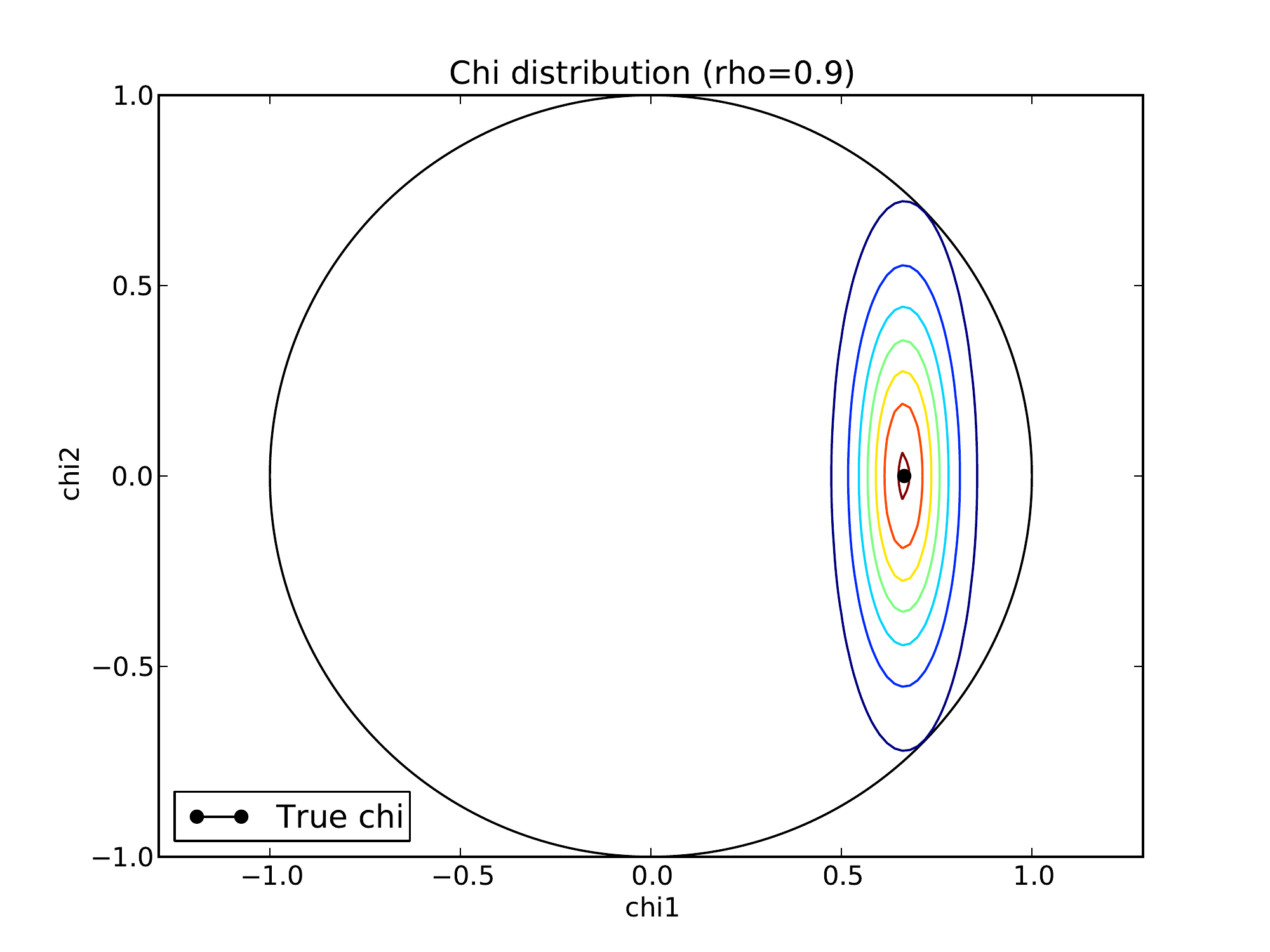}
\end{tabular}
\caption{Marsaglia-Tin distribution for a galaxy with $\nu=5$, size $s=1$, weighted with a Gaussian weight with width $s_{w}=1.5$. The correlation between the Stokes parameters is changed in each panel between 0 and 0.9 to illustrate the impact that the correlation coefficients have in shaping the Marsaglia-Tin distribution. }
\label{fig:CorrMarsaglia}
\end{figure*}

As we previously discussed and showed in Figure \ref{fig:CorrSize}, the effect of employing in the measurements a weighting function which has a different size and ellipticity than the underlying object is to modify the correlation between the Stokes parameters. The amplitude of the correlation is quite important to define the actual shape of the Marsaglia-Tin distribution as we show in Figure \ref{fig:CorrMarsaglia}. In particular, the more the Stokes parameters are correlated the more the Marsaglia-Tin distribution shrinks in the ellipticity direction for fixed signal-to-noise level and galaxy properties.
 
In the following we investigate the bias in the measured unweighted ellipticity coming from employing a weighting function with a size that is defined as a fraction of the objects semi-major axis. This particular choice of the weighting function is motivated by the fact that we do not want to truncate the object by applying a weighting function that is too small compared to the galaxy size. We warn the reader that the results presented in the rest of the paper depend significantly on the choice of the weighting function. The way they depend is however easy to understand and we will take great care in explaining it.

\begin{figure*}
\includegraphics[width=\columnwidth, angle=0]{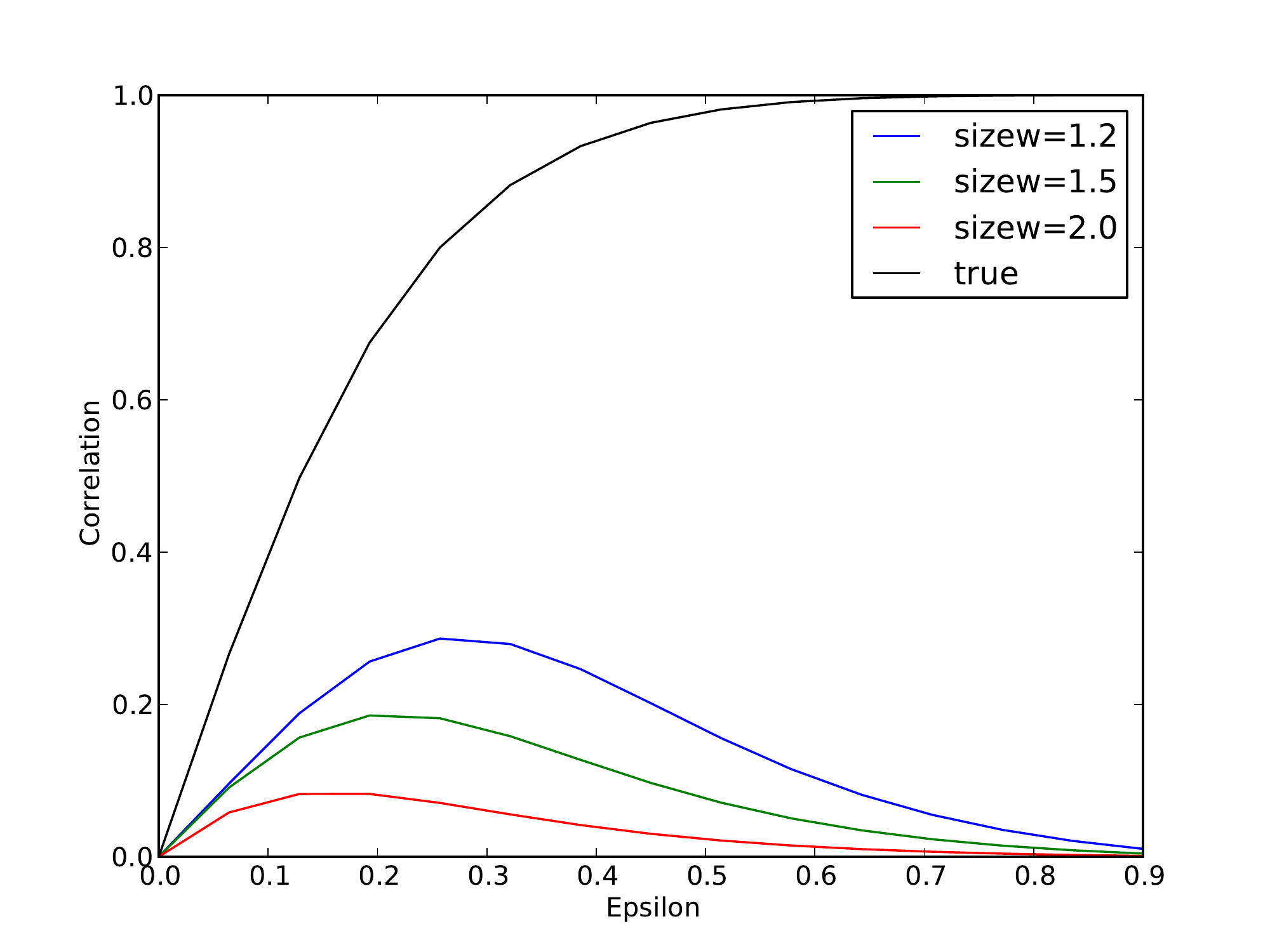}
\includegraphics[width=\columnwidth, angle=0]{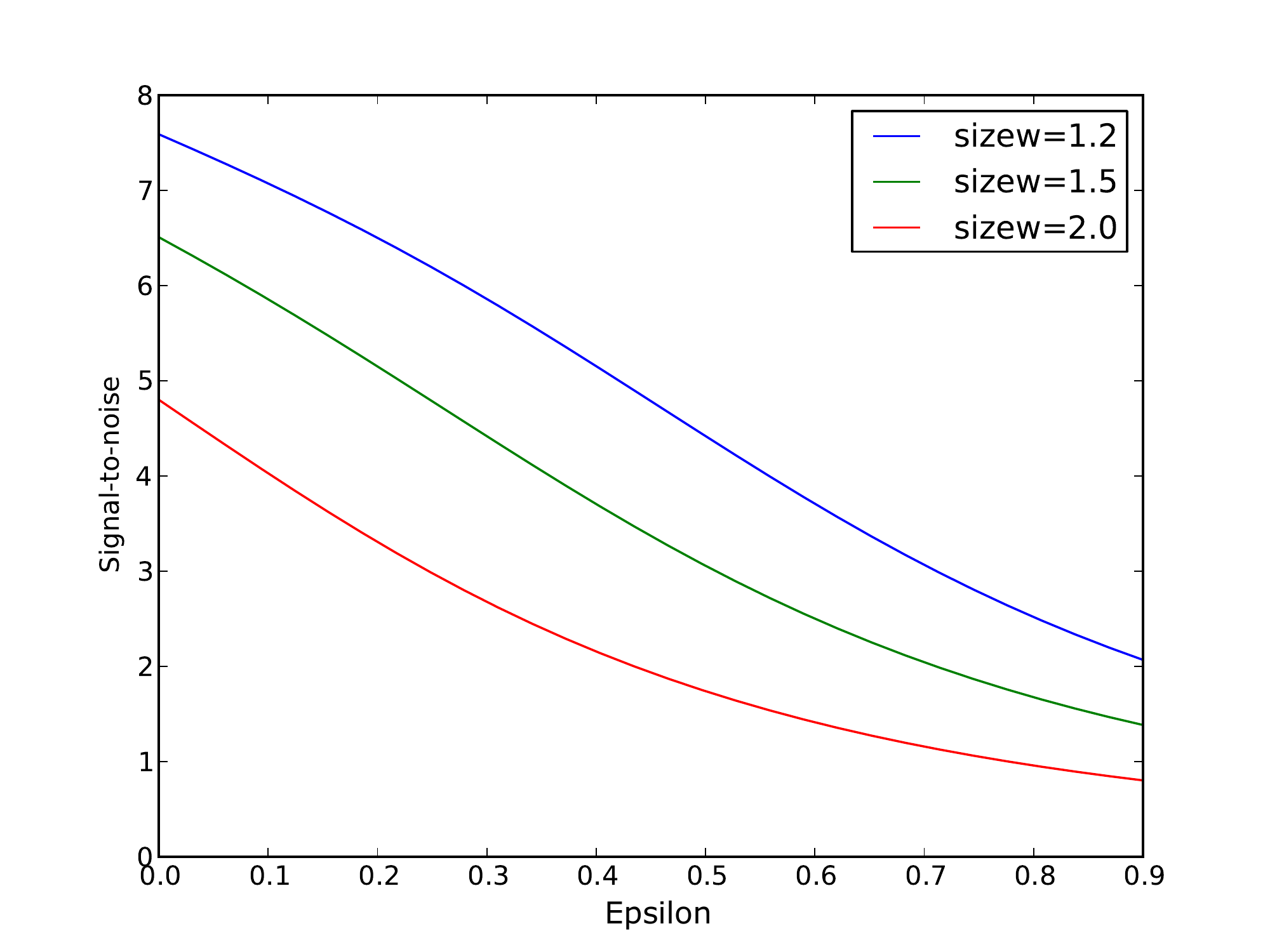}
\caption{\textit{Left panel:} Correlation between $u$ and $s$ as a function of the object $\epsilon$-ellipticity. The black curve corresponds to the true correlation, while the blue, red and green curves to the case in which the moments are measured with a circular weighting function having size 1.2, 1.5 and 2.0 times the object semi-major axis. \textit{Right panel:} Signal-to-noise of $Q_{20}$ as a function of $\epsilon$-ellipticity. The three curves corresponds to the same choices of the weighting function size as in the left panel. The flux signal-to-noise is here $\nu=10$.}
\label{fig:SizeWeight}
\end{figure*}

We start investigating the variation of measured correlation between the Stokes parameters for this particular choice of the weighting function.
The results are shown in Figure \ref{fig:SizeWeight} as a function of $\epsilon$-ellipticity and for three different values of the size of the weighting function (1.2,1.5,2.0 times the object semi-major axis). We note that the correlation is always lower than the true one (defined as the correlation measured at infinite signal-to-noise or in the case the weighting function is matched to the galaxy profile), and it tends to zero as the size of the weighting function becomes larger.

Furthermore another consequence of our choice of the weighting function is that the signal-to-noise of the quadrupole moments decreases with increasing ellipticity (bottom panel of Figure \ref{fig:SizeWeight}). This is simply a consequence that for an elliptical object most of the area inside the weighting function is filled by noise. 

Different choices of the size or shape of the weighting function would lead to similar plots as Figure \ref{fig:SizeWeight}. In particular, as the weighting function becomes close to the galaxy profile the measured correlation becomes closer to the intrinsic one, and the bias will be lower. 

\begin{figure*}
\includegraphics[width=\columnwidth, angle=0]{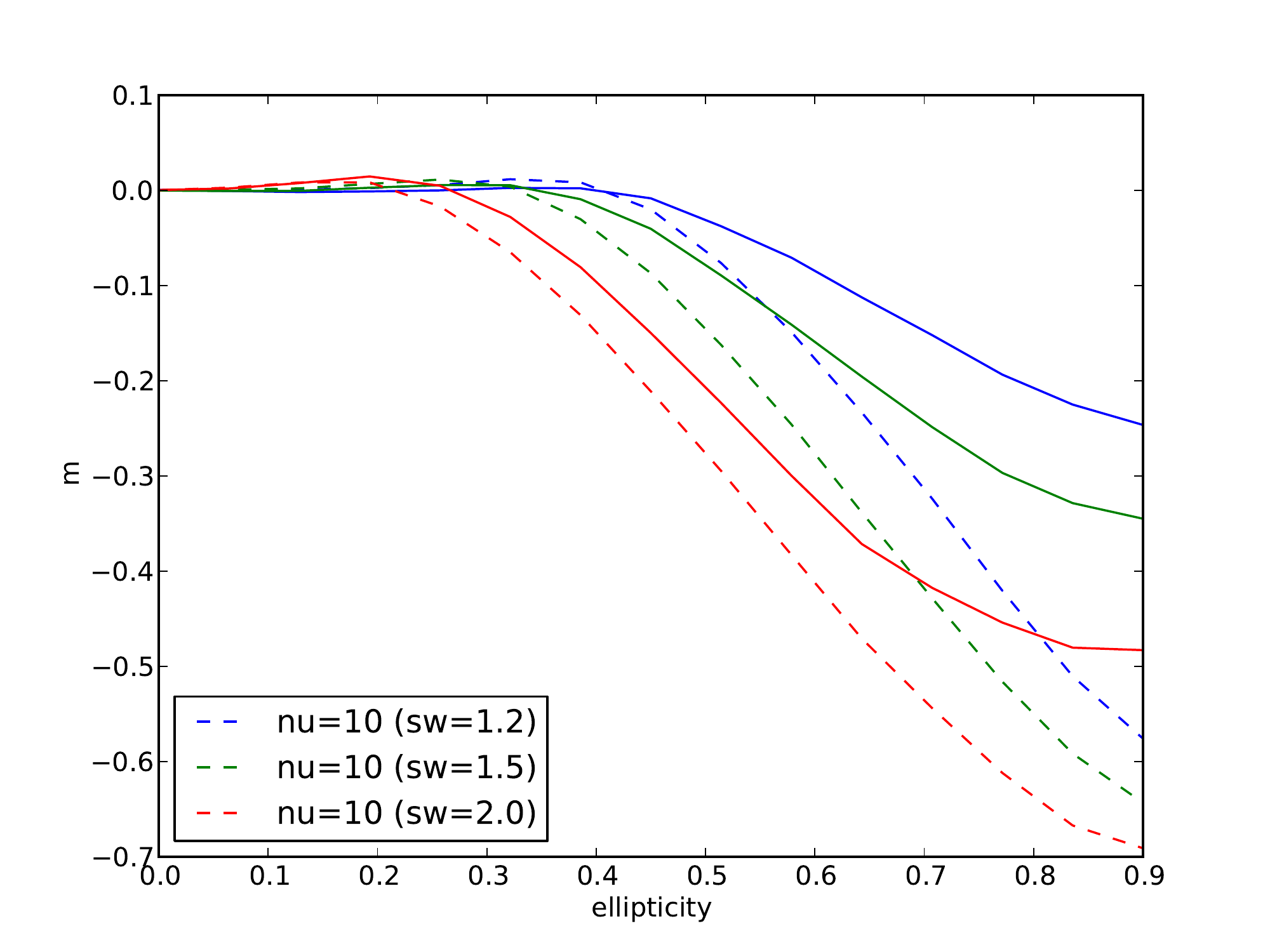}
\includegraphics[width=\columnwidth, angle=0]{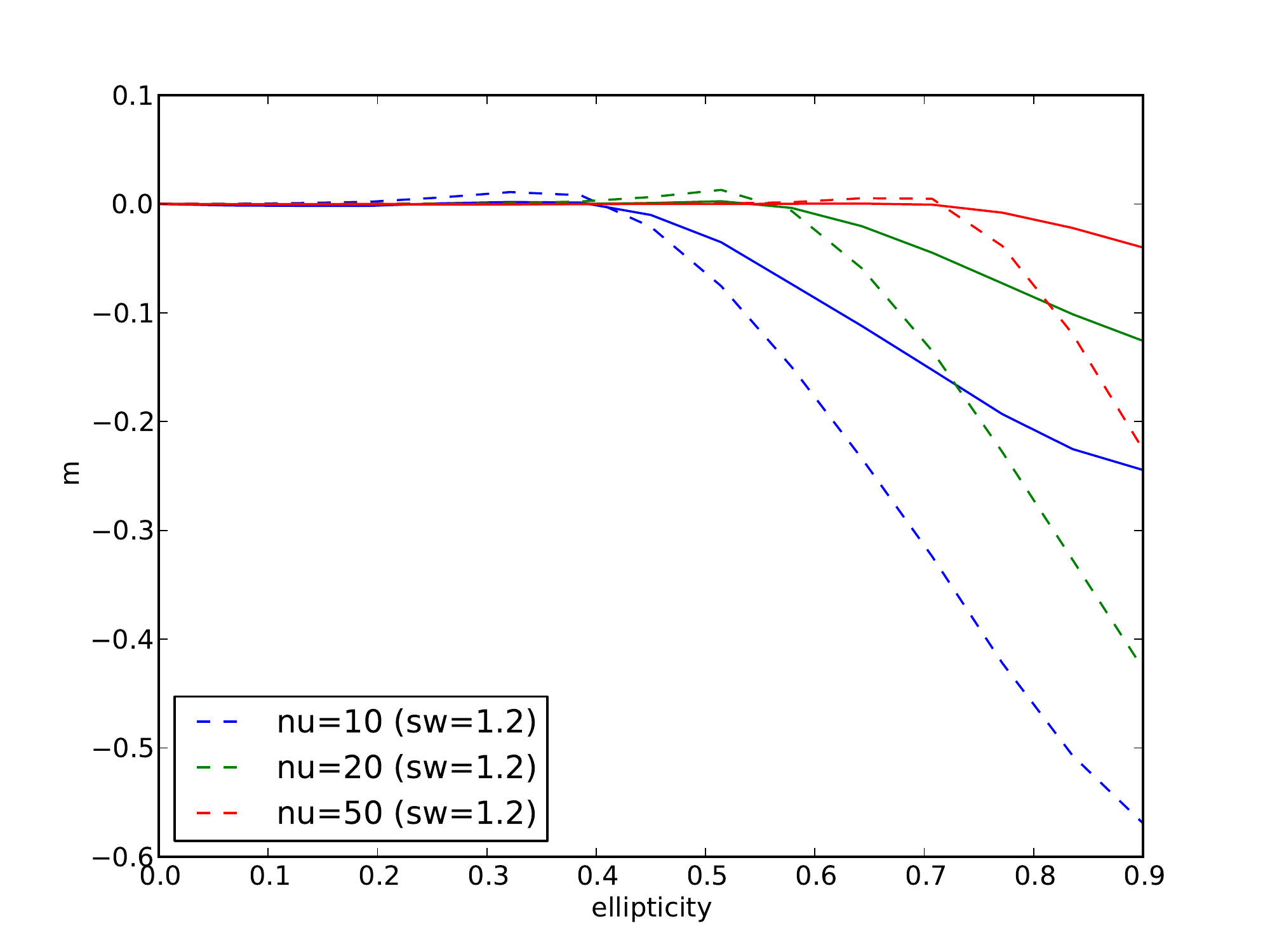}
\caption{\textit{Left panel:} Multiplicative bias $m$ as a function of the object $\epsilon$-ellipticity for three different sizes of the weighting function (1.2, 1.5, 2.0 times the object semi-major axis). The signal-to-noise is here $\nu=10$. \textit{Right panel:} Multiplicative bias $m$ as a function of the object $\epsilon$-ellipticity for three values of the object signal-to-noise ($\nu=10,20,50$). The solid lines correspond to the case when the normalised polarisation $\chi$ is used as an ellipticity estimator, while the dashed lines to the case when $\epsilon$ is defined as an ellipticity estimator. The bias is a strong function of the object signal-to-noise and it depends on the choice of the size of the weighting function.}
\label{fig:BiasWeight}
\end{figure*}

Finally we quantify the amplitude of the bias as a function of ellipticity and signal-to-noise. The result is shown in Figure \ref{fig:BiasWeight}.
For a fixed value of signal-to-noise the amplitude of the bias is driven by two factors: the correlation between the Stokes parameters and the signal-to-noise on the quadrupole moments (or the ellipticity of the object). 
This is not a surprising behaviour. In the case that the ellipticity is low, the Marsaglia-Tin distribution is almost entirely confined inside the unit-circle. Hence the bias is driven by the skewness of the distribution, and as we explained in Section \ref{The probability distribution of ellipticity} 
the Marsaglia-Tin distribution tends to be skewed towards large ellipticities in case the correlation between the Stokes parameters is lower than the true one. In case the ellipticity is high, and the signal-to-noise is low, the Marsaglia-Tin distribution is not confined inside the unit circle anymore. Hence the bias is driven by the truncation of the distribution at the unit circle. In the first case the bias tends to be positive, while in the second case it tends to be negative.

\subsection{The effect of the PSF}
We have so far neglected the fact that the object is convolved with the PSF. Working in moment-space has the advantage 
that the effect of the PSF convolution (and deconvolution) can be analytically accounted for in each moment. 
In fact the moments of  the galaxy $\{Q\}_{i,j}$, the PSF $\{P\}_{i,j}$ and the convolved object $\{Q^{\star}\}_{i,j}$ are related:
\begin{equation}
\label{eq:Deimos}
\{Q^{\star}\}_{i,j}=\sum_{k}^{i}\sum_{l}^{j}\dbinom {i} {k}\dbinom{j}{l}\{Q\}_{k,l}\{P\}_{i-k,j-l}
\end{equation} 
as shown by \cite{Flusser98} and \cite{Melchior11}. 
One remarkable aspect of this equation is that the convolved moments of order $i+j$ are 
only a function of the unconvolved moments and the PSF moments of at most the same order. 

\begin{figure*}
\includegraphics[width=\columnwidth, angle=0]{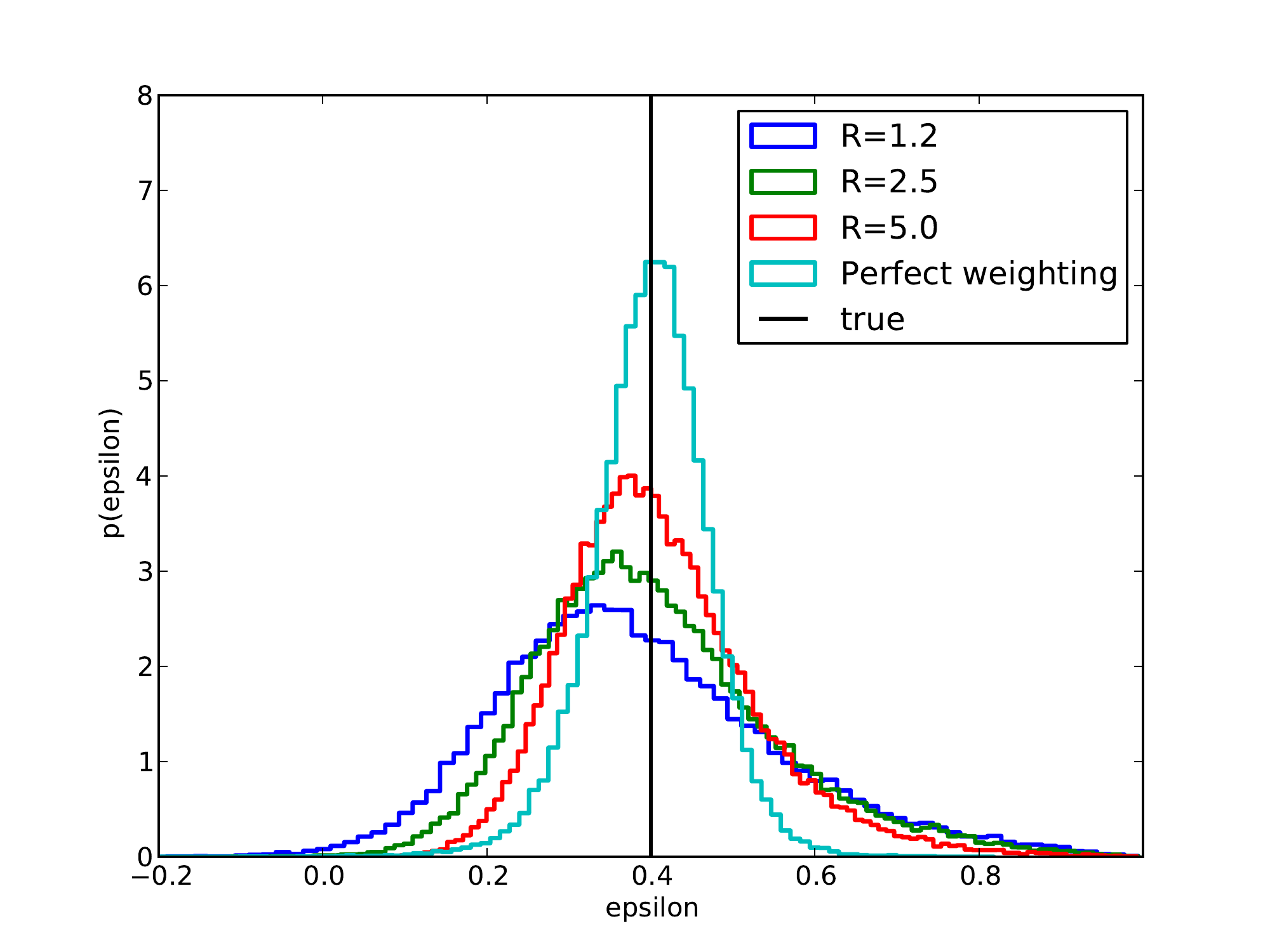}
\includegraphics[width=\columnwidth, angle=0]{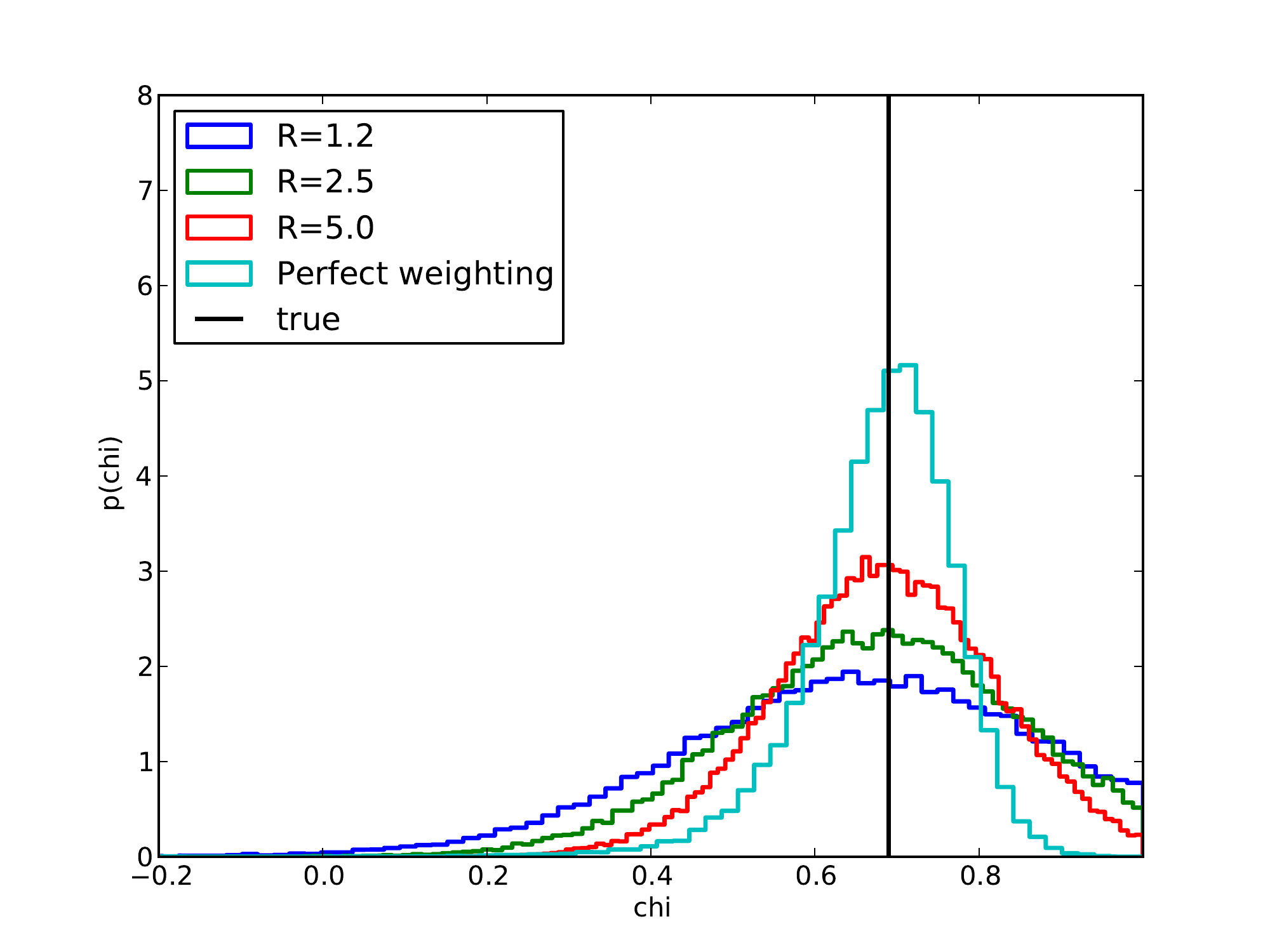}
\caption{\textit{Left panel:} Probability distribution of the absolute value of the $\epsilon$-ellipticity  given a true galaxy ellipticity of $\epsilon=(0.4,0.0)$ and zero PSF ellipticity. The cyan line corresponds to the case of using a weighting function which is matched perfectly to the galaxy profile and no PSF convolution, the red line to the case of using a circular weighting function with size 1.2 times the galaxy semi-major axis and a resolution of $R=5.0$, the green line to a resolution of 2.5 and the blue line to a resolution of 1.2. Note that in all cases we removed objects having an unphysical combination of second-order moments ($Q_{20}Q_{02}-Q_{11}^2<0$). \textit{Right panel:} as in the left panel but for the normalised polarisation $\chi$.In this case we removed objects with an ellipticity larger than unity. The galaxy signal-to-noise is in both cases $\nu=10$. We note that the signal-to-noise level determines the minimal width of the distribution (cyan histogram), while the resolution of the galaxies causes the distribution to become broader and skewed (red, green and blue histograms).  } 
\label{fig:HistoEll}
\end{figure*}

The effect of the convolution is to bias the galaxy ellipticity towards the ellipticity of the PSF. In the case of a 
circular PSF the galaxy will appear rounder. The ability to properly deconvolve the PSF from the observed galaxy image 
depends on the galaxy signal-to-noise and the amplitude of this effect scales with the resolution of the object, 
defined as the ratio of the object and PSF areas:
\begin{equation}
R=\Bigg(\frac{\{Q\}_{20}+\{Q\}_{02}}{\{P\}_{20}+\{P\}_{02}}\Bigg)\frac{\{P\}_{00}}{\{Q\}_{00}}
\label{eq:resolution}
\end{equation}
where $\{Q\}_{ij}$ are the galaxy moments and $\{P\}_{ij}$ the moments of the PSF. It is straightforward, using equation (\ref{eq:Deimos}), to show that:
\begin{equation}
\chi^{\star}=\frac{\chi}{1+(1/R)} +\frac{\chi^{\mathrm{PSF}}}{1+R}.
\label{eq:chicon}
\end{equation}
Hence the probability distribution of the deconvolved normalised polarisation $\chi$ can be derived by combining equations (\ref{eq:MarsagliaTin}) and (\ref{eq:chicon}):
\ba
&p(\chi^{dec}_{1},\chi^{dec}_{2})=\int p_{\chi}\left(\frac{\chi_1}{1+(1/R)} +\frac{\chi_1^{\mathrm{PSF}}}{1+R},\frac{\chi_2}{1+(1/R)} +\frac{\chi_2^{\mathrm{PSF}}}{1+R}\right) \nn
&\frac{1}{1+(1/R)}p(R)\mathrm{d}R
\label{eq:MarsagliaPSF}
\ea
where $p(R)$ is the probability distribution of the resolution as defined in equation (\ref{eq:resolution}). 
This latter probability distribution can also be described as a Marsaglia-Tin distribution, given the fact that $R$ is defined as a ratio between the trace of the quadrupole tensor and the flux of the object (assuming that the PSF moments are perfectly known). 
In practice we evaluate the above probability distribution numerically following the same procedure as described in the previous section, but including a convolution and deconvolution step. 
The deconvolution is done in moment space, following equation (\ref{eq:Deimos}), for each quadrupole moment. The convolved moments are noisy quantities 
and the deconvolution involves the ratio of the sum of quadrupole moments and the flux. Consequently the ellipticity probability distribution in presence of the PSF is broader and the mean of the distribution shifts towards the ellipticity of the PSF as the resolution $R$ goes to zero as we show in Figure \ref{fig:HistoEll}.

\begin{figure}
\includegraphics[width=9cm, angle=0]{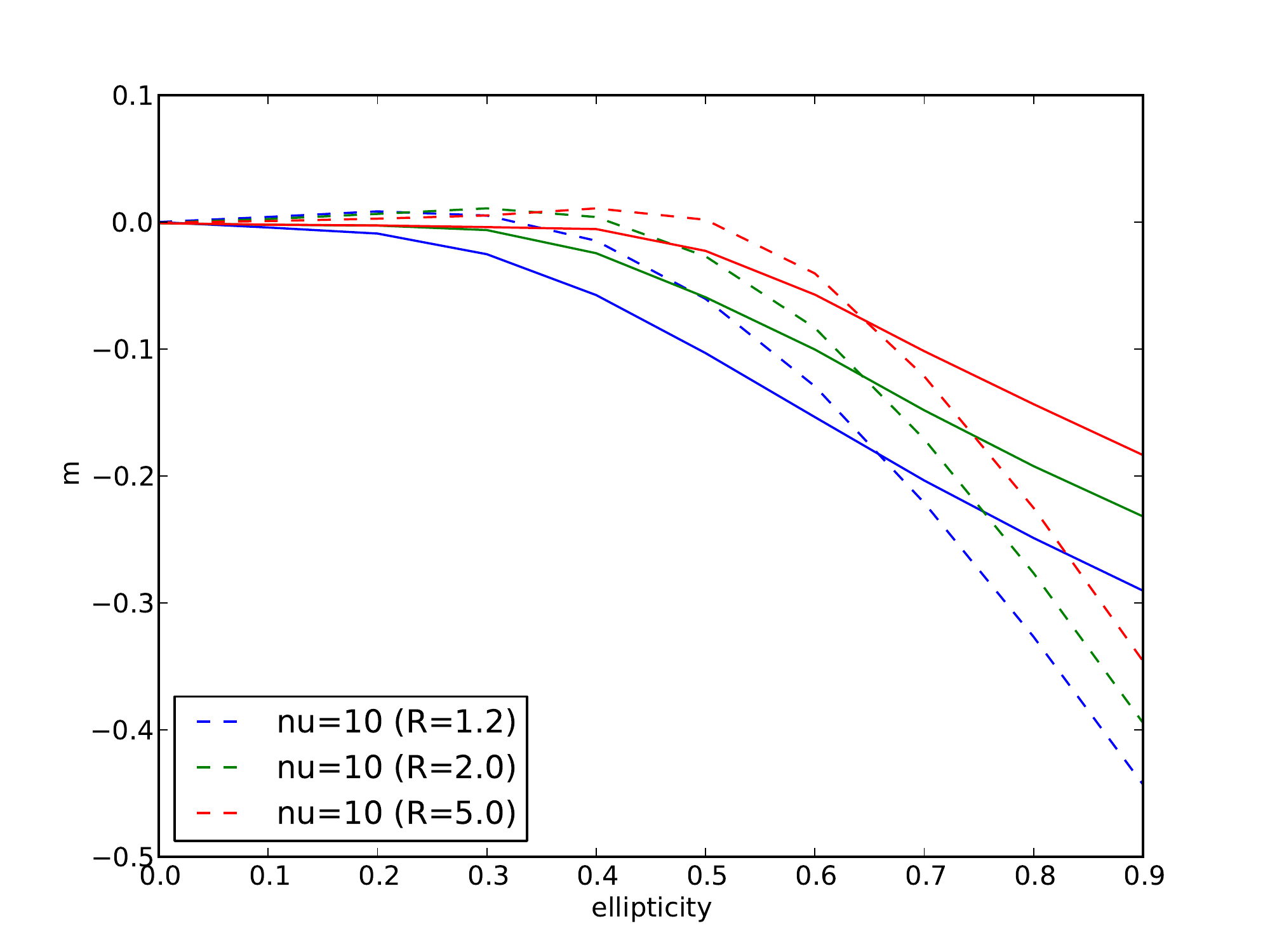}
\caption{Multiplicative bias $m$ as a function of the object ellipticity for three different resolutions $R=(1.2, 2.0, 5.0)$.  The signal-to-noise is $\nu=10$ and the weighting function has been chosen to be 1.2 times the object semi-major axis. The solid lines correspond to the case when $\chi$ is used as an ellipticity estimator, while the dashed lines to the case when $\epsilon$ is defined as an ellipticity estimator. }
\label{fig:BiasWeightPSF}
\end{figure}

The bias of the observed $\epsilon$-ellipticity as a function of the object resolution is shown in Figure \ref{fig:BiasWeightPSF}. 
When the object is much larger than the PSF the ellipticity bias is the same as the one discussed in the previous section and its amplitude and direction can be understood from the properties of the Marsaglia-Tin distribution. 
If the size of the object becomes comparable to the size of the PSF then the probability distribution of the convolved ellipticity is the convolution of the Marsaglia-Tin distribution with the probability distribution of the resolution $R$ as shown in equation (\ref{eq:MarsagliaPSF}). 

\subsection{The observed noisy ellipticity distribution}

For a moment-based approach each measured ellipticity can be seen as one possible realisation of a Marsaglia-Tin distribution, defined with respect to some true ellipticity (which is unknown), and a signal-to-noise level.

The observed noisy ellipticity distribution measured from the data is in fact a combination of the Marsaglia-Tin distribution and the sheared intrinsic ellipticity distribution convolved with the PSF. 
To be more precise, it is a marginal joint probability distribution where the marginalisation is performed over the intrinsic ellipticity distribution:
\begin{equation}
p(e^{obs}|\nu)=\int p_M(e^{obs} | e^{s},\nu) p_{c}(e^{s})  \mathrm{d} e^{s} 
\label{eq:obsDis}
\end{equation}
where $p_M(x)$ is the Marsaglia-Tin distribution (either in its original form in the case the galaxy is well resolved, or modified according to Eq. \ref{eq:MarsagliaPSF}) and $p_{c}(e^{s})$ the convolved (sheared) 
intrinsic ellipticity distribution. 

In this sense the Marsaglia-Tin distribution is the most fundamental probability function in lensing measurement since the observed ellipticity distribution and eventually the shear are derived from it. 
In the case of infinite signal-to-noise, the Marsaglia-Tin distribution reduces to a delta-function and therefore 
the observed ellipticity distribution corresponds to the convolved sheared intrinsic ellipticity distribution.

Given the likelihood distributions for $\chi$ and $\epsilon$, is it straightforward to generalise this to a Bayesian framework by including appropriate prior distributions. A comprehensive review of how to include such priors in polarisation distributions is given in \cite{Quinn12} (see also \cite{Vaillancourt06}). Polarisation measurements assume that the Stokes parameters are uncorrelated, which is not the case in weak lensing, however
a similar exercise can be performed. Given an expected posterior distribution for $\chi$ 
from the measurements, one approach would be to include this distribution directly in theoretical estimates of the cosmic shear power spectrum; we reserve this for future work.
 
As the mean of the Marsaglia-Tin distribution does not correspond in general to the true input ellipticity value, 
the average of the marginalised joint probability of the Marsaglia-Tin distribution and an intrinsic ellipticity distribution does not correspond to the true shear. This has been already shown for example in Figure \ref{fig:BiasWeightPSF} of \cite{Melchior12}. In the rest of the paper we will refer to this bias in shear as the \textit{Marsaglia bias}.

\subsection{Intrinsic ellipticity distribution and shear bias}

We have studied so far how much the measurement of an object's ellipticity is affected by noise. Here we propagate the bias in the $\epsilon$-ellipticity into a bias in the shear. The shear is always computed as an average of the ellipticity over a population of objects in order to remove the effect of their random orientation. Since the Marsaglia bias is a function of the ellipticity, the bias on the shear will depend (among other things) on the form of intrinsic ellipticity distribution. This is a crucial point that has often been neglected in weak lensing analyses (see \cite{Kitching08} for an investigation into its effect on lensfit shape measurement). In order to illustrate the effect that the ellipticity distribution has on the amplitude of the bias, we consider the same three cases as illustrated in Figure \ref{fig:BiasWeightPSF}, namely a bias in $\epsilon$-ellipticity coming from employing a circular weighting function in the moment measurements and coming from the deconvolution of the PSF.

\begin{figure}
\includegraphics[width=9cm, angle=0]{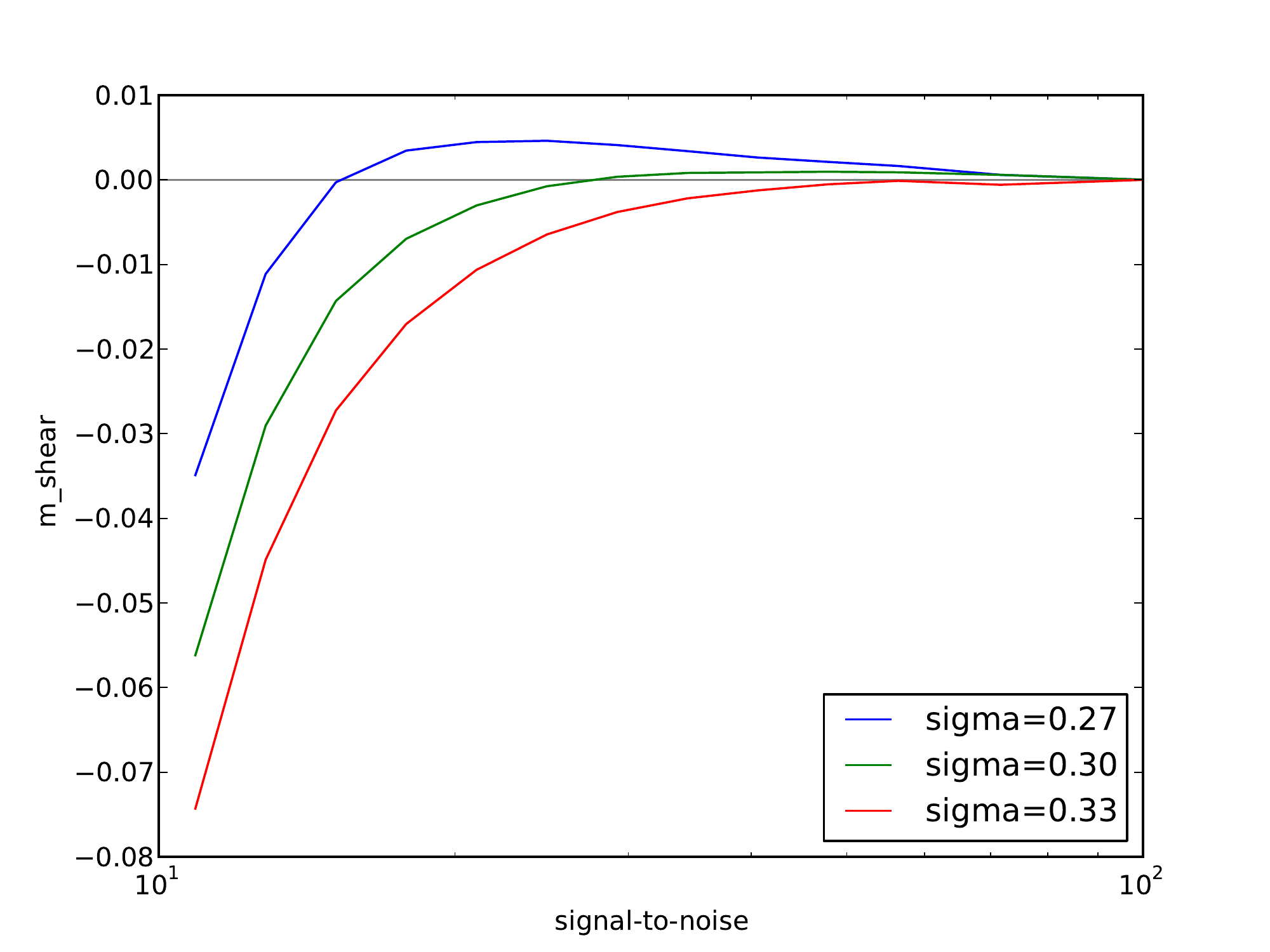}
\caption{Shear multiplicative Marsaglia bias as a function of signal-to-noise. The three curves represent the case of galaxies with intrinsic ellipticities following a Rayleigh distribution with $\sigma_{\epsilon}=0.27$  (blue), $\sigma_{\epsilon}=0.3$ (green) and $\sigma_{\epsilon}=0.33$ (red). The width of the weighting function has been chosen to be 1.2 times the object semi-major axis. This plot highlights the importance of knowing the ellipticity distribution in order to calibrate the shear bias.}
\label{fig:MarsagliaIe}
\end{figure}

We assume that the intrinsic $\epsilon$-ellipticity distribution can be described in terms of a Rayleigh distribution:
\begin{equation}
\label{eq:intrinsiceps}
p(|\epsilon|, \sigma_{\epsilon}) = \frac{|\epsilon|}{\sigma_{\epsilon}^2}\exp{\left(-\frac{|\epsilon|^2}{2\sigma_{\epsilon}^2}\right)}\;.
\end{equation} 
This particular function form is motivated by fitting the $\epsilon$-ellipticity distribution as measured in the 
GEMS catalogue \citep{Haussler09}. Typically $\sigma_{\epsilon} \sim 0.3$.
The multiplicative bias in the shear is then computed by averaging the multiplicative bias in the $\epsilon$-ellipticity over this distribution.\footnote{The fact that the ellipticity multiplicative bias depends on the ellipticity is not only responsible for a shear multiplicative bias but also for an additive bias scaling with $\langle m(\epsilon)\epsilon \rangle$, where the average is taken over the intrinsic ellipticity distribution.}
The results are shown in Figure \ref{fig:MarsagliaIe} for three different values of the intrinsic $\epsilon$-ellipticity dispersion, $\sigma_{\epsilon}=0.27$ and $\sigma_{\epsilon}=0.3$ and $\sigma_{\epsilon}=0.33$. 
The bias is typically negative at low signal-to-noise and it tends to zero at high signal-to-noise. It is interesting to notice that it can also be positive for some intermediate range of $\nu$ and low intrinsic elliptcity dispersion. This is due to the fact that the $\epsilon$-ellipticity bias can be positive for low values of the ellipticity as discussed in the previous section.

\section{Propagation of requirements}

The Marsaglia bias, in combination with the model bias and the method bias contributes to the total multiplicative and additive bias present in an ellipticity catalogue. How large this bias can be depends on the width and the depth of a given survey, or more generally it depends on its statistical power. In short, the requirements on the amplitude on the bias are always defined such that the systematic bias is lower than the statistical error. In this way the full statistical power of the survey can be exploited. For a more specific discussion on how the requirements on the multiplicative bias are set we refer to \cite{Massey13}.
Current surveys typically require the noise (multiplicative) bias to be lower than $10^{-2}$, for upcoming surveys, like KiDS\footnote{\url{http://kids.strw.leidenuniv.nl/}} or DES\footnote{\url{http://www.darkenergysurvey.org/}}, lower than $2\times 10^{-3}$ and for future surveys, 
such as Euclid\footnote{\url{http://euclid-ec.org}} \citep{Laureijs11}, lower than $5 \times 10^{-4}$. 

It is important to notice that these are actually not requirements on the absolute amplitude on the bias, but rather on its knowledge. 

Typically, the shear multiplicative bias is a function of galaxy morphology, signal-to-noise, resolution, PSF size, as investigated in detail by e.g. \cite{Bridle09, Kitching12}. Furthermore the bias depends also on the intrinsic ellipticity distribution of the galaxies. We stress here again that this is a consequence of the fact that the bias in the ellipticity is a function of the ellipticity itself. 
We can Taylor-expand the shear multiplicative bias as: 
\begin{equation}
m(\vec{\theta}(\sigma_{\epsilon}))\simeq m_{0}(\vec{\theta})+\frac{\partial m(\vec{\theta})}{\partial \sigma_{\epsilon}}\mathrm{d}\sigma_{\epsilon}
\label{eq:shearM}
\end{equation}
where $\vec{\theta}=(\nu,R,...)$ is a vector of the object properties. 
We denote with $m_0$ the multiplicative bias function is computed for a fiducial intrinsic ellipticity distribution. $m_0$ was the only multiplicative term estimated in \citep{Heymans06, Massey07b} and \citep{Bridle09}. In \cite{Kitching12} both the $m_0$ multiplicative term and also the variation (first derivative) of $m$ as a function of PSF ellipticity and size were evaluated (referred to as $\alpha$ in that paper).

The first-order term indicates the variation of the multiplicative bias function as a function of the width of the intrinsic ellipticity distribution. The amplitude of this term is crucial to quantify how calibratable a method is, as it will become evident in the following.

It is then apparent that the requirements that the multiplicative bias after calibration ($m-m_{0}$) has to be smaller than a given number, translates immediately into requirements on the knowledge of the intrinsic ellipticity distribution. A larger dependence of the multiplicative bias on the intrinsic ellipticity distribution (i.e. larger the ellipticity bias as a function of the ellipticity) translates into a tighter requirement on the knowledge of $\sigma_{\epsilon}$.

From equation (\ref{eq:shearM}) it is clear that in the limiting case of a perfectly realistic simulation in all points in the $\vec{\theta}$-parameter space or of an extremely deep version of the data \textit{any} method can be perfectly calibrated independently of the amplitude of the original bias. In fact in this case all the partial derivatives will vanish and the only quantity that has to be measured is $m_0$. 

In all other cases it is crucial to propagate the requirements on the multiplicative bias to requirements either on the knowledge of some physical quantities describing galaxy morphology, such as intrinsic ellipticity and size, which have a strong influence on the amplitude of the bias, or on the requirements on the observational depth of some fields that might be used for calibration. 
 
\subsection{Requirements on the intrinsic ellipticity distribution}

\begin{figure}
\centering
\includegraphics[width=9cm, angle=0]{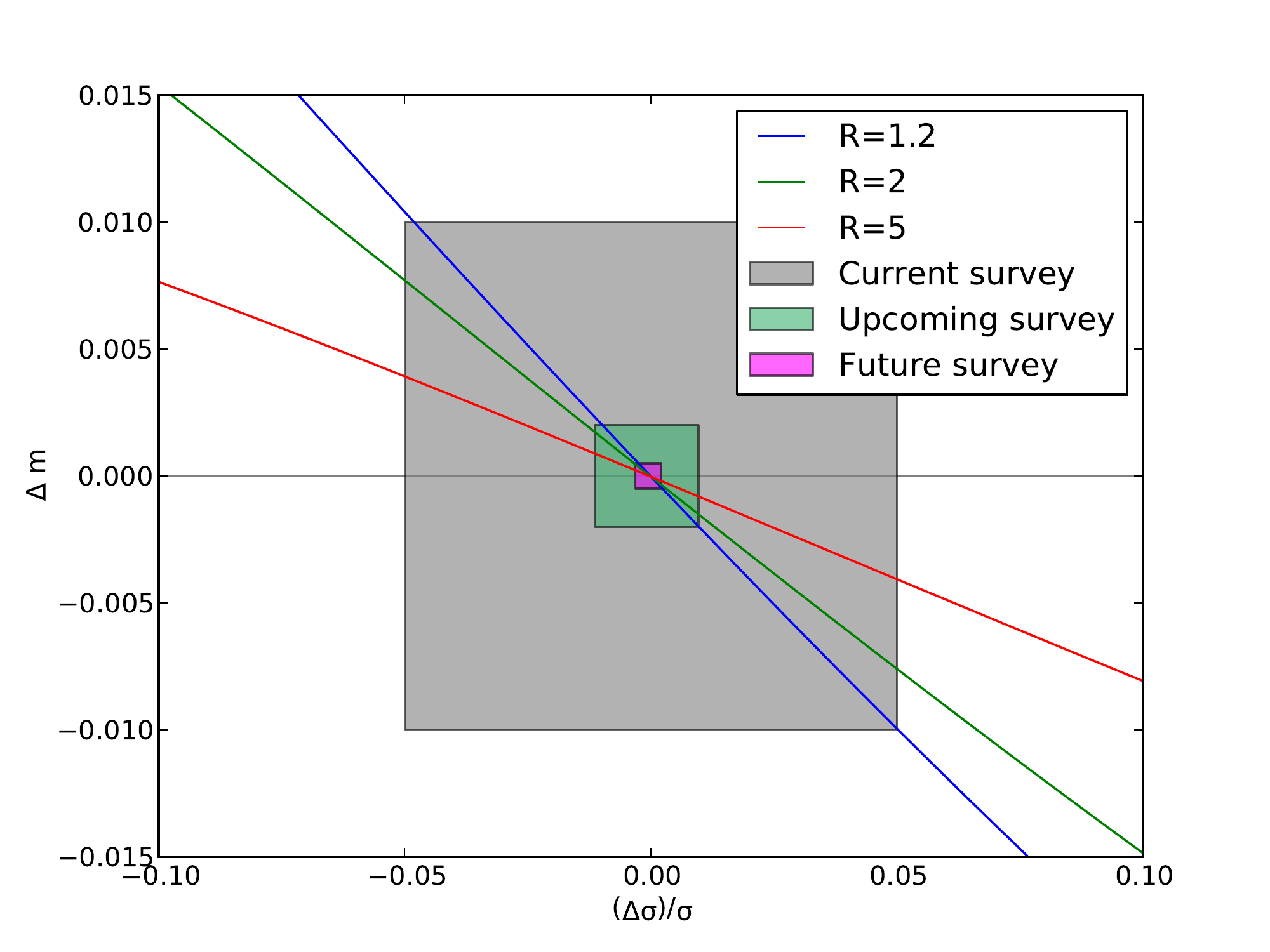}
\caption{Variation of the shear multiplicative bias $m$ as a function of the relative variation of the width of the intrinsic $\epsilon$-ellipticity distribution. We assume here as a fiducial value $\sigma_{\epsilon}=0.30$ and a signal-to-noise of $\nu=10$. The PSF is circular and the object resolution is $R=1.2, 2.0, 5$.  The shaded areas correspond to the allowed region in the $\Delta m/\Delta \sigma_{\epsilon}$ plane such that the residual multiplicative bias $\Delta m$ is lower than $10^{-2}$ (current survey), $2\times 10^{-3}$ (upcoming survey) and $5\times 10^{-4}$ (future survey) for the case $\nu=10$ and $R=1.2$.}
\label{fig:deltaMdeltaS}
\end{figure}

We propagate here the requirements on the amplitude of the multiplicative bias into requirements on the knowledge of the intrinsic $\epsilon$-ellipticity distribution. This can be done by looking at the variation of the multiplicative bias $\Delta m=m(\sigma_{\epsilon})-m(\sigma_{\epsilon_{0}})$ as a function of the variation of the $\epsilon$-ellipticity dispersion $\Delta \sigma_{epsilon}=\sigma_{\epsilon}-\sigma_{\epsilon_{0}}$. In Figure \ref{fig:deltaMdeltaS} we show the result for the cases described in the previous section at signal-to-noise $\nu=10$. 

The galaxy resolution is fixed to be $R=1.2$ and the PSF is assumed to be circular symmetric (hence the additive bias is zero). Given the fact that the multiplicative Marsaglia bias does not depend on the ellipticity of the PSF, the latter assumption is not restrictive.
  
\begin{figure}
\centering
\includegraphics[width=9cm, angle=0]{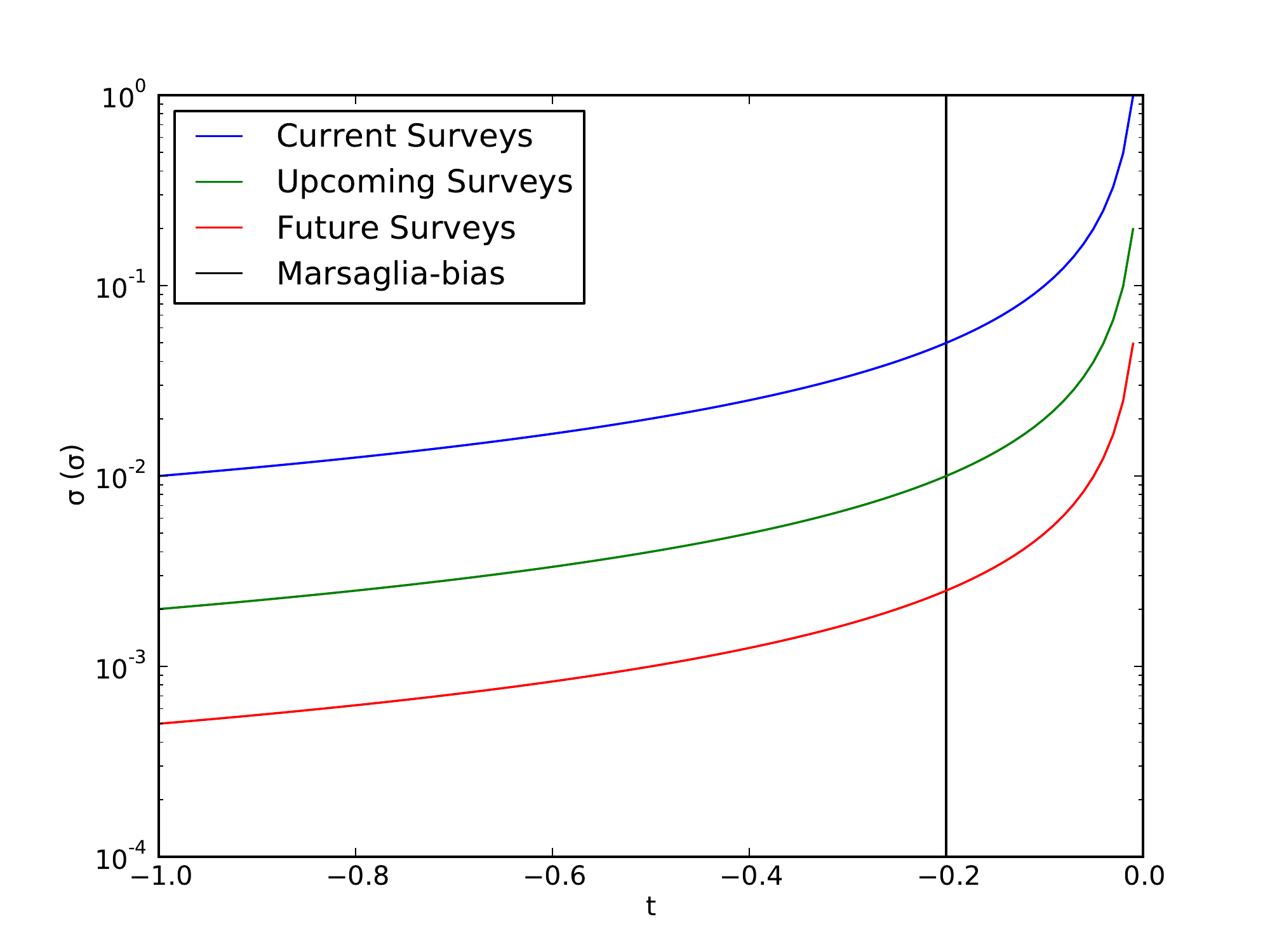}
\caption{Maximum tolerable relative uncertainty in the measurements of the intrinsic $\epsilon$-ellipticity distribution dispersion $\sigma_{\epsilon}$ as a function of the steepness of the variation of the multiplicative bias $m$ with the variation of $\sigma_{\epsilon}$ (equation \ref{eq:steep}) for current, upcoming and future surveys. The straight vertical line correspond to the case of the Marsaglia bias shown in Figure \ref{fig:deltaMdeltaS}. }
\label{fig:propReq}
\end{figure}

In general the variation of $m$ as a function of variation of $\sigma_{\epsilon}$ is well approximated by a linear relation:
\begin{equation}
\Delta m \simeq  t \frac{\Delta \sigma_{\epsilon}}{\sigma_{\epsilon}}
\label{eq:steep}
\end{equation}
The steepness of the line $t$ depends in general on signal-to-noise, galaxy properties and the method employed in the measurement. In Figure \ref{fig:propReq} we show the requirements on the knowledge of the intrinsic $\epsilon$-ellipticity distribution as a function of $t$ assuming that the only source of bias is the Marsaglia-bias. 
We find that if $t \simeq -0.2$ (this is the case for the Marsaglia bias discussed in this paper),  $\sigma_{\epsilon}$ has to be known  with a precision of $\sim 5\%$ in order to properly calibrate shear estimates for current surveys, for upcoming surveys with a precision of $\sim 1 \%$ and for future surveys with a precision of $\sim 0.3 \%$. 

Similar numbers can be derived for any method once the sensitivity of the calibration to variation of the intrinsic ellipticity distribution is known.

\section{Correcting for the noise bias}

In the previous sections we discussed the origin of the Marsaglia bias and we quantified its amplitude as a function of the intrinsic galaxy ellipticity, signal-to-noise, object resolution and size of the weighting function employed to measure the moments of an object.  Furthermore we derived requirements on the knowledge of the intrinsic ellipticity distribution from requirements on the multiplicative bias.

The focus of this section is to explore possible ways to meet those requirements and to derive reliable calibrations  to correct for the Marsaglia bias. 

In particular we explore two scenarios:
\begin{itemize}
\item \textbf{Using synthetic images for the calibration}: in this case the focus is the realism of the simulation. The requirements (for example) on the knowledge of the intrinsic ellipticity distribution have to be met in some external data set. This information is then used to make the simulation in combination with specific details about the PSF, the pixel size, the galaxy morphology etc. The shape measurement method is eventually calibrated on this simulation. 
\item \textbf{Self-calibration using a deeper version of the data}: in this case a representative sample of the galaxies in the survey is observed at higher signal-to-noise. The focus here is on how much deeper the observations have to be in order to accurately derive the noise calibration. 
\end{itemize}

\subsection{Approaches to calibration}
\label{sec:calibrationapproaches}

\subsubsection{Using synthetic data}

Image simulations are often used to derive a calibration for shape measurement methods. Typically, fitting formulae are derived that depend on signal-to-noise, galaxy size, PSF properties and so forth \citep[e.g.][]{Miller13}. 
Those fitting formulae are then applied (in a statistical sense) to the data. We already argued that those calibrations will be reliable only if the simulations are close to the real Universe. In particular we quantified in the previous section (see Figure \ref{fig:deltaMdeltaS}) the requirements on the knowledge of the intrinsic $\epsilon$-ellipticity dispersion in order for the shear calibration to meet the multiplicative-bias requirements for current, upcoming and future surveys.

The knowledge of the intrinsic ellipticity distribution with arbitrary precision is eventually limited by the signal-to-noise of the objects in the calibration sample. The pixel noise has the effect of rendering the ellipticity distribution  broader than it is in reality. Therefore we have to investigate the minimum signal-to-noise at which galaxies should be observed, in some data set, such that the difference between the true ellipticity dispersion and the measured one is lower than the requirements we derived in the previous section.

In order to address this question we simulate different $\epsilon$-ellipticity distributions having a fiducial dispersion $\sigma_{\epsilon_{0}}=0.3$ but with different values of the signal-to-noise. Each time we compute the `noisy' $\epsilon$-ellipticity dispersion and compare it with the fiducial value. We repeat the same experiment, changing also the object's resolution $R$.
The minimum signal-to-noise at which the galaxies should be observed is then the one for which $\Delta \sigma_{\epsilon}$ is just below the requirements.

\begin{figure}
\includegraphics[width=9cm, angle=0]{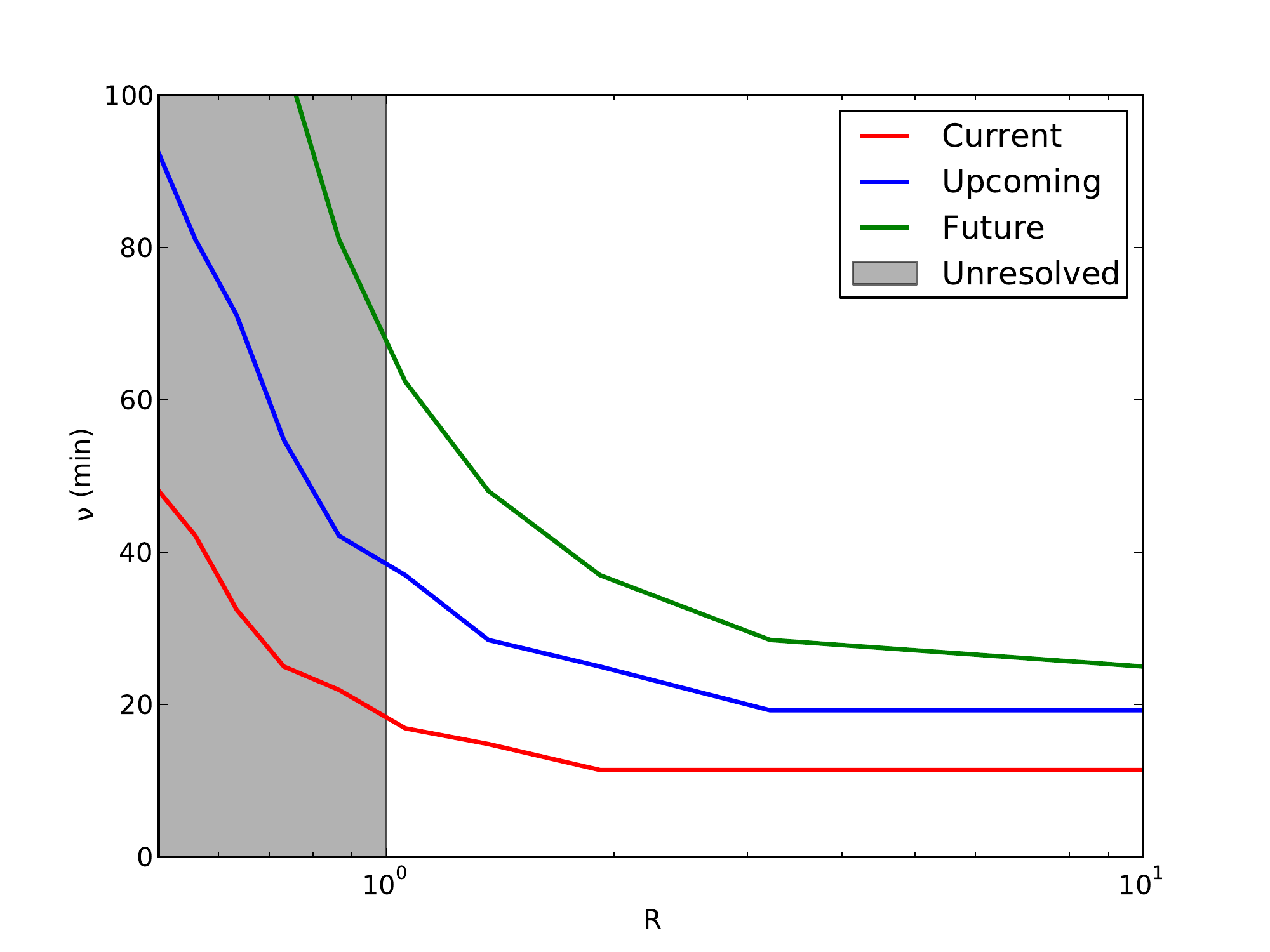}
\caption{Minimum signal-to-noise $\nu$ as a function of the object resolution $R$, as defined in equation (\ref{eq:resolution}), for which the relative difference between the measured intrinsic $\epsilon$-ellipticity distribution and the true one is lower than 0.05 (red curve), 0.01 (blue curve) and 0.003 (green curve). Those are the requirements derived in Section 4.1 for current, upcoming and future surveys. The PSF is assumed here to be circular.}
\label{fig:PrecisionGauss}
\end{figure}

The result is shown in Figure \ref{fig:PrecisionGauss} as a function of galaxy resolution. 
As expected, in order to infer the intrinsic $\epsilon$-ellipticity dispersion from a population of barely resolved objects (low $R$), observations at higher signal-to-noise are required. The reason is that the PSF deconvolution degrades with the inverse of the object resolution.
For quite well resolved objects ($R \sim 1.5$) we conclude that for current surveys observations of galaxies having signal-to-noise of 15 are enough to know the intrinsic $\epsilon$-ellipticity distribution with a precision of 5\%, for upcoming surveys the signal-to-noise has to be at least 30 in order to get to a percent precision. Finally for future surveys and a target of 0.3\%  in the precision of the $\epsilon$ ellipticity, the ellipticity dispersion of galaxies has to have been observed at least at a signal-to-noise of 60. 

Once the intrinsic ellipticity has been measured on some data set, it can be used as one of the inputs to generate synthetic data, which eventually could be used to calibrate shape measurement algorithms.

\subsubsection{Using deep observations}

Another possibility to calibrate, or at least to correct for, the Marsaglia bias is to have deeper observations of a significant sub-sample of the data. In this way the same object will be observed two times: one at low signal-to-noise where the bias is higher and one at high signal-to-noise where the bias is lower.  
The advantage of this approach is that the intrinsic ellipticity distribution in the low signal-to-noise and in the high signal-to-noise version of the data is the same. Moreover the PSF, the pixel size, the band width etc. are exactly the same in the two observations. Therefore in principle one does not have to know the intrinsic ellipticity distribution beforehand. 

This is an advantage since in principle the ellipticity distribution might have different shapes in different bands or its measurements might be harmed by the telescope PSF or by other detector effects.

The important question to answer is whether the high signal-to-noise version of the data (e.g. a deeper survey) is deep enough to remove completely the bias, or at least whether it is deep enough to calibrate the method such that the residual systematic is below some requirements. 
This information can be read off Figure \ref{fig:PrecisionGauss} for current, upcoming and future surveys.

\subsection{Requirements on calibration data}

Both of these approaches require observations of a certain number of galaxies at higher signal-to-noise, in the first case to determine the intrinsic ellipticity distribution, to be used as input of image simulations, and in the second case to suppress the Marsaglia bias. The need for a certain signal-to-noise and the maximum tolerable statistical uncertainty on the measurement of the variance of the intrinsic ellipticity leads to requirements on the depth and size of the calibration sample. These requirements are identical for both approaches, with the key difference that the calibration of the synthetic data could be done with external data, as long as one is able to apply the same selection criteria as for the main survey, including for example observing galaxies in the same filter\footnote{We expect the intrinsic ellipticity distribution to be colour-dependent \citep{vanUitert12}.}, whereas the direct calibration using deeper observations has to be done with the same telescope, instrument, observing conditions, etc.

\subsubsection{Survey area}

Here we address the number of galaxies used for a calibration of the intrinsic ellipticity dispersion. We refer to 
\cite{Taylor13} who investigated errors on covariance estimates with a focus on simulations (in particular we refer to Section 5.3) that describes how the number of samples of the true data vector (referred to as $N_S$) relates to the desired fractional error on the covariance, $\Delta\sigma_{\epsilon}/\sigma_{\epsilon}$ in our notation. This implies
\be 
\label{req}
N_S\simeq N_{\rm bin}\left(\frac{2}{(\Delta\sigma_{\epsilon}/\sigma_{\epsilon})^2}\right)\;,
\ee
where we have introduced $N_{\rm bin}$ which is the number of independent bins (e.g. in redshift) in which the variance of the ellipticity needs to be measured. $N_S$ can be thought of as the number of galaxies observed, at high signal-to-noise, in the calibration sample.  

\begin{figure}
\includegraphics[width=9cm, angle=0]{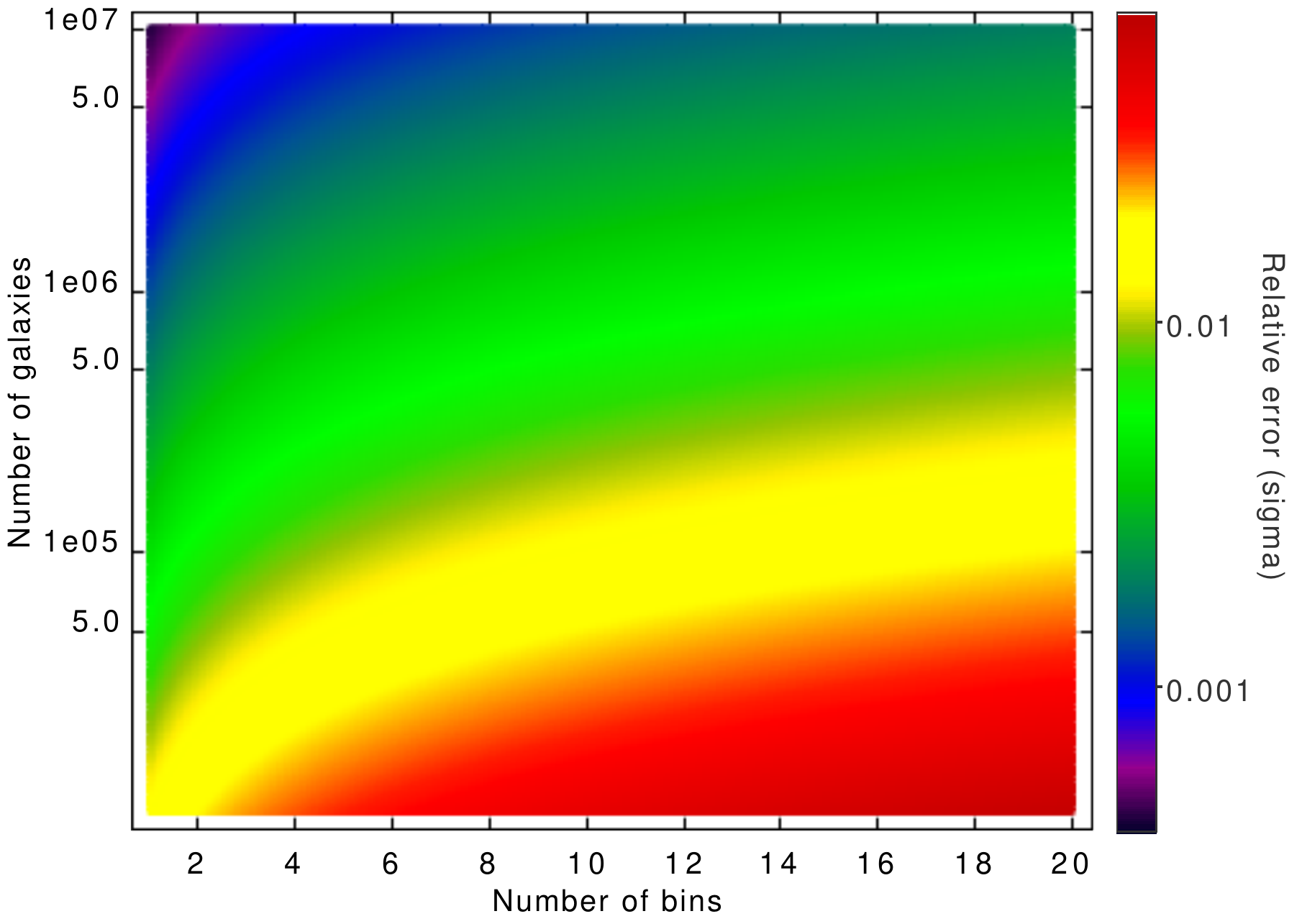}
\caption{Relative error $\Delta\sigma_{\epsilon}/\sigma_{\epsilon}$ as a function of the number of independent redshift bins $N_{\rm{bin}}$ and of the number of observed galaxies $N_S$. The relative error on the knowledge of the intrinsic $\epsilon$-ellipticity distribution can be computed by propagating requirements on the precision of the shear measurements. }
\label{fig:NsNb}
\end{figure}

In Figure \ref{fig:NsNb} we show the relative error $\Delta\sigma_{\epsilon}/\sigma_{\epsilon}$ as a function of the number of independent redshift bins $N_{\rm bin}$ and of the number of observed galaxies $N_S$. Fixing $N_{\rm bin}$ we can read off the number of galaxies we need to observe in order to meet the requirements on $\Delta\sigma_{\epsilon}/\sigma_{\epsilon}$. 
This number can be further propagated into a requirement on the area that a calibration survey has to have as $\mathrm{Area} = N_S/n_{\mathrm{eff}}$, where $n_{\mathrm{eff}}$ is the effective number density of galaxies.

For example, a survey like Euclid would require $N_S \simeq 5\times 10^6$ to be observed in a calibration survey (assuming 10 independent redshift bins). Assuming a galaxy number density of $30\,{\rm arcmin}^{-2}$, a calibration survey should have an area of $\simeq 45 \mathrm{deg^2}$. This number is quite similar to the area of the Euclid deep survey \citep{Laureijs11}, which is planned to cover $40 {\rm deg}^2$. Similar calculations can be done for other upcoming weak lensing surveys, the results being summarised in Table 2. 

\begin{table*}
\begin{minipage}{\linewidth}
\caption{Requirements for calibration data of upcoming and future weak-lensing surveys. For four on-going and planned surveys we list the expected effective galaxy number density $n_{\mathrm{eff}}$, the relative error on the width of the intrinsic ellipticity distribution $\Delta \sigma_{\epsilon}/\sigma_{\epsilon}$, the derived number of galaxies required for calibration $N_S$, the corresponding survey area, the minimum signal-to-noise required in the calibration survey $\nu_{\rm min}^{\rm deep}$, and the magnitude difference $\Delta \mathrm{Mag_{lim}}$ between the calibration data and the main survey. Note that for KiDS, DES, and HSC we assumed $N_{\rm bin}=5$ redshift bins, while for Euclid $N_{\rm bin}=10$ was chosen. The minimum signal-to-noise in the wide survey was set to $\nu_{\rm min}^{\rm wide}=10$ throughout.}
\label{tab:summary}
\begin{tabular}{lllll}
\hline
Quantity & KiDS & DES & HSC & Euclid\\
\hline
$n_{\mathrm{eff}}$ $\mathrm{[arcmin^{-2}]}$& 9\footnote{Preliminary measurement (KiDS team, priv. comm.). This is the number density of objects having a \textit{reliable} shape measurement.} & 12\footnote{Prediction taken from https://www.darkenergysurvey.org/reports/proposal-standalone.pdf} & 15\footnote{From Figure 6 of \cite{Chang13}} & 30\footnote{Prediction taken from \cite{Laureijs11}}\\[1ex]
$\Delta \sigma_{\epsilon}/\sigma_{\epsilon}$ & $7\times 10^{-3}$ &$7\times 10^{-3}$   & $7\times 10^{-3}$ & $2.2\times 10^{-3}$\\
$N_S$ & $2 \times 10^{5}$ & $2 \times 10^{5}$  & $2\times 10^{5}$ & $5 \times 10^{6}$ \\
Area of calibration field [${\rm deg}^2$] & 6.1 & 4.6 & 3.7 & 45\\
$\nu_{\rm min}^{\rm deep}$ & 30 & 30 & 30 & 60\\
$\Delta \mathrm{Mag_{lim}}$& 1.2 & 1.2& 1.2 & 1.9\\
\hline
\end{tabular}
\end{minipage}
\end{table*}
  
\subsubsection{Survey depth}


The calibration sample needs to be deeper than the main survey to reach the minimum signal-to-noise requirements determined from Figure \ref{fig:PrecisionGauss}. We can use this information to compute a requirement on the depth of the calibration data:\footnote{We assumed here that $\nu$ scales linearly with flux which is fair if the object is faint, so that Poisson noise is negligible.}
\begin{equation}
\Delta \mathrm{Mag_{lim}} \equiv 
m_\mathrm{lim}^{\mathrm{deep}} - m_\mathrm{lim}^{\mathrm{wide}} = -2.5[\log (\nu^{\mathrm{wide}}_{\mathrm{min}})-\log(\nu^{\mathrm{deep}}_{\mathrm{min}})]
\end{equation}
where $\nu^{\mathrm{wide}}_{\mathrm{min}}$ is the signal-to-noise of a source with magnitude $m_\mathrm{lim}^{\mathrm{wide}}$ in the wide survey and $\nu^{\mathrm{deep}}_{\mathrm{min}}$ is the minimum signal-to-noise in the calibration data, such that the noise-bias is below requirements.

For a Euclid-like survey, for which $\nu^{\mathrm{wide}}_{\mathrm{min}}=60$ is read off from Figure \ref{fig:PrecisionGauss}, the calibration sample should be $\simeq 1.9$ magnitudes deeper than the wide survey. The Euclid deep survey again fulfils this requirement by reaching 26.5 mag in the optical band, compared to 24.5 mag for the wide survey. Therefore this survey is well designed to be used for both calibration approaches outlined in Section \ref{sec:calibrationapproaches}. Again we list the corresponding results for other current and upcoming surveys in Table 2.

\section{Stokes parameters}

The Marsaglia bias we have discussed so far is a consequence of using the ellipticity as an estimator of the shear. Defining an ellipticity requires performing a non-linear operation on the image's pixels which causes the bias in the presence of noise.
The question we want to address in this final section is: Is it possible to avoid performing non-linear operations on the pixels but still be able to measure the shear? Or in other terms: is it possible to avoid using the ellipticity as a shear estimator? 
The answer to this question is affermative. Rather than the two components of the ellipticity and the object's size, we propose to constrain three parameters, the \textit{Stokes parameters} that were defined earlier.

The definition of these three numbers requires only linear operations on the pixels (a weighted sum in this case). Therefore, if the noise in the pixels is Gaussian, we expect the distribution of the Stokes parameters to also be Gaussian. An unbiased measurement of the ellipticity would be the ratio of the average of the Stokes parameters over many noise realisations,
\begin{equation}
\label{eq:chinew}
\chi=\frac{\langle u\rangle + i\langle v\rangle}{\langle s\rangle}.
\end{equation}
Note that this is not the same as the average of the ratio of the Stokes parameters over many noise realisations, which would be the common way to proceed. An early attempt of using the Stokes parameters to measure the shear can be found in \cite{Zhang11}. 

\begin{figure}
\includegraphics[width=9cm, angle=0]{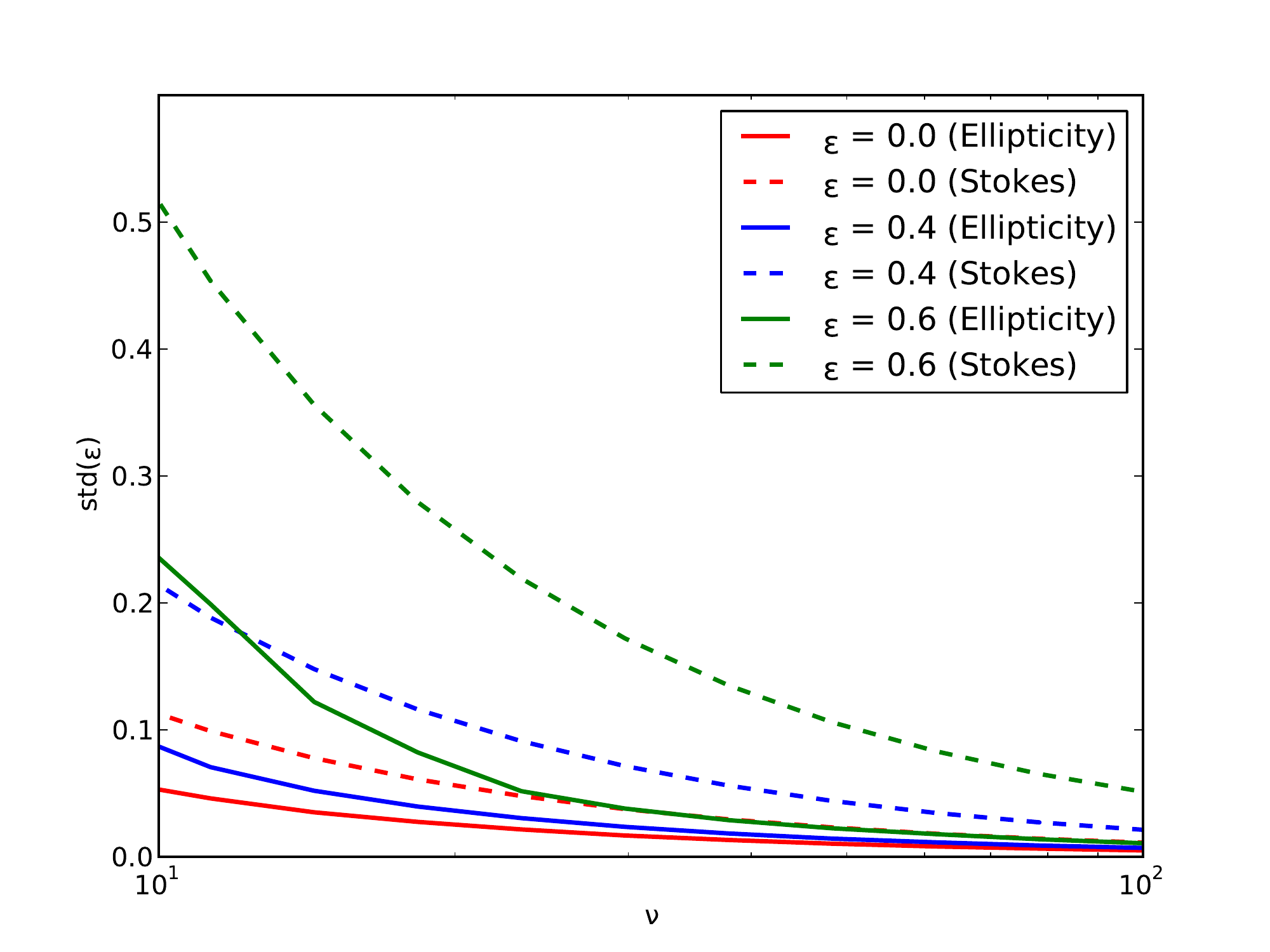}
\caption{Standard deviation of the $\epsilon$-ellipticity computed by averaging many noise realisations of the same object as a function of signal-to-noise $\nu$. The size and the flux of the object are here fixed. Solid lines correspond to the case in which the final  $\epsilon$-ellipticity is computed by averaging the $\epsilon$-ellipticities measured in each noise realisation, while dashed lines correspond to the case in which the final ellipticity is computed by the ratio of the average of the Stokes parameters measured in each noise realisation. Different colours correspond to different ellipticity values. }
\label{fig:VarianceComp}
\end{figure}

However, as can be seen in Figure \ref{fig:VarianceComp}, the standard deviation associated with the measurement according to equation (\ref{eq:chinew}) is much larger than the standard deviation associated with the $\epsilon$-ellipticity computed by averaging the ellipticities measured in each noise realisation. 
In the latter case the standard deviation has been computed numerically from $10^6$ objects sampled from the Marsaglia-Tin distribution, while in the case of the Stokes parameters the standard deviation can be analytically computed in the case of objects with elliptical Gaussian light profiles 
\begin{equation}
\sigma_{\frac{\langle u\rangle}{\langle s \rangle}}=\frac{\langle u \rangle}{\langle s \rangle}\sqrt{\left(\frac{\sigma^2_u}{\langle u\rangle^2}+\frac{\sigma^2_s}{\langle s \rangle^2}-\rho \frac{\sigma_u\sigma_s}{\langle u\rangle \langle s \rangle}\right)},
\end{equation} 
where $\sigma_u$ and $\sigma_s$ are defined in Section 2.3 via Eq. 14, $\rho$ is the correlation coefficient between $u$ and $s$, which is almost unity (and can also be analytically computed in the Gaussian case). Analogous formulae can be derived for  $\langle v \rangle / \langle s \rangle$.

The ratio between the standard deviation computed in the two cases depends on the ellipticity itself and it is in the range $[2..4]$ for ellipticities between $\chi=0$ and $0.6$. This means that, in order to get the same error in the final ellipticity measurement, roughly 10 times more objects have to be used in the case of the Stokes parameters. However the advantage of the Stokes parameters is that the noise bias vanishes.

\subsection{Lensing transformation}

We have shown that the Stokes parameters can be used in order to get an unbiased measurement of the ellipticity in case of noisy images, even though at the price of a larger variance. We show in this section how the Stokes parameters can be used to define a shear estimator.
We start by deriving their transformations under a lensing transformation: 
\begin{subequations}
\label{eq:StokesLensing}
\begin{align}
u^{s}=[u-2g_1 s+(g_1^2-g_2^2)u+2g_1g_2v]\times C \\
v^s=[v-2g_2s+(g_2^2-g_1^2)v+2g_1g_2u]\times C \\
s^s=[s(1+|g|^2)-2g_1u-2g_2v]\times C,
\end{align}
\end{subequations}
where $C\equiv(1-|g|^2)(1-\kappa)^4$ and $\kappa$ is the convergence. The superscript $s$ again denotes intrinsic quantities. The relation between the Stokes parameters and the shear looks more complicated than the one between the shear and the ellipticity (that can be derived from the above equations). However in the case of weak shear $g \ll 1$ the above equations can be linearised and the shear can be computed as follows,
\begin{subequations}
\label{eq:StokesLinear}
\begin{align}
g_1 \simeq \frac{\langle u \rangle}{2\, \langle s \rangle} \\
g_2 \simeq \frac{\langle v \rangle}{2\, \langle s \rangle}.
\end{align}
\end{subequations}
Note that in the usual approach we would have collapsed the Stokes parameters into two numbers (the ellipticity components) for any object and then we would have taken the average to get the shear:
\begin{equation}
(u,v,s)_{i} \Rightarrow \left(\frac{u}{s}, \frac{v}{s}\right)_{i} \Rightarrow \left(\left\langle\frac{u}{s}\right\rangle, \left\langle\frac{v}{s}\right\rangle\right).
\end{equation}
We propose instead to compute the average of the Stokes parameters and then to use this to compute the shear:
\begin{equation}
(u,v,s)_{i} \Rightarrow \left(\frac{\langle u \rangle}{\langle s \rangle}, \frac{\langle v \rangle}{\langle s \rangle} \right),
\end{equation}
it is clear that the two operations do not commute.

\subsection{Two-point statistics}

We define estimators of the two-point correlation functions of the Stokes parameter as follows,
\begin{equation}
\label{eq:xigeneraldef}
\hat{\xi}_A(x) \equiv \frac{1}{N_p(x)} \sum_{i \neq j} A_i A_j \Delta_x(i,j)\;,
\end{equation}
where the sum runs over all galaxies in a sample, and where $\Delta_x(i,j)$ is a selection function that is unity if galaxies $i$ and $j$ have an angular separation that falls into a bin centred on $x$ with width $\Delta x$. Then the number of galaxy pairs is given by 
\begin{equation}
N_p(x)= \sum_{i \neq j} \Delta_x(i,j)\;.
\end{equation}
Definition (\ref{eq:xigeneraldef}) holds for $A=(u,v,s)$ and is readily extended to incorporate weights for each galaxy, in analogy to the corresponding definition of the ellipticity correlation function; see e.g. \cite{Schneider02a}. Using equations (\ref{eq:StokesLinear}), one can calculate the ensemble average for the correlation functions, yielding 
\begin{subequations}
\label{eq:Stokeslll}
\begin{align}
\xi_s(x) \equiv \langle \hat{\xi}_s(x) \rangle = \langle s^2(x) \rangle = \bar{s}^2 \\
\label{eq:xis}
\xi_u(x) \equiv \langle \hat{\xi}_u(x) \rangle = 4 \langle g_1^2(x) \rangle \bar{s}^2 \\
\xi_v(x) \equiv \langle \hat{\xi}_v(x) \rangle = 4 \langle g_2^2(x) \rangle \bar{s}^2 ,
\end{align}
\end{subequations}
where $\bar{s}$ denotes the mean of the distribution of $s$. The shear acting on one galaxy and $s$ of another galaxy are uncorrelated to first order, so that the correlators can be split up. Note that noise terms originating from the intrinsic Stokes parameters do not appear in the averages because auto-correlations of galaxies have been excluded in Equation (\ref{eq:xigeneraldef}). A combination of these ensemble averages yields the equivalent of the ellipticity correlation function $\xi_+$,
\begin{equation}
\label{eq:xipdef}
\xi_+(x) = \frac{\xi_u(x) + \xi_v(x)}{4\, \xi_s(x)} = \langle g g^*(x) \rangle\;,
\end{equation}
where we inserted Equations (\ref{eq:Stokeslll}) to arrive at the second equality. In the following we will restrict ourselves to the analysis of $\xi_+$, but note in passing that an equivalent of $\xi_-$ can be defined in analogy to the ellipticity case by constructing tangential and cross components of the Stokes parameters $u$ and $v$ via
\begin{equation}
\nu_+ + \ic \nu_\times \equiv - (u + \ic v) \expo{-2 \ic \varphi}\;, 
\end{equation}
where $\varphi$ denotes the polar angle of the line connecting a pair of galaxies.

Following the work of \cite{Schneider02a}, it is possible to derive an analytic expression for the covariance of the Stokes parameter correlation function in Equation (\ref{eq:xipdef}), assuming a simple survey geometry, uniformly distributed galaxies, and a Gaussian-distributed shear field. This calculation is detailed in Appendix \ref{sec:cov}. In summary, we find that, after neglecting clearly subdominant terms originating from the scatter in $s$, the covariance can be cast into the same form as the covariance of the ellipticity correlation function $\xi_+$ when replacing the ellipticity dispersion per component $\sigma_\epsilon/\sqrt{2}$ with $\sigma_u/(2 \bar{s})$. Comparing with Equations (\ref{eq:StokesLinear}), this result is intuitive if errors on $s$ are negligible.

We conclude this section highlighting the differences and the similarity between the use of the Stokes parameters and stacking technique methods proposed for weak lensing measurements \citep{Lewis09}. Stacking techniques require averaging over pixels values from many galaxies. This is a linear operation and hence does not introduce bias in the limit that the centroid and the PSF are known exactly. The shear is then estimated from a high-signal-to-noise stacked image and hence it is not affected by noise bias. The limitation of this technique is that the stacking has to be done over regions where both the shear and the PSF are constant. The Stokes parameters, like standard stacking methods, combine information from many galaxy in a linear fashion before estimating a shear. However, as shown in this section, a shear correlation function can be defined in terms of correlation functions of the Stokes parameters. Hence the method can be used in case of a varying shear field. 

\subsection{Performance of estimators}

In order to test the performance of the shear estimator based on the Stokes parameters we simulate 50 Gaussian random fields of size $400\,{\rm deg}^2$ each, containing a total of about $2 \times 10^8$ grid points. The shear fields are determined by an input convergence power spectrum of the form
\begin{equation}
P_{\kappa}=\frac{9H_0^4\Omega_m^2}{4c^2}\int^{w_{H}}_{0}\mathrm{d}w \frac{W^2(w)}{a^2(w)}P_{\delta}\left(\frac{l}{w},w\right)
\end{equation}
where $W(w)$ is the lensing kernel which depends on the redshift distribution of source galaxies \citep{Bartelmann01}, $w$ is the comoving distance, $a$ the scale factor, $P_{\delta}$ the matter power spectrum, $\Omega_m$ the matter density of the universe and $H_0=h\times 100$Kms$^{-1}$Mpc$^{-1}$ the Hubble constant.   We assume a spatially flat $\Lambda$CDM cosmology with parameters $\Omega_{\rm m}=1 - \Omega_\Lambda=0.27$, $\Omega_{\rm b}=0.045$, $h=0.73$, $n_{\rm s}=1$, and $\sigma_8=0.8$. The matter power spectrum is computed using the transfer function by \cite{Eisenstein99} and non-linear corrections according to the fit function by \cite{Smith03}. The redshift distribution is of the form $p(z) \propto z^2 \exp \bbc{-(z/z_0)^{1.5}}$, truncated at $z=2$, and with $z_0$ chosen such that the median redshift is 0.7.

We then assign a size, a magnitude and an intrinsic ellipticity to each grid point and compute a noise realisation of this particular object as described in Section 3.3.2. In particular, we associate five numbers with each grid point: the two components of the ellipticity (drawn from the Marsaglia-Tin distribution) and the three Stokes parameters (jointly drawn from a Gaussian distribution). 
The PSF is assumed to be circular and the objects well resolved ($R=2$). The intrinsic ellipticity distribution is of the form given in Equation (\ref{eq:intrinsiceps}) with dispersion $\sigma_{\epsilon}=0.3$. The true size and the magnitude of the objects are jointly drawn from the distributions measured in the DEEP2 survey\footnote{http://deep.ps.uci.edu/DR4/zcatalog.html} \citep{Newman13}.

\begin{figure}
\includegraphics[width=9cm, angle=0]{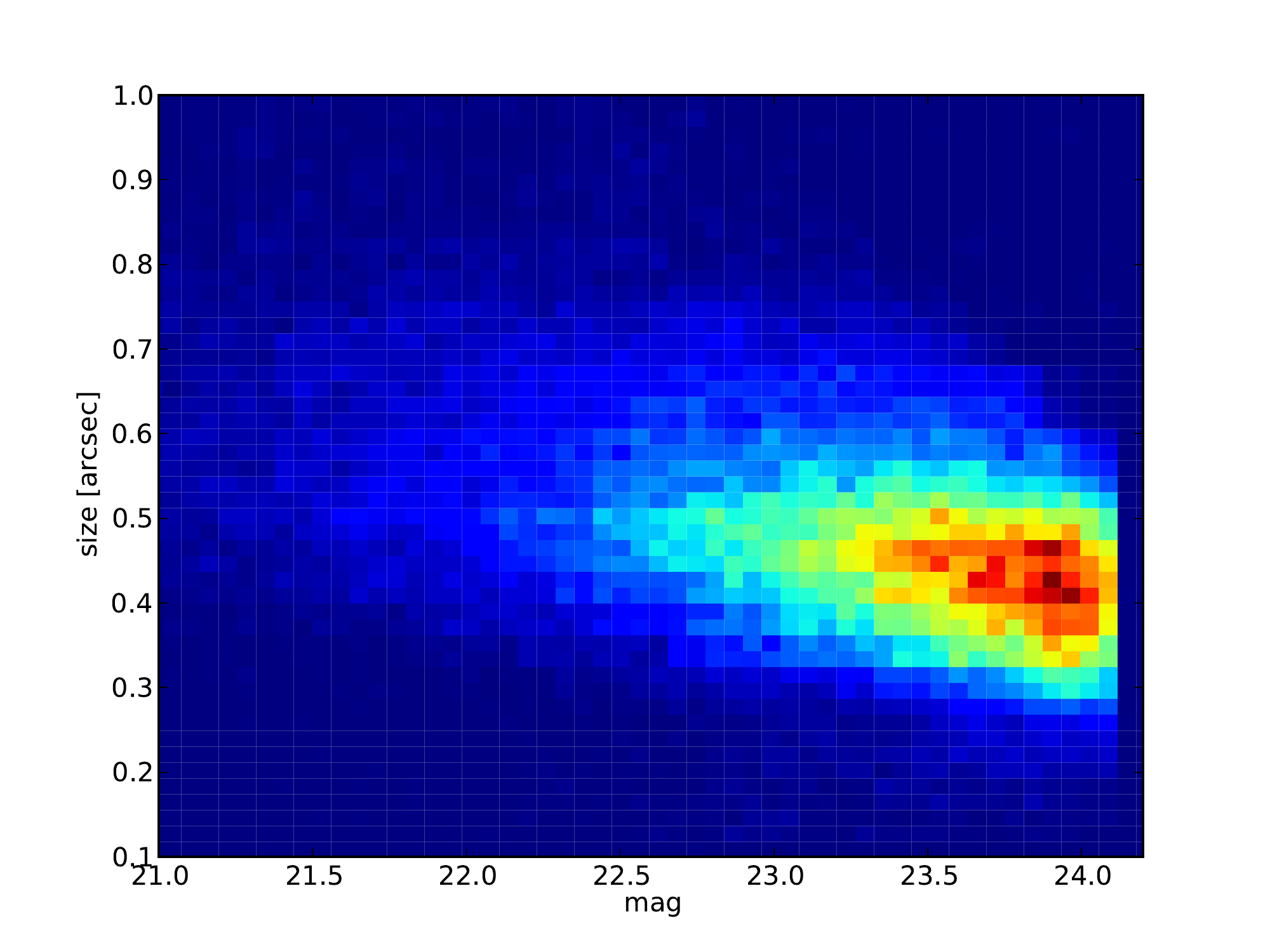}
\caption{Size-magnitude distribution for objects from the DEEP2 survey \citep{Newman13}}
\label{fig:deep2}
\end{figure}

The joint size-magnitude distribution is shown in figure \ref{fig:deep2}.

\begin{figure}
\includegraphics[width=9cm, angle=0]{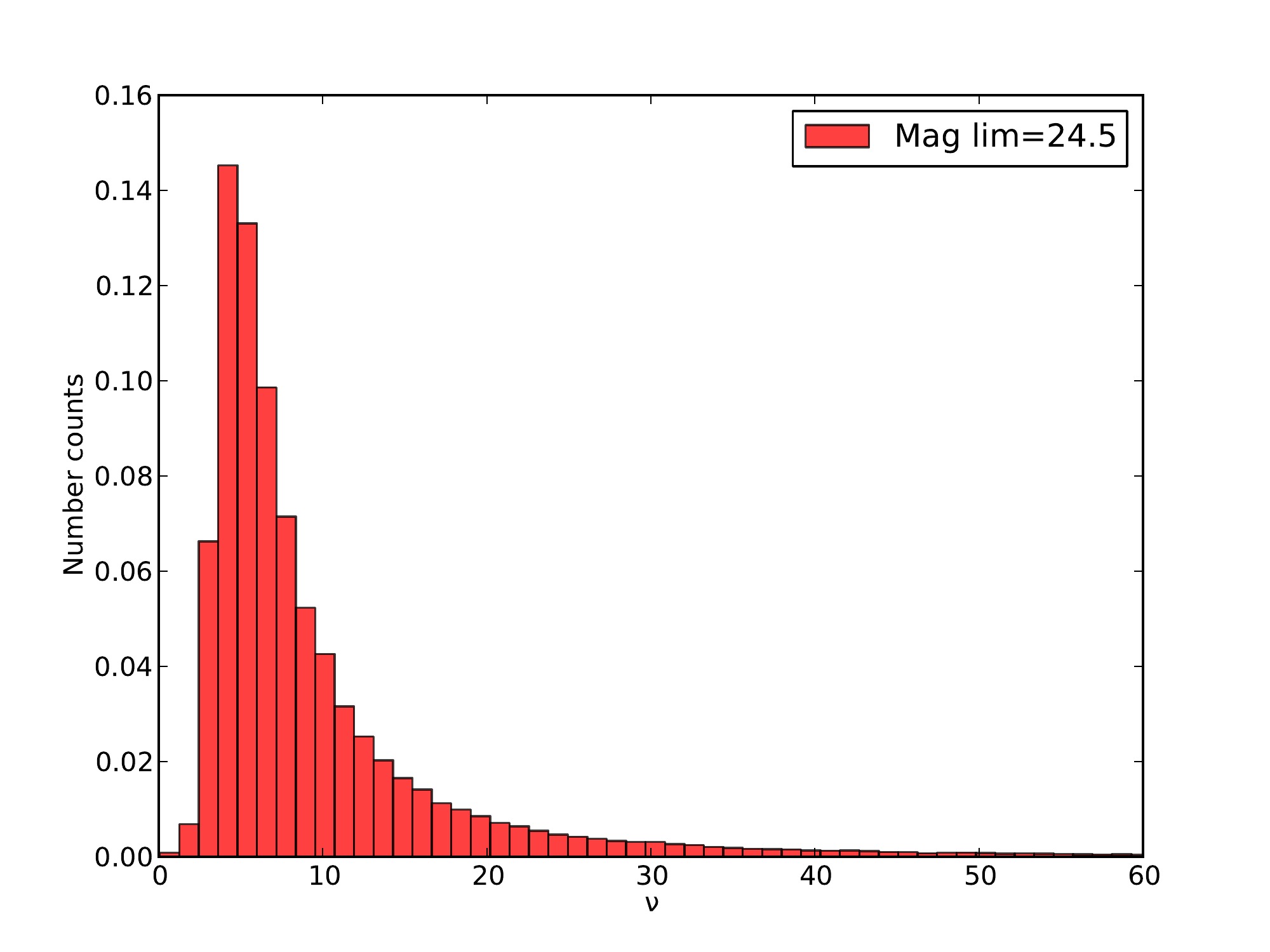}
\caption{Intrinsic signal-to-noise distribution for the mock galaxies in the 50 Gaussian Random Fields. This is derived by sampling the size and the magnitude from the DEEP2 galaxy catalogue and setting the magnitude of the background to be 26.5. }
\label{fig:signal2noise}
\end{figure}

To this end we select all galaxies with quality flag $\geq 3$ (secure redshifts) and create a two-dimensional histogram from the parameters RG (dispersion of Gaussian fit to the light distribution of a galaxy in the $R$-band) and MAGR (CFHT $R$-band magnitude).
We fix the magnitude limit, corresponding to a 10 $\sigma$ source detection, to be $\mathrm{mag}_{\lim}=24.5$. We use this information to compute the magnitude of the background, corresponding to 1 $\sigma$ source detection. This number is then converted into a flux and used to compute the signal-to-noise $\nu$ . The final signal-to-noise distribution is shown in Figure \ref{fig:signal2noise}.

In a real analysis all the objects at signal-to-noise lower than 5 would be eliminated from the catalogue, or they would be heavily down-weighted in the final analysis. On the contrary in what follows we decide to keep all the objects in the catalogue. The reason is that in this way the Marsaglia bias is larger and can in fact be measured in a statistically significant way using only 50 Gaussian random fields. Hence the following results should not be considered as a prediction of the expected bias in a given survey but rather are meant to illustrate the gains and the losses of using the Stokes parameters instead of the ellipticity to compute the shear correlation function. 

The ellipticity correlation function and the correlation functions of the Stokes parameters are computed using the publicly available tree code ATHENA\footnote{http://www2.iap.fr/users/kilbinge/athena/}.
Before computing the correlation function we introduce some cuts in the Stokes parameters catalogue. As previously discussed, the dispersion of the Stokes parameters is very large, in particular for large and bright objects. Those objects preferentially sit in the long tails of the distributions, and hence increase the scatter. Moreover in presence of noise it is very easy to produce outliers with values orders of magnitude larger than the true one (this is the case especially for highly elliptical objects).

 If those objects are included in the catalogue the variance increases dramatically. We find that two cuts help in getting rid of those outliers: $s>0$ and $\sqrt{u^2+v^2}<M$, where $M$ has to be chosen according to the flux values (we remind the reader again that $u,v,s$ are not normalised by flux in this analysis in order to preserve the Gaussianity of their distribution in presence of noise). In our case we try several cuts in the range $M \in [10^5..10^7]$ and we report the results for three specific cuts $M=[10^{5}, 10^{6}, 3\times 10^6]$. 
The cuts helps to get rid to the large variance in the correlation function at the expense of introducing a bias, since the original distribution is altered. 
\begin{figure*}
\includegraphics[width=\columnwidth, angle=0]{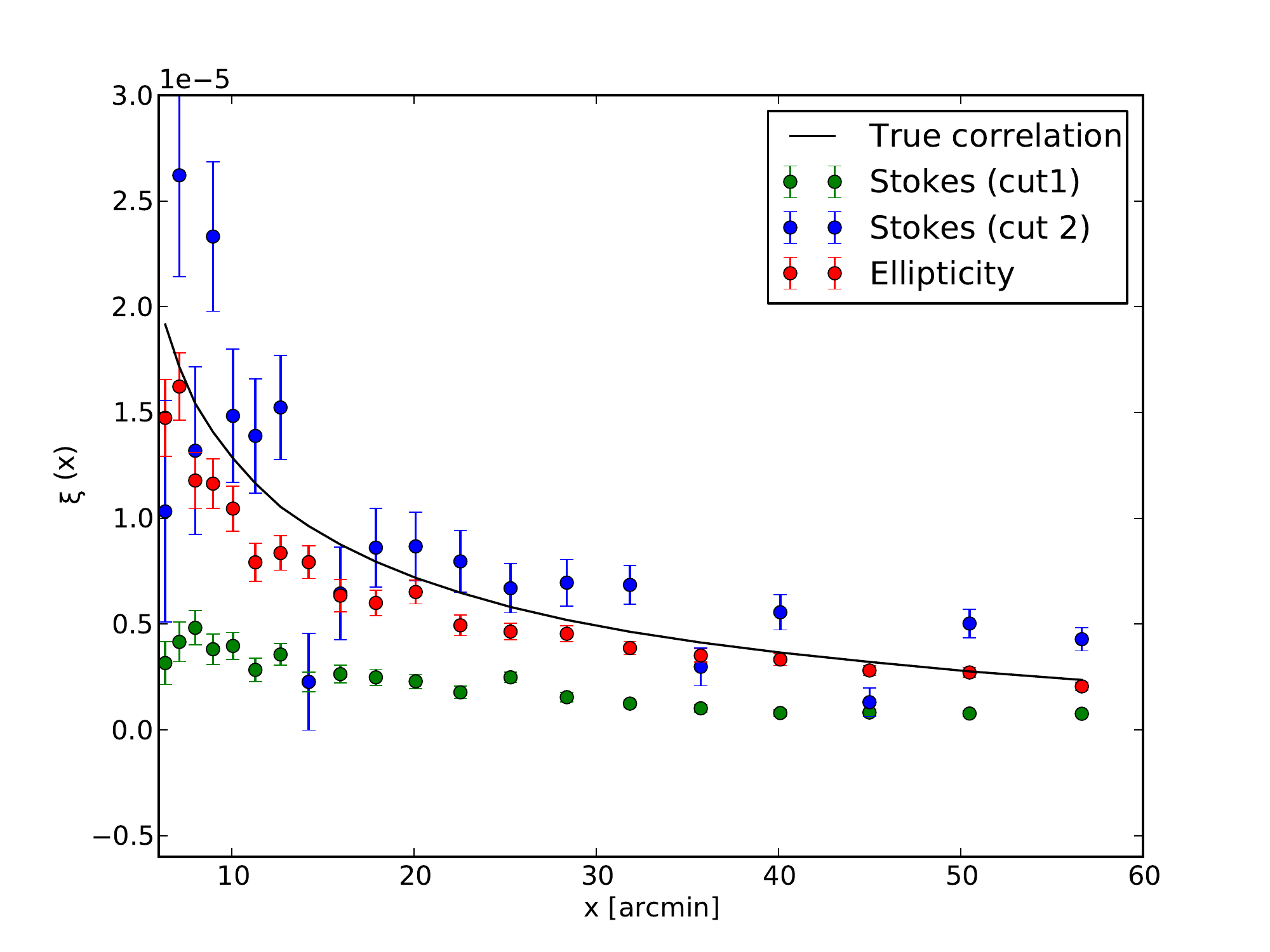}
\includegraphics[width=\columnwidth, angle=0]{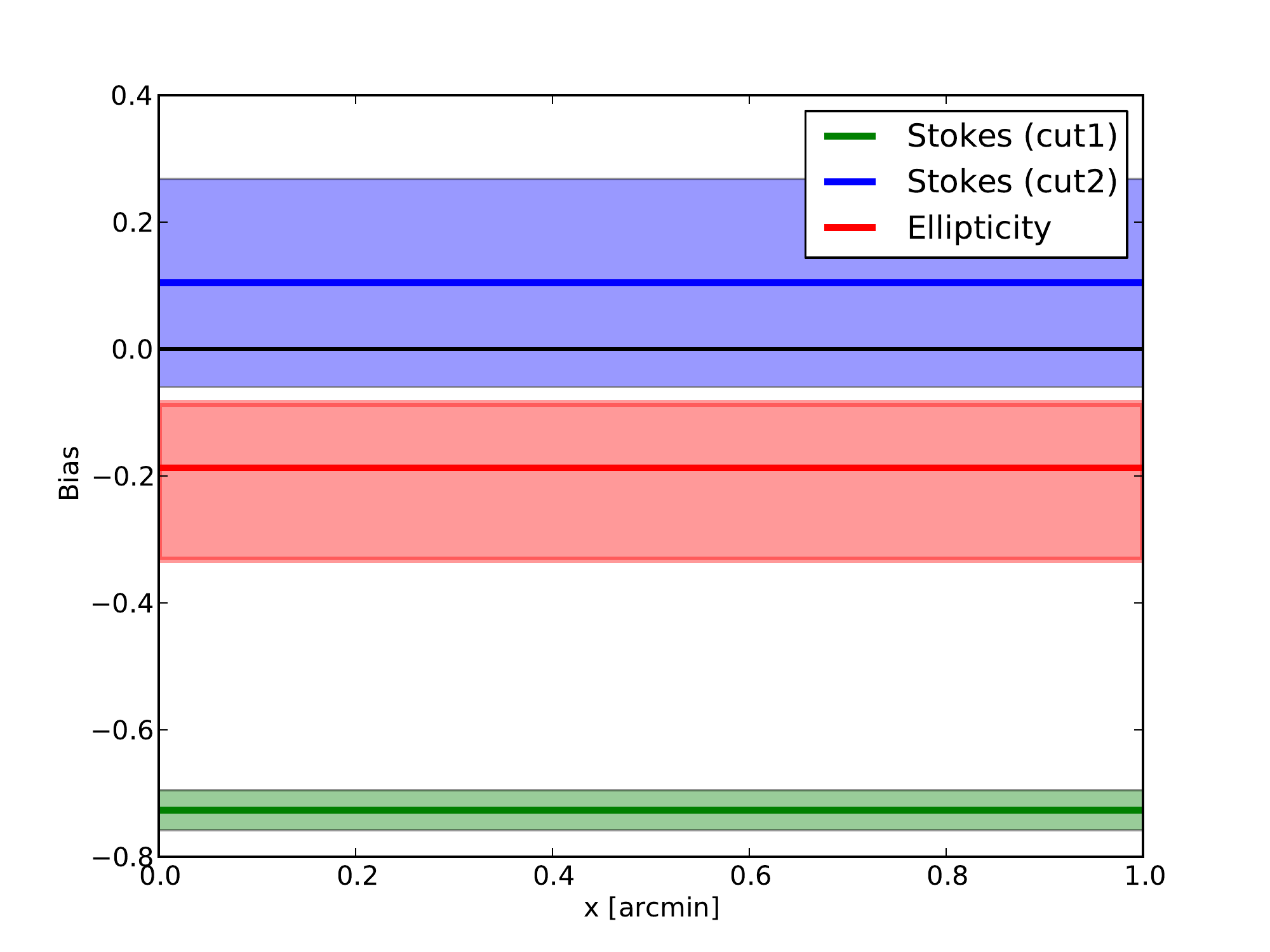}
\includegraphics[width=\columnwidth, angle=0]{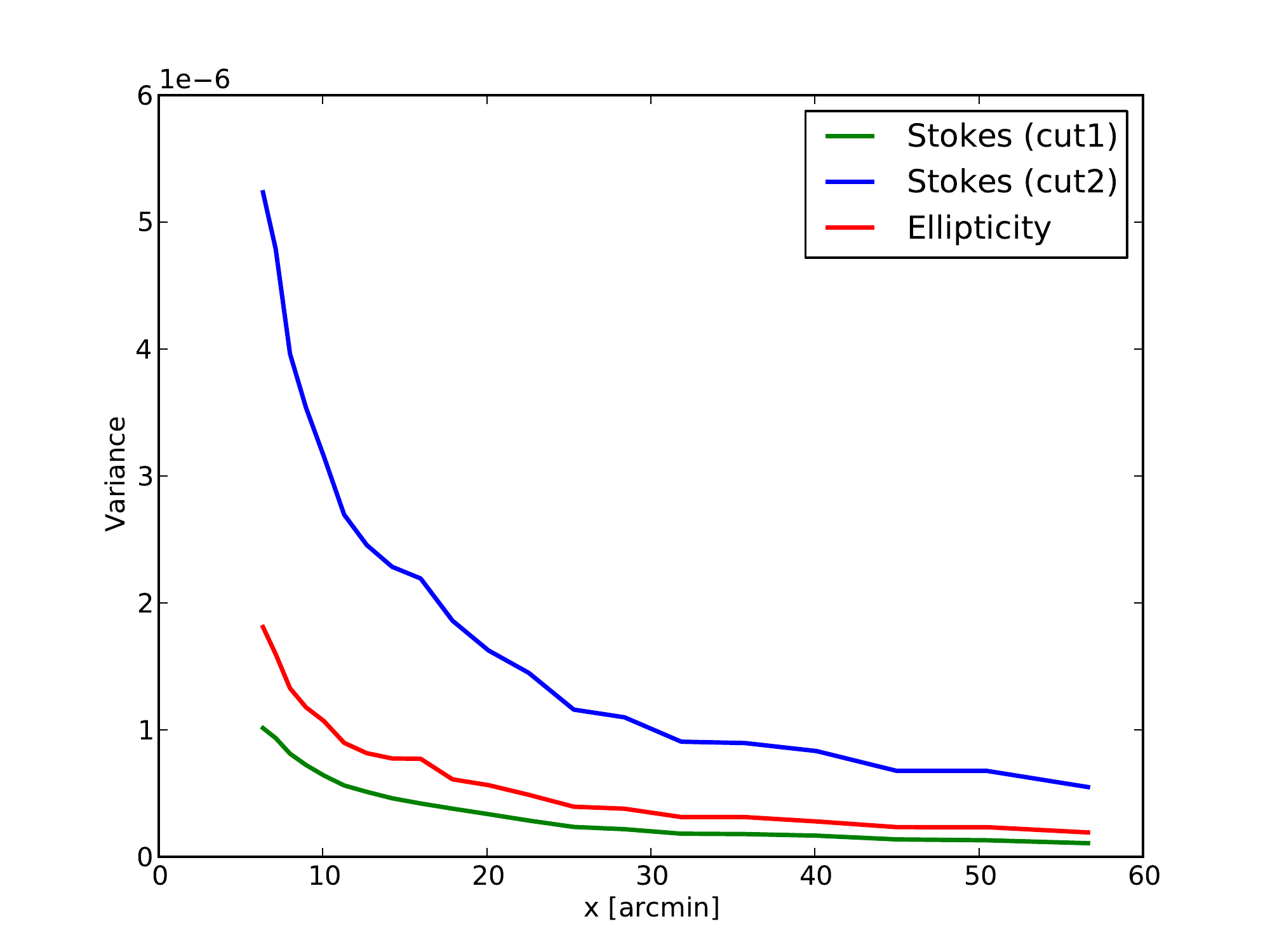}
\caption{\textit{Top left panel:} Shear correlation function measured using Stokes parameters (green and blue curves) and elliptcity (red curve) as shear estimate. \textit{Top right panel:}  Averaged bias over all angular scales in the shear correlation function and 1-$\sigma$ confidence region when the correlation function is computed using ellipticity (red curve) and the Stokes parameters (green and blue). \textit{Bottom panel:} Variance on the shear correlation function as a function of angular separation for the same three cases.
In all panels the green and blue curves correspond to different cuts in the catalogue in the case the Stokes parameters are used. In particular blue corresponds to the case $\sqrt{u^2+v^2} < 3\times 10^6$ and green $\sqrt{u^2+v^2} < 10^5$.   }
\label{fig:Stokes}
\end{figure*}

The effect of the cuts in the catalogue can be seen by comparing the green, blue and red lines in Figure \ref{fig:Stokes}. The more stringent the cut, the larger is the bias and the smaller is the variance. 
We finally compare the shear correlation functions as derived by using the Stokes parameters and by using the ellipticity as a shear estimator. We notice that it is possible to reduce the bias in the shear correlation function by using the Stokes parameters, but at the expense of a larger variance. In particular, using $\sqrt{u^2+v^2} <  10^6$ as a cut in the catalogue the final bias is lower by a factor of two but the variance is larger by almost the same factor. 
The bottom line here is that by choosing different shear estimators it might be possible to trade the bias for the variance, or vice versa, depending on the application. 

The fact that the distribution of the (un-normalised) Stokes parameters is very broad might pose similar problems, in terms of precision of shear measurements, to all methods following this route to go from the pixels to a shear estimate. For example the recent Bayesian approach proposed by \cite{Bernstein13}, at least in the version using moments, might have to deal with very broad likelihood functions in moments space. Hence in order to have precise shear measurements a quite aggressive prior might have to be employed. Further investigations in this direction are required.

\section{Conclusions}

In this paper we generalise the results of \cite{Marsaglia65} and \cite{Tin65} to the case of multivariate correlated ratios of Gaussian distributed variables and derive an analytic expression for their joint probability. Within the context of weak gravitational lensing we explore the impact of the non-Gaussianity of this distribution on the measurement of ellipticity, defined as a correlated pair of Stokes parameters measurement from the surface brightness distributions of image pixel.  

In weak gravitational lensing shear estimates based on an object's ellipticity are inescapably biased because of pixel noise. In particular the cause of the bias is that the ellipticity, independently on how it is defined, requires one to perform some non-linear operation on the pixels. Therefore even if the noise in the pixels is Gaussian and uncorrelated, the noisy ellipticity distribution is not Gaussian. 

We showed in this paper how bias caused by the non-Gaussian nature of the Marsaglia-Tin distribution, so called  Marsaglia bias, depends primarily on the intrinsic ellipticity distribution, the object signal-to-noise and the galaxy resolution. We investigated how requirements in the amplitude of such bias $m \leq 10^{-2}$ (current surveys, like CFHTLS), $m \leq 2\times 10^{-3}$ (upcoming surveys, like KiDS, DES, HSC) and $m \leq 5\times 10^{-4}$ (future surveys, like Euclid) can be translated into requirements on the knowledge of the intrinsic ellipticity distribution. 
We found that the intrinsic ellipticity dispersion $\sigma_\epsilon$ has to be known with a precision of $\sim 5\%$ in order to properly calibrate shear estimates for current surveys, for upcoming surveys with a precision of $\sim 1 \%$ and for future surveys with a precision of $\sim 0.3 \%$. These numbers have been derived assuming a signal-to-noise 10 galaxy with a resolution $R=1.2$ (ratio between the convolved area of the object and the area of the PSF).

We then explored two possible scenarios to correct for the noise bias: using numerical simulations and using a deeper version of the data. 
In the first case the ellipticity distribution (among many other galaxy properties) has to be known from some external data set. We estimated the area and depth required in order to measure the intrinsic ellipticity distribution with a given accuracy and precision. For upcoming surveys like DES or KiDS the area has to be of $\sim 5 \deg$ while for future survey, like Euclid, the area has to be of  $\sim 50 \deg$.

In the second case we found that for future survey the depth of the observations required to correct for the Marsaglia has to be several magnitudes larger than the magnitude limit of the wide survey. This number is quite similar to the Euclid deep field.

Finally we explore the possibilities of using the Stokes parameters (the polarisation and the area of the object) to define a shear estimator. We showed that the shear estimator defined in this way is unbiased even in presence of noise, but the price to pay is a variance which is a factor of 3 larger (or more, depending on the cuts in the catalogue). 

The multivariate Marsaglia-Tin distribution is likely to have applications beyond that of weak lensing. In the astronomical context any data consisting  
of correlated polarisation measurements, for example from the Cosmic Microwave Background, will be described by such a distribution.

\section*{Acknowledgements}
We thank Henk Hoekstra, Lance Miller and Andy Taylor for useful discussions and a careful reading of the manuscript. We thank the anonymous referee for her/his useful comments which helped the presentation of this work. MV is funded by grant 614.001.103 from the Netherlands Organisation for Scientific Research (NWO) and from the European Research Council under FP7 grant number 279396. TDK is supported by a Royal Society URF grant. BJ acknowledges support by an STFC Ernest Rutherford Fellowship, grant reference ST/J004421/1. 

\bibliographystyle{mn2e}
\bibliography{/Users/massimoviola/Work/Bibliography/biblio}

\onecolumn
\appendix

\section{Posterior shear distribution}
Here we write down an expression for the shear posterior in terms of the observed and intrinsic ellipticity posteriors. We do not use this expression explicitly in this  paper, but we include it here for completeness. 
We start from equation (\ref{eq:shear2el}), linking the observed ellipticity $\epsilon$, the intrinsic ellipticity $\epsilon^{s}$ and the shear $g$. We rearrange it to get 
\be
g=\frac{\epsilon-\epsilon^s-|\epsilon|^2\epsilon^s+|\epsilon^s|^2\epsilon}{1+|\epsilon|^2|\epsilon^s|^2}
\ee
using the algebra of random variables we can write down the
probability of $g$ using the quotient distribution and the product distribution. 
In the following we do not expand integrals over variables of the form
$|x|^2$ we leave this for future work. 
Defining the numerator as $U=\epsilon-\epsilon^{s}-|\epsilon|^2\epsilon^{s}+|\epsilon^{s}|^2$ we have 
\ba
p_U(U=\epsilon-\epsilon^s-|\epsilon|^2\epsilon^s+|\epsilon^s|^2\epsilon|\epsilon,\epsilon^s)&=&[p_o(\epsilon)*p_e(-\epsilon^s)]\nn
&*&\left[-\int\frac{1}{|(\epsilon^s)^{\prime}|}p_e((\epsilon^s)^{\prime})p_{|\epsilon|^2}\left(\frac{|\epsilon|^2\epsilon^s}{(\epsilon^s)^{\prime}}\right){\rm d}(\epsilon^s)^{\prime}\right]\nn
&*&\left[\int\frac{1}{|\epsilon^{\prime}|}p_{\epsilon}(\epsilon^{\prime})p_{|\epsilon^s|^2}\left(\frac{|\epsilon^s|^2\epsilon}{\epsilon^{\prime}}\right){\rm d}\epsilon^{\prime}\right],
\ea
where a $*$ represents a convolution, and for the denominator $Z=1+|\epsilon|^2|e|^2$ we have 
\ba
p_Z(Z=1+|\epsilon|^2|\epsilon^s|^2|\epsilon,\epsilon^s)&=&\left[\int\frac{1}{|\epsilon^{s\prime}|^2}p_{|\epsilon^s|^2}(|\epsilon^{s\prime}|^2)p_{|\epsilon|^2}\left(\frac{|\epsilon|^2|\epsilon^s|^2}{|\epsilon^{s\prime}|^2}\right){\rm
    d}|\epsilon^{s\prime}|^2\right]
\ea
which, using the quotient distribution, gives the probability of the shear to be 
\ba
p(g)=\int
|(1+|\epsilon^{\prime}|^2|\epsilon^{s\prime}|^2)|p_U(U=g(1+|\epsilon^{\prime}|^2|\epsilon^{s\prime}|^2)|\epsilon^{\prime},\epsilon^{s\prime})p_Z(Z=1+|\epsilon^{\prime}|^2|\epsilon^{s\prime}|^2|\epsilon^{\prime},\epsilon^{s\prime}){\rm
  d}(1+|\epsilon^{\prime}|^2|\epsilon^{s\prime}|^2)
\ea
or, in its complete form 
\ba
p(g)&=&\int|(1+|\epsilon^{\prime}|^2|\epsilon^{s\prime}|^2)|\nn
&\huge\{&[p_o(\epsilon^{\prime})*p_{\epsilon^s}(-\epsilon^{s\prime})]\nn
&*&\left[-\int\frac{1}{|\epsilon^{s\prime}\prime|}p_{\epsilon^s}(\epsilon^{s\prime\prime})p_{|\epsilon|^2}\left(\frac{|\epsilon^{\prime}|^2\epsilon^{s\prime}}{\epsilon^{s\prime\prime}}\right){\rm d}\epsilon^{s\prime\prime}\right]\nn
&*&\left[\int\frac{1}{|\epsilon^{\prime \prime}|}p_{\epsilon}(\epsilon^{\prime \prime})p_{|\epsilon^s|^2}\left(\frac{|\epsilon^{s\prime}|^2\epsilon^{\prime}}{\epsilon^{\prime \prime}}\right){\rm d}\epsilon^{\prime\prime}\right]\huge\}\nn
&\huge\{&\left[\int\frac{1}{|\epsilon^{s\prime\prime}|^2}p_{|\epsilon^{s}|^2}(|\epsilon^{s\prime\prime}|^2)p_{|\epsilon|^2}\left(\frac{|\epsilon^{\prime}|^2|\epsilon^{s\prime}|^2}{|\epsilon^{s\prime\prime}|^2}\right){\rm
    d}|\epsilon^{s\prime\prime}|^2\right]\huge\}{\rm d}(1+|\epsilon^{\prime}|^2|\epsilon^{s\prime}|^2).
\ea
To complete the convolutions one needs the Jacobian between observed and intrinsic
ellipticities this has components
\ba
\frac{\partial \epsilon_1}{\partial \epsilon^{s}_1}&=&(1+2g_1\epsilon^{s}_1+g_1^2-g_2^2)D^{-1}-N_R(2g_1+2g_1^2\epsilon^{s}_1+2g_2^2\epsilon^{s}_1)D^{-2} \nn
\frac{\partial \epsilon_2}{\partial \epsilon^{s}_2}&=&(1+2g_2\epsilon^{s}_2+g_2^2-g_1^2)D^{-1}-N_I(2g_2+2g_1^2\epsilon^{s}_2+2g_2^2\epsilon^{s}_2)D^{-2} \nn
\frac{\partial \epsilon_1}{\partial \epsilon^{s}_2}&=&(2g_1\epsilon^{s}_2+2g_1g_2)D^{-1}-N_R(2g_2+2g_1^2\epsilon^{s}_2+2g_2^2\epsilon^{s}_2)D^{-2} \nn
\frac{\partial \epsilon_2}{\partial \epsilon^{s}_1}&=&(2g_1\epsilon^{s}_1+2g_1g_2)D^{-1}-N_I(2g_1+2g_1^2\epsilon^{s}_1+2g_2^2\epsilon^{s}_1)D^{-2}
\ea
where 
\ba
D&=&1+2g_1\epsilon^{s}_1+2g_2\epsilon^{s}_2+g_1^2(\epsilon^{s}_1)^2+g_1^2(\epsilon^{s}_2)^2+g_2^2(\epsilon^{s}_1)^2+g_2^2(\epsilon^{s}_2)^2 \nn
N_R&=&\epsilon^{s}_1+g_1+g_1(\epsilon^{s}_1)^2+g_1(\epsilon^{s}_2)^2+g_1^2\epsilon^{s}_1-g_2^2\epsilon^{s}_1+2g_1g_2\epsilon^{s}_2 \nn
N_I&=&\epsilon^{s}_2+g_2+g_2(\epsilon^{s}_1)^2+g_2(\epsilon^{s}_2)^2-g_1^2\epsilon^{s}_2+g_2^2\epsilon^{s}_2+2g_1g_2\epsilon^{s}_1
\ea
the determinant of the Jacobian $J(\epsilon,\epsilon^{s},g)$ is then 
\ba
|J(\epsilon,\epsilon^{s},g)|=\left(\frac{\partial \epsilon_1}{\partial \epsilon^{s}_1}\frac{\partial
  \epsilon_2}{\partial \epsilon^{s}_2}-\frac{\partial \epsilon_1}{\partial
  \epsilon^{s}_2}\frac{\partial \epsilon_2}{\partial \epsilon^{s}_1}\right).
\ea
Note that if $\epsilon=0$ and $\epsilon^{s}=0$ then the determinant of the Jacobian
reduces to 
\ba
|J(0,0,g)|=[1-|g|^2]^2,
\ea
which is similar to the shear responsivity or sensitivity factor used in some shape measurement methods.
In the linearised case $g=\epsilon-\epsilon^{s}$ the shear posterior is a much more simple expression $p(g)=p_o(\epsilon|g,\epsilon^{s})*p(-\epsilon^{s})$.

\section{Ellipticity de-weighting}
To compute the unweighted ellipticity given a weighted ellipticity (drawn from the Marsaglia-Tin distribution), and a known circular Gaussian weighting function, we need the weighted quadrupole moments for an elliptical Gaussian object. These are given by:
\begin{equation}
Q_{11}^{w}=\frac{2\pi s^4s_w^4(s^2+(1+2\epsilon_1 + |\epsilon|^2)s_w^2}{s^4+2(1+|\epsilon|^2)s^2s_w^2+(1-|\epsilon|^2)^2s_w^4}
\end{equation}
\begin{equation}
Q_{22}^{w}=\frac{2\pi s^4s_w^4(s^2+(1-2\epsilon_1 + |\epsilon|^2)s_w^2}{s^4+2(1+|\epsilon|^2)s^2s_w^2+(1-|\epsilon|^2)^2s_w^4}
\end{equation}
\begin{equation}
Q_{12}=\frac{4\epsilon_2 \pi s^4s_w^6}{s^4+2(1+|\epsilon|^2)s^2s_w^2+(1-|\epsilon|^2)^2s_w^4}
\end{equation}
where $s_w$ is the size of the weighting function and $s$ and $\epsilon$ are the unweighted size and the ellipticity of the object. The unweighted moments are then computed by numerically solving the system of equations above; this is done in practice by using a standard root finding algorithm.

\section{Covariance of the Stokes parameters correlation function}
\label{sec:cov}

The covariance of $\xi_+$ as given by equation (\ref{eq:xipdef}) can be expressed in terms of the covariances of the correlation functions $\xi_{u,v,s}$ via standard error propagation,
\eqa{
\nn
\cov{\xi_+(x_1)}{\xi_+(x_2)} &=& \frac{1}{16\, \bar{s}^4}\; \left\{ \cov{\xi_u(x_1)}{\xi_u(x_2)} + \cov{\xi_u(x_1)}{\xi_v(x_2)}\right.\\ \nn
&& \hspace*{-3.5cm} \left. +\; \cov{\xi_v(x_1)}{\xi_u(x_2)} + \cov{\xi_v(x_1)}{\xi_v(x_2)} \right\} + \frac{\xi_+(x_1)\; \xi_+(x_2)}{\bar{s}^4}\; \cov{\xi_s(x_1)}{\xi_s(x_2)}\\ 
\label{eq:errorprop}
&& \hspace*{-3.5cm} +\; \frac{\xi_+(x_1)}{4\, \bar{s}^4}\; \bbc{ \cov{\xi_s(x_1)}{\xi_u(x_2)} + \cov{\xi_s(x_1)}{\xi_v(x_2)} } + \frac{\xi_+(x_2)}{4\, \bar{s}^4}\; \bbc{ \cov{\xi_u(x_1)}{\xi_s(x_2)} + \cov{\xi_v(x_1)}{\xi_s(x_2)} }\;,
}
where we made use of Equation (\ref{eq:xis}). Using the definition of the correlation function in Equation (\ref{eq:xigeneraldef}), their covariance is given by
\eq{
\label{eq:covgeneraldef}
\cov{\xi_A(x_1)}{\xi_B(x_2)} = \frac{\sum_{i,j,k,l} \Delta_{x_1}(i,j)\; \Delta_{x_2}(k,l)\; \bba{A_i A_j B_k B_l}}{N_p(x_1)\; N_p(x_2)} - \xi_A(x_1)\; \xi_B(x_2)\;,
}
where $A,B \in \bbc{u,v,s}$. To evaluate the four-point correlator, we shall assume that all fields follow Gaussian statistics, so that the correlator can be expressed in terms of two-point correlators via Wick's theorem. In what follows we will make frequent use of the following correlators,
\eqa{
\label{eq:u2ptcorr}
\bba{ u_i^s u_j^s } &=& \bba{ v_i^s v_j^s } = \delta_{ij} \sigma_u^2\;;\\
\bba{ s_i s_j } &=& \bar{s}^2 + \delta_{ij} \sigma_s^2\;;\\
\label{eq:remaining2ptcorr}
\bba{ u_i^s g_{\mu j} } &=& \bba{ u_i^s s_j } = \bba{ s_i g_{\mu j} } = 0\;, ~~~\mu=1,2\;,
}
and analogous for $v$. Note that $u^s$ and $v^s$ have the same dispersion $\sigma_u$ due to isotropy. In contrast to $u$ and $v$, $s$ has non-vanishing mean $\bar{s}$. The last suite of equalities holds because the gravitational shear acting on a galaxy is generally not correlated with the intrinsic size or elongation of that galaxy, nor does one expect a correlation between intrinsic sizes and elongations of galaxies on average.

To compute the covariance of $\xi_s$, we evaluate the correlator by applying Wick's theorem,
\eq{
\label{eq:scorr}
\bba{s_i s_j s_k s_l} = \bar{s}^4 + \bar{s}^2 \sigma_s^2 \br{ \delta_{ik} + \delta_{jl} + \delta_{il} + \delta_{jk} } + \sigma_s^4 \br{\delta_{ik} \delta_{jl} + \delta_{il} \delta_{jk} }\;.
}
Note that the distribution of $s$ is not symmetric in realistic cases, so that three- and four-point connected correlators do not vanish in general if all indices are equal. However, we can assume $i \neq j$ and $k \neq l$ throughout, as the correlation functions are not considered at zero lag, so that three or more equal indices are never encountered.

Following the approach of \cite{Schneider02a} (S02 hereafter), we determine the ensemble average over all galaxy positions, assuming a uniform distribution, to arrive at covariance formulae that are independent of the actual galaxy positions in a survey. The ensemble average operator is given by (S02)
\eq{
E \equiv \prod_{i=1}^{n_{\rm g} A_{\rm s}} \frac{1}{A_{\rm s}} \int_{A_{\rm s}} \dd^2 x_i\;,
}
where the product runs over all galaxies in the survey, which has size $A_{\rm s}$ and galaxy number density $n_{\rm g}$. Applying $E$ in complete analogy to S02, we obtain
\eq{
\frac{E\br{\sum_{i,j,k} \Delta_{x_1}(i,j)\; \Delta_{x_2}(j,k)}}{N_p(x_1)\; N_p(x_2)} = \frac{1}{n_{\rm g} A_{\rm s}}\;,
}
so that the full covariance of $\xi_s$ reads
\eq{
\cov{\xi_s(x_1)}{\xi_s(x_2)} = \delta_{x_1 x_2}\; \frac{2 \sigma_s^4}{N_p(x_1)} + \frac{4 \bar{s}^2 \sigma_s^2}{A_{\rm s} n_{\rm g}}\;,
}
where $\delta_{x_1 x_2}$ is a Kronecker symbol, so that the first term only contributes to the diagonal of the covariance.

Next we turn to $\cov{\xi_u(x_1)}{\xi_s(x_2)}$ for which we have to process the correlator
\eq{
\label{eq:uscorr}
\bba{u_i u_j s_k s_l} = \bba{u^s_i u^s_j s_k s_l} + 4 \bba{g_{1i} g_{1j} s_i s_j s_k s_l} = 4 \bba{g_{1i} g_{1j}} \bba{s_i s_j s_k s_l}\;,
}
where the remaining four-point correlator can be expanded via Equation (\ref{eq:scorr}). The correlators of shear appearing here and in similar correlators can be replaced with shear correlation functions as follows (S02)
\eqa{
\label{eq:g11}
\bba{g_{1i} g_{1j}} &=& \frac{1}{2} \bb{\xi_+(ij) + \xi_-(ij)\, \cos(4 \varphi_{ij})}\;;\\
\bba{g_{2i} g_{2j}} &=& \frac{1}{2} \bb{\xi_+(ij) - \xi_-(ij)\, \cos(4 \varphi_{ij})}\;;\\
\label{eq:g12}
\bba{g_{1i} g_{2j}} &=& \frac{1}{2}\, \xi_-(ij)\, \sin(4 \varphi_{ij})\;,
}
where $\xi_\pm(ij) \equiv \xi_\pm(|\vek{x}_i-\vek{x}_j|)$. Here, $\varphi_{ij}$ denotes the polar angle between $\vek{x}_i$ and $\vek{x}_j$. Inserting Equation (\ref{eq:uscorr}) into Equation (\ref{eq:covgeneraldef}), one needs to evaluate the following ensemble averages,
\eqa{
\label{eq:ensav1}
\frac{E\br{\sum_{i,j} \Delta_{x_1}(i,j) \bba{g_{1i} g_{1j}} }}{N_p(x_1)^2} = \frac{\xi_+(x_1)}{2\, N_p(x_1)}\;;\\
\label{eq:ensav2}
\frac{E\br{\sum_{i,j,k} \Delta_{x_1}(i,j)\; \Delta_{x_2}(i,k) \bba{g_{1i} g_{1j}}}}{N_p(x_1)\; N_p(x_2)} =  \frac{\xi_+(x_1)}{2\, n_{\rm g} A_{\rm s}}\;,
}
where we made use of Equation (\ref{eq:g11}). With these expressions at hand the full covariance simplifies to
\eqa{
\cov{\xi_u(x_1)}{\xi_s(x_2)} &=& \delta_{x_1 x_2}\; \frac{4 \sigma_s^4 \xi_+(x_1)}{N_p(x_1)} + \frac{8 \bar{s}^2 \sigma_s^2 \xi_+(x_1)}{A_{\rm s} n_{\rm g}}\\ \nn
&=& \cov{\xi_v(x_1)}{\xi_s(x_2)} = \cov{\xi_s(x_2)}{\xi_u(x_1)} = \cov{\xi_s(x_2)}{\xi_v(x_1)}\;,
}
where the equalities in the second line are determined analogously.

To calculate the covariance of $\xi_u$, one requires the correlator
\eqa{
\label{eq:ucorr}
\bba{u_i u_j u_k u_l} &=& \bba{u^s_i u^s_j u^s_k u^s_l} + 16 \bba{ g_{1i} g_{1j} g_{1k} g_{1l} s_i s_j s_k s_l} + 4 \bba{u^s_i u^s_j g_{1k} g_{1l} s_k s_l} + \mbox{5 perm.}\\ \nn
&=& \sigma_u^4 \br{\delta_{ik} \delta_{jl} + \delta_{il} \delta_{jk} } + 16 \bbc{ \bba{ g_{1i} g_{1j}} \bba{ g_{1k} g_{1l}} + \bba{ g_{1i} g_{1k}} \bba{ g_{1j} g_{1l}} + \bba{ g_{1i} g_{1l}} \bba{ g_{1j} g_{1k}} } \bba{s_i s_j s_k s_l}\\ \nn
&& +\; 4 \sigma_u^2 \bar{s}^2 \bbc{ \delta_{ik}  \bba{ g_{1j} g_{1l}}  + \delta_{il}  \bba{ g_{1j} g_{1k}}  + \delta_{jk}  \bba{ g_{1i} g_{1l}}  + \delta_{jl}  \bba{ g_{1i} g_{1k}} }\\ \nn
&& +\; 4 \sigma_u^2 \sigma_s^2 \bbc{ \delta_{ik} \delta_{jl} \bba{g_{1j}^2} +  \delta_{il} \delta_{jk} \bba{g_{1j}^2} + \delta_{jk} \delta_{il} \bba{g_{1i}^2} + \delta_{jl} \delta_{ik} \bba{g_{1i}^2} }\;,
}
where, in order to arrive at the last equality, we made repeated use of Equations (\ref{eq:u2ptcorr}), (\ref{eq:remaining2ptcorr}), and (\ref{eq:scorr}). Inserting this result into Equation (\ref{eq:covgeneraldef}) leads to
\eqa{
\nn
\cov{\xi_u(x_1)}{\xi_u(x_2)} &=& \delta_{x_1 x_2}\; \bbc{ \frac{2 \sigma_u^4}{N_p(x_1)} + \frac{16 \sigma_u^2 \sigma_s^2}{N_p(x_1)^2} \sum_{i,j} \Delta_{x_1}(i,j) \bba{g_{1i}^2} + \frac{32 \sigma_s^4}{N_p(x_1)^2} \sum_{i,j} \Delta_{x_1}(i,j) \bb{ \bba{g_{1i}^2} \bba{g_{1j}^2} + 2 \bba{g_{1i} g_{1j}}^2 } }\\ \nn
&& \hspace*{-3.5cm} +\; \frac{16 \sigma_u^2 \bar{s}^2}{N_p(x_1) N_p(x_2)} \sum_{i,j,k} \Delta_{x_1}(i,j) \Delta_{x_2}(i,k) \bba{g_{1j} g_{1k}} + \frac{64 \sigma_s^2 \bar{s}^2}{N_p(x_1) N_p(x_2)} \sum_{i,j,k} \Delta_{x_1}(i,j) \Delta_{x_2}(i,k) \bb{ \bba{g_{1i}^2} \bba{g_{1j} g_{1k}}  + 2 \bba{g_{1i} g_{1j}} \bba{g_{1i} g_{1k}} }\\ 
\label{eq:uucov}
&& \hspace*{-3.5cm} +\; \frac{32 \bar{s}^4}{N_p(x_1) N_p(x_2)} \sum_{i,j,k,l} \Delta_{x_1}(i,j) \Delta_{x_2}(k,l) \bba{g_{1i} g_{1k}} \bba{g_{1j} g_{1l}}\;.
}
Subsequent calculations simplify if from here onwards one considers the sum of the $\xi_u$ and $\xi_v$ covariances that appears in Equation (\ref{eq:errorprop}). These covariances are obtained in full analogy to Equations (\ref{eq:ucorr}) and (\ref{eq:uucov}). After inserting Equation (\ref{eq:g11}) into Equation (\ref{eq:g12}), one can proceed to compute the ensemble averages of the various terms by employing the expressions
\eqa{
\frac{E\br{\sum_{i,j} \Delta_{x_1}(i,j)\; F_a(\varphi_{ij})\; F_b(\vek{x}_j-\vek{x}_i)}}{N^2_p(x_1)} &=& \frac{1}{2 \pi N_p(x_1)} \int_0^{2\pi} \dd \varphi\; F_a(\varphi)\; F_b\br{x_1 \cos \varphi; x_1 \sin \varphi}\;;\\
&& \hspace*{-6.5cm} \frac{E\br{\sum_{i,j,k} \Delta_{x_1}(i,j)\; \Delta_{x_2}(i,k)\; F_a(\varphi_{ij})\; F_b(\vek{x}_j-\vek{x}_i)\; F_c(\varphi_{ik})\; F_d(\vek{x}_k-\vek{x}_i)}}{N_p(x_1)\; N_p(x_2)}\\ \nn
&& \hspace*{-3.5cm} = \frac{1}{(2 \pi)^2 n_{\rm g} A_{\rm s}} \int_0^{2\pi} \dd \varphi_1\; F_a(\varphi_1)\; F_b\br{x_1 \cos \varphi_1; x_1 \sin \varphi_1} \int_0^{2\pi} \dd \varphi_2\; F_c(\varphi_2)\; F_d\br{x_2 \cos \varphi_2; x_2 \sin \varphi_2}\;,
}
where $F_{a,b,c,d}$ are arbitrary smooth functions, together with Equations (\ref{eq:ensav1}) and (\ref{eq:ensav2}). In addition we apply Equations (29) and (30) from S02. This results in
\eqa{
\nn 
&& \cov{\xi_u(x_1)}{\xi_u(x_2)} + \cov{\xi_u(x_1)}{\xi_v(x_2)} + \cov{\xi_v(x_1)}{\xi_u(x_2)} + \cov{\xi_v(x_1)}{\xi_v(x_2)}\\ \nn
&=& \delta_{x_1 x_2}\; \frac{4 \sigma_u^4 + 16 \sigma_u^2 \sigma_s^2 \xi_+(0) + 8 \sigma_s^4 \xi_+^2(0)}{N_p(x_1)} + \delta_{x_1 x_2}\; \frac{16 \sigma_s^4}{N_p(x_1)}\; \bb{3 \xi_+^2(x_1) + \xi_-^2(x_1)}\\ \nn
&& + \frac{96\bar{s}^2 \sigma_s^2 \xi_+(x_1) \xi_+(x_2)}{A_{\rm s} n_{\rm g}} + \frac{32 \bar{s}^2 \sigma_s^2 \xi_+(0) + 16 \bar{s}^2 \sigma_u^2}{\pi A_{\rm s} n_{\rm g}}\; \int_0^\pi \dd \varphi\; \xi_+\br{\sqrt{x_1^2 + x_2^2 - 2 x_1 x_2 \cos \varphi}}\\
&& + \frac{32 \bar{s}^4}{\pi A_{\rm s}}\; \int_0^\infty \dd \phi\, \phi \int_0^\pi \dd \varphi_1 \int_0^\pi \dd \varphi_2\; \bbc{\xi_+(\psi_a) \xi_+(\psi_b) + \xi_-(\psi_a) \xi_-(\psi_b) \cos \bb{4(\varphi_a -\varphi_b)} }\;,
}
where $\psi_a$ and $\psi_b$ are defined in equation (31) of S02, and where $\varphi_a$ and $\varphi_b$ are their polar angles. Inserting all covariances computed in this way into Equation (\ref{eq:errorprop}), one finally arrives at
\eqa{
\label{eq:fullcovplus}
\cov{\xi_+(x_1)}{\xi_+(x_2)} &=& \delta_{x_1 x_2}\; \frac{\sigma_u^4 + 2 \sigma_s^4 \xi_+^2(0) + 4 \sigma_u^2 \sigma_s^2 \xi_+(0) + 36 \sigma_s^4 \xi_+^2(x_1) + 4 \sigma_s^4 \xi_-^2(x_1)}{8 \pi \bar{s}^4 A_{\rm s} n_{\rm g}^2 x_1 \Delta x}\\ \nn
&& \hspace*{-3.5cm} +\; \frac{18 \sigma_s^2 \xi_+(x_1) \xi_+(x_2)}{\bar{s}^2 A_{\rm s} n_{\rm g}} + \frac{2 \sigma_s^2 \xi_+(0) + \sigma_u^2}{\pi \bar{s}^2 A_{\rm s} n_{\rm g}}\; \int_0^\pi \dd \varphi\; \xi_+\br{\sqrt{x_1^2 + x_2^2 - 2 x_1 x_2 \cos \varphi}}\\ \nn
&& \hspace*{-3.5cm} +\; \frac{2}{\pi A_{\rm s}}\; \int_0^\infty \dd \phi\, \phi \int_0^\pi \dd \varphi_1 \int_0^\pi \dd \varphi_2\; \bbc{\xi_+(\psi_a) \xi_+(\psi_b) + \xi_-(\psi_a) \xi_-(\psi_b) \cos \bb{4(\varphi_a -\varphi_b)} }\;.
}

Under realistic and weak assumptions, the expression in Equation (\ref{eq:fullcovplus}) can be dramatically simplified. For actual galaxy samples $\sigma_s$ should not exceed $\sigma_u$ by more than factors of a few, so $\sigma_s \sim \sigma_u$ holds. Moreover, $\xi_-(x) \lesssim \xi_+(x) \lesssim \xi_+(0) \lesssim 10^{-3}$, where the first inequality ceases to hold only on very large scales. Then $\sigma_s^2\, \xi_\pm(x) \ll \sigma_u^2$, so that to high accuracy 
\eqa{\nn
\cov{\xi_+(x_1)}{\xi_+(x_2)} &\approx& \delta_{x_1 x_2}\; \frac{\sigma_u^4}{8 \pi \bar{s}^4 A_{\rm s} n_{\rm g}^2 x_1 \Delta x}  + \frac{\sigma_u^2}{\pi \bar{s}^2 A_{\rm s} n_{\rm g}}\; \int_0^\pi \dd \varphi\; \xi_+\br{\sqrt{x_1^2 + x_2^2 - 2 x_1 x_2 \cos \varphi}}\\ 
&& \hspace*{-3.5cm} +\; \frac{2}{\pi A_{\rm s}}\; \int_0^\infty \dd \phi\, \phi \int_0^\pi \dd \varphi_1 \int_0^\pi \dd \varphi_2\; \bbc{\xi_+(\psi_a) \xi_+(\psi_b) + \xi_-(\psi_a) \xi_-(\psi_b) \cos \bb{4(\varphi_a -\varphi_b)} }\;.
}
Thus, the covariance formula reduces to the result for the standard definition of $\xi_+$ as given by S02 when identifying $\sigma_{\epsilon,i}$ with $\sigma_u/(2 \bar{s})$, where $\sigma_{\epsilon,i}$ is the ellipticity dispersion per component. This is equivalent to assuming that the noise in $s$ is negligible, i.e. $\sigma_s \rightarrow 0$, so that $u$ and $v$ can be treated as rescaled versions (by $2 \bar{s}$) of $\epsilon_{1,2}$. Using this argument, it is then straightforward to write down the covariance of $\xi_-$, as well as the cross-covariance between $\xi_+$ and $\xi_-$, based on Equations (35) to (38) of S02.

\end{document}